\documentclass{ws-rv9x6}

\begin{document}

\setcounter{chapter}{0}

\chapter{JORDAN-WIGNER FERMIONIZATION
         AND THE THEORY OF LOW-DIMENSIONAL QUANTUM SPIN MODELS.\\
         DYNAMIC PROPERTIES}

\markboth{O. Derzhko}{Jordan-Wigner fermionization}

\author{Oleg Derzhko}

\address{Institute for Condensed Matter Physics\\
         of the National Academy of Sciences of Ukraine\\
         1 Svientsitskii Street, L'viv-11, 79011, Ukraine\\
         E-mail: derzhko@icmp.lviv.ua}

\begin{abstract}
The Jordan-Wigner transformation is known as a powerful tool in 
condensed matter theory, especially in the theory of low-dimensional
quantum spin systems. The aim of this chapter is to review the
application of the Jordan-Wigner fermionization technique for
calculating dynamic quantities of low-dimensional quantum spin
models. After a brief introduction of the Jordan-Wigner
transformation for one-dimensional spin one-half systems and some of
its extensions for higher dimensions and higher spin values we focus
on the dynamic properties of several low-dimensional quantum spin
models. We start from the famous $s=1/2$ $XX$ chain. As a first step
we recall well-known results for dynamics of the $z$-spin-component
fluctuation operator and then turn to the dynamics of the dimer and
trimer fluctuation operators. The dynamics of the trimer
fluctuations involves both the two-fermion (one particle and one
hole) and the four-fermion (two particles and two holes)
excitations. We discuss some properties of the two-fermion and
four-fermion excitation continua. The four-fermion dynamic
quantities are of intermediate complexity between simple two-fermion
(like the $zz$ dynamic structure factor) and enormously complex
multi-fermion (like the $xx$ or $xy$ dynamic structure factors)
dynamic quantities. Further we discuss the effects of dimerization,
anisotropy of $XY$ interaction, and additional Dzyaloshinskii-Moriya
interaction on various dynamic quantities. Finally we consider the
dynamic transverse spin structure factor $S_{zz}({\bf{k}},\omega)$
for the $s=1/2$ $XX$ model on a spatially anisotropic square lattice
which allows one to trace a one-to-two-dimensional crossover in
dynamic quantities.
\end{abstract}

\section{Introduction (Spin Models, Dynamic Probes etc.)
         \label{secdk1}}

The subject of quantum magnetism dates back to 1920s.
E.~Ising\cite{001} suggested a simplest model of a magnet
as a collection of $N$ spins
which may acquire two values $\sigma=\pm 1$
and interact with nearest neighbors on a lattice
as $\sum J\sigma_i\sigma_j$
and with an external magnetic field
as $-h\sum\sigma_i$.
To explain the properties of the model we have to calculate
the partition function
$Z={\rm{Tr}}\exp(-\beta H)$
which yields the Helmholtz free energy per site
$ f=\lim_{N\to\infty}\left(-T\ln Z/N\right)$
(in what follows we set $k_{{\rm{B}}}=1$ to simplify the notations).
In one dimension the problem was solved by E.~Ising.
Later L.~Onsager solved the square-lattice Ising model\cite{002}
and we know the solution in two dimensions\cite{baxter}.
There is no solution of the Ising model in three dimensions until now.

Another version of interspin interaction
was suggested by P.~A.~M.~Dirac and W.~Heisenberg.
The Heisenberg exchange interaction reads
$\sum J\vec{\sigma}_i\cdot \vec{\sigma_j}
=\sum J\left(\sigma_i^x\sigma_j^x+\sigma_i^y\sigma_j^y+\sigma_i^z\sigma_j^z\right)$
where the Pauli matrices
$\vec{\sigma}=(\sigma^x,\sigma^y,\sigma^z)$
are defined as
\begin{eqnarray}
\sigma^x
=
\left(
\begin{array}{cc}
0 & 1 \\
1 & 0
\end{array}
\right),
\;\;\;
\sigma^y
=
\left(
\begin{array}{cc}
0          & -{\mbox{i}} \\
{\mbox{i}} & 0
\end{array}
\right),
\;\;\;
\sigma^z
=
\left(
\begin{array}{cc}
1 &  0 \\
0 & -1
\end{array}
\right).
\label{1.01}
\end{eqnarray}
Denoting the halves of the Pauli matrices as $s^{\alpha}=\sigma^{\alpha}/2$
(in what follows we set $\hbar=1$ to simplify the notations)
we consider the following Hamiltonian
\begin{eqnarray}
H
=\sum_{\langle i,j \rangle}
\left(
J^xs_{i}^xs_{j}^x+J^ys_{i}^ys_{j}^y+J^ys_{i}^zs_{j}^z
\right)
-h\sum_{i}s^z_i.
\label{1.02}
\end{eqnarray}
We note that the Hamiltonian of the anisotropic $XYZ$ Heisenberg model (\ref{1.02})
covers in the limiting cases some specific models like
the Ising model ($J^x=J^y=0$),
the isotropic $XY$ (or $XX$ or $XX0$) model ($J^x=J^y$, $J^z=0$),
the anisotropic $XY$ model ($J^x\ne J^y$, $J^z=0$),
the isotropic ($XXX$) Heisenberg model ($J^x=J^y=J^z$),
and
the Heisenberg-Ising ($XXZ$) model ($J^x=J^y=J$, $J^z=\Delta\, J$).

Again we would like to calculate the partition function $Z$
of the spin-1/2 model (\ref{1.02}).
Unfortunately, this task is very complicated even in one dimension.
Due to H.~Bethe
we know how to find the eigenstates
of the spin-1/2 linear chain Heisenberg model\cite{003}.
The famous Bethe ansatz for the wave function has the form
\begin{eqnarray}
\vert \psi\rangle
=\sum_{1\le n_1<\ldots<n_r\le N}
a(n_1,\ldots,n_r)s^-_{n_1}\ldots s^{-}_{n_r}
\vert\uparrow\uparrow\ldots\uparrow\rangle,
\nonumber\\
a(n_1,\ldots,n_r)
=\sum_{{\cal{P}}\in S_r}
\exp
\left({\rm{i}}\sum_{j=1}^r k_{{\cal{P}}_j}n_j
+\frac{{\rm{i}}}{2}\sum_{i<j}\theta_{{\cal{P}}_i{\cal{P}}_j}\right),
\nonumber\\
2\cot\frac{\theta_{ij}}{2}
=
\cot\frac{k_i}{2}-\cot\frac{k_j}{2},
\;\;\;
Nk_i=2\pi\lambda_i+\sum_{j\ne i}\theta_{ij},
\label{1.03}
\end{eqnarray}
where the sum in the definition of coefficients $a(n_1,\ldots,n_r)$,
${\cal{P}}\in S_r$,
runs over all $r!$ permutations of the labels $\{1,2,\ldots,r\}$,
${\cal{P}}_j$ is the image of $j$ under the permutation ${\cal{P}}$.
For further details see, e.g., Ref.~\cite{muller}.

Let us briefly recall the quantities of interest in the statistical mechanical studies
of the spin models.
As we have mentioned already the thermodynamic quantities
like the entropy, the specific heat, the magnetization etc.
follow from the partition function
$Z=\sum_{\lambda}\exp\left(-E_{\lambda}/T\right)={\rm{Tr}}\exp\left(-\beta H\right)$,
the sum runs over all states $\lambda$ of the system  
with energy $E_{\lambda}$.
Usually we are also interested in the equal-time spin correlation functions,
e.g. $\langle\vec{s}_i\cdot\vec{s}_j\rangle$,
$\langle\left(\ldots\right)\rangle
={\rm{Tr}}\left(\exp\left(-\beta H\right)\left(\ldots\right)\right)/Z$;
their nonzero limiting values,
(e.g., $\lim_{\vert i-j\vert\to\infty}\langle\vec{s}_i\cdot\vec{s}_j\rangle$)
may indicate the existence of long-range order in the system.

Within a linear response regime we add to the Hamiltonian $H_0$
a small perturbation
$H_0\to H_0-b(t) B$,
where the external field $b(t)$ couples to the dynamical variable  $B$,
and observe a response of a dynamical variable $A$,
$\langle A(t)\rangle-\langle A\rangle_0
=\int_{-\infty}^{\infty}{\rm{d}}t^{\prime}\chi_{AB}(t-t^{\prime})b(t^{\prime})$
with
$\chi_{AB}(t-t^{\prime})
={\rm{i}}\theta(t-t^{\prime})\langle\left[A(t),B(t^{\prime})\right]\rangle_0$
(here $\theta(x)$ is the Heaviside step function).
The Fourier-transform of the dynamic susceptibility
$\chi_{AB}(t-t^{\prime})$,
$\Re\chi_{AB}(\omega)+{\rm{i}}\Im\chi_{AB}(\omega)$,
is the quantity which can be measured experimentally.
We note that the real and imaginary parts of the dynamic susceptibility
are connected via the dispersion (or Kramers-Kronig) relation.
On the other hand,
the imaginary part of the dynamic susceptibility can be expressed
with the help of the fluctuation-dissipation theorem
through another dynamic quantity, the dynamic structure factor.
Thus,
$S_{AA}(\omega)=\int_{-\infty}^{\infty}{\rm{d}}t\exp\left({\rm{i}}\omega t\right)
\langle A(t) A\rangle
=2\Im\chi_{AA}(\omega)/\left(1-\exp(-\beta\omega)\right)$.

Usually,
the operator $A$ is constructed from the local operator of the considered system $A_n$ as follows:
$A_k=(1/\sqrt{N})\sum_{n=1}^N\exp({\rm{i}}kn)A_n$.
We can also rewrite the dynamic structure factor in the following forms
\begin{eqnarray}
S_{AA}(k,\omega)
=\sum_{l=1}^N\exp\left(-{\rm{i}}kl\right)
\int_{-\infty}^{\infty}{\rm{d}}t\exp\left({\rm{i}}\omega t\right)
\langle A_n(t) A_{n+l}(0)\rangle
\nonumber\\
=2\pi\sum_{\lambda,\lambda^{\prime}}
\frac{\exp\left(-\beta E_{\lambda^{\prime}}\right)}{Z}
\left\vert
\langle\lambda^{\prime}\vert A_k\vert\lambda \rangle
\right\vert^2
\delta\left(\omega-E_{\lambda}+E_{\lambda^{\prime}}\right).
\label{1.04}
\end{eqnarray}
Sometimes it is convenient to make the following change in the first line in Eq. (\ref{1.04}):
$A_n(t)\to A_n(t)-\langle A\rangle$,
$A_{n+l}(0)\to A_{n+l}(0)-\langle A\rangle$.
We also note that in the zero-temperature limit $T=0$ (or $\beta\to\infty$)
the second line in Eq. (\ref{1.04}) becomes simpler,
$S_{AA}(k,\omega)
=2\pi\sum_{\lambda}\left\vert\langle{\rm{GS}}\vert A_k\vert\lambda\rangle\right\vert^2
\delta\left(\omega-\omega_{\lambda}\right)$,
$\omega_{\lambda}=E_{\lambda}-E_{{\rm{GS}}}$.

In what follows we discuss mainly the dynamic properties
of spin-1/2 $XY$ chains;
just for this class of spin models
application of the Jordan-Wigner fermionization approach is most fruitful.
We notice here that recently it has been found
that Cs$_2$CoCl$_4$ is a good realization of the spin-1/2 $XX$ chain\cite{004}
and calculations of the dynamic quantities for the corresponding spin models
might be important for the interpretation 
of the data from dynamic experiments\cite{005}.
As an example of earlier studies
we may mention dynamic experiments on the spin-1/2 $XX$ chain compound PrCl$_3$\cite{006}.

The rest of this chapter is organized as follows.
At first we briefly introduce the Jordan-Wigner transformation
(Sec.~\ref{secdk2})
and concisely discuss some of its generalizations
(Sec.~\ref{secdk3}).
Then we consider in detail the dynamic structure factors
for the spin-1/2 isotropic $XY$ chain in a transverse field
distinguishing the quantities which probe two-fermion, four-fermion and many-fermion excitations
(Sec.~\ref{secdk4}).
Next we examine the dynamics for two slightly more complicated chains:
the dimerized isotropic $XY$ chain
(Sec.~\ref{secdk5})
and the $XY$ chains with the Dzyaloshinskii-Moriya interaction
(Sec.~\ref{secdk6}).
The results obtained for one-dimensional $XY$ spin models do not involve any approximation.
This is not true in the two-dimensional case
for which the Jordan-Wigner approach provides only approximate expressions for dynamic quantities.
We illustrate the Jordan-Wigner fermionization approach in two dimensions
examining some dynamic quantities for the square-lattice spin-1/2 isotropic $XY$ model
(Sec.~\ref{secdk7}).
We end up with a brief summary
(Sec.~\ref{secdk8}).

\section{The Jordan-Wigner Transformation
         \label{secdk2}}

To be specific,
we consider the one-dimensional spin $s=1/2$ $XXZ$ Heisenberg chain with the Hamiltonian
\begin{eqnarray}
H=\sum_{n=1}^NJ\left(s_n^xs_{n+1}^x+s_n^ys_{n+1}^y
+\Delta s_n^zs_{n+1}^z\right)
-h\sum_{n=1}^Ns_n^z;
\label{2.01}
\end{eqnarray}
we imply either periodic or open boundary conditions in Eq. (\ref{2.01}).
Here the spin operators $s^{\alpha}_i$ satisfy the commutation relations
$\left[s_i^{\alpha},s_j^{\beta}\right]
={\rm{i}}\delta_{ij}\epsilon_{\alpha\beta\gamma}s_i^{\gamma}$,
$\epsilon_{\alpha\beta\gamma}$ is the totally antisymmetric Levi-Civita tensor with
$\epsilon_{xyz}=1$.
In particular,
$\left[s_i^{x},s_j^{y}\right]
={\rm{i}}\delta_{ij}s_i^z$ etc.
Obviously, $s^\alpha$ can be viewed as the halves of the Pauli matrices (\ref{1.01}).
After introducing the spin raising and lowering operators
(or the ladder operators)
$s^{\pm}_n=s_n^x\pm {\mbox{i}}s_n^y$
($s_n^x=\left(s_n^++s_n^-\right)/2$,
$s_n^y=\left(s_n^+-s_n^-\right)/2{\rm{i}}$)
the Hamiltonian (\ref{2.01}) becomes
\begin{eqnarray}
H=\sum_nJ\left(\frac{1}{2}\left(s_n^+s_{n+1}^-+s_n^-s_{n+1}^+\right)
\right.
\nonumber\\
\left.
+\Delta \left(s_n^+s_n^--\frac{1}{2}\right)\left(s_{n+1}^+s_{n+1}^--\frac{1}{2}\right)\right)
-h\sum_n\left(s_n^+s_n^--\frac{1}{2}\right).
\label{2.02}
\end{eqnarray}
We note that the spin raising and lowering operators
satisfy commutation relations of Fermi type
at the same site,
i.e.
\begin{eqnarray}
\left\{s_n^-,s_n^+\right\}=1,
\;\;\;
\left\{s_n^-,s_n^-\right\}
=\left\{s_n^+,s_n^+\right\}=0
\label{2.03}
\end{eqnarray}
and of Bose type at different sites
\begin{eqnarray}
\left[s_n^-,s_m^+\right]
=\left[s_n^-,s_m^-\right]
=\left[s_n^+,s_m^+\right]=0,
\;\;\;
n\ne m.
\label{2.04}
\end{eqnarray}

We may use the Jordan-Wigner transformation\cite{007}
to introduce Fermi operators
according to the following formulas
\begin{eqnarray}
c_1=s_1^-,
\;\;\;
c_n=\left(-2s_1^z\right)\left(-2s_2^z\right)
\ldots \left(-2s_{n-1}^z\right)s^-_n,
\;
n=2,\ldots,N,
\label{2.05}
\end{eqnarray}
\begin{eqnarray}
c_1^{\dagger}=s_1^+,\;\;\;
c_n^{\dagger}=\left(-2s_1^z\right)\left(-2s_2^z\right)
\ldots \left(-2s_{n-1}^z\right)s^+_n,
\;
n=2,\ldots,N.
\label{2.06}
\end{eqnarray}
(Sometimes one can find in Eqs. (\ref{2.05}), (\ref{2.06})
instead of $-2s^z$ the identical expressions
$1-2s^+s^-=\exp\left(\pm{\rm{i}}\pi s^+s^-\right)$.)
Really,
the operators introduced always satisfy the Fermi commutation relations
\begin{eqnarray}
\left\{c_n,c_m^{\dagger}\right\}=\delta_{nm},
\;\;\;
\left\{c_n,c_m\right\}
=\left\{c_n^{\dagger},c_m^{\dagger}\right\}=0.
\label{2.07}
\end{eqnarray}
(To check this one has to note that $\left(-2s^z\right)^2=1$
and that $s^zs^{\pm}=-s^{\pm}s^z$.)
The inverse transformation to the one given by Eqs. (\ref{2.05}), (\ref{2.06})
reads
\begin{eqnarray}
s_1^-=c_1,
\;\;\;
s_n^-=\exp\left(\pm{\rm{i}}\pi\sum_{j=1}^{n-1}c_j^{\dagger}c_j\right)c_n,
\;
n=2,\ldots,N,
\label{2.08}
\end{eqnarray}
\begin{eqnarray}
s_1^+=c_1^{\dagger},\;\;\;
s_n^+=\exp\left(\pm{\rm{i}}\pi\sum_{j=1}^{n-1}c_j^{\dagger}c_j\right)c^{\dagger}_n,
\;
n=2,\ldots,N.
\label{2.09}
\end{eqnarray}
Moreover,
the Hamiltonian (\ref{2.02}) in terms of the Fermi operators
(\ref{2.05}), (\ref{2.06})
has the following form
\begin{eqnarray}
H=\sum_n
J\left(\frac{1}{2}\left(c^{\dagger}_nc_{n+1}-c_nc_{n+1}^{\dagger}\right)
\right.
\nonumber\\
\left.
+\Delta\left(c_n^{\dagger}c_n-\frac{1}{2}\right)\left(c_{n+1}^{\dagger}c_{n+1}-\frac{1}{2}\right)
\right)
-h\sum_{n}\left(c_n^{\dagger}c_n-\frac{1}{2}\right)
\label{2.10}
\end{eqnarray}
(we use $c^{\dagger}_jc^{\dagger}_{j+1}
=s_j^+\left(-2s_j^z\right)s_{j+1}^+
=s_j^+s_{j+1}^+$ etc.).
In the case of periodic boundary conditions implied for the spin Hamiltonian (\ref{2.02})
the transformed Hamiltonian (\ref{2.10}) obeys either periodic or antiperiodic boundary conditions
depending on the parity of the number of fermions.
However,
in what follows the calculated quantities in the thermodynamic limit $N\to\infty$
will be insensitive to the boundary conditions implied
(for further details see Ref.~\cite{008}).

From Eq. (\ref{2.10}) it becomes clear that the spin-1/2 isotropic $XY$ chain
in a transverse ($z$) magnetic field with the Hamiltonian
\begin{eqnarray}
H=\sum_nJ\left(s_n^xs_{n+1}^x+s_n^ys_{n+1}^y\right)
+\Omega\sum_{n}s_n^z
\label{2.11}
\end{eqnarray}
in the Jordan-Wigner picture is represented by the Hamiltonian
\begin{eqnarray}
H=\sum_n\frac{J}{2}\left(c_n^{\dagger}c_{n+1}-c_nc_{n+1}^{\dagger}\right)
+\Omega\sum_{n}\left(c_n^{\dagger}c_n-\frac{1}{2}\right)
\label{2.12}
\end{eqnarray}
and therefore is an exactly solvable model\cite{009,010}.
Moreover, the $XY$ exchange interaction may be anisotropic;
then the intersite interaction has the form
\begin{eqnarray}
J^xs_n^xs_{n+1}^x+J^ys_n^ys_{n+1}^y
\nonumber\\
\to
\frac{J}{2}
\left(c_n^{\dagger}c_{n+1}-c_nc_{n+1}^{\dagger}\right)
+\frac{\gamma}{2}
\left(c_n^{\dagger}c_{n+1}^{\dagger}-c_nc_{n+1}\right)
\label{2.13}
\end{eqnarray}
with $J=(J^x+J^y)/2$, $\gamma=(J^x-J^y)/2$.
We can also consider an additional intersite interaction,
the so-called Dzyaloshinskii-Moriya interaction,
which does not spoil a simple fermionic Hamiltonian\cite{011}
\begin{eqnarray}
D\left(s_n^xs_{n+1}^y-s_n^ys_{n+1}^x\right)
\to
\frac{{\rm{i}}D}{2}
\left(c_n^{\dagger}c_{n+1}+c_nc_{n+1}^{\dagger}\right).
\label{2.14}
\end{eqnarray}
Moreover,
within the Jordan-Wigner fermionization approach we can examine rigorously
some types of multi-spin interactions\cite{012},
for example,
\begin{eqnarray}
s_n^xs_{n+1}^zs^x_{n+2}+s_n^ys_{n+1}^zs^y_{n+2}
\to
-\frac{1}{4}
\left(c_n^{\dagger}c_{n+2}-c_nc_{n+2}^{\dagger}\right);
\nonumber\\
s_n^xs_{n+1}^zs^y_{n+2}-s_n^ys_{n+1}^zs^x_{n+2}
\to
-\frac{{\rm{i}}}{4}
\left(c_n^{\dagger}c_{n+2}+c_nc_{n+2}^{\dagger}\right).
\label{2.15}
\end{eqnarray}
Within the frames of the Jordan-Wigner approach
we can also generalize simple spin-1/2 $XY$ chains
assuming
regularly alternating Hamiltonian parameters\cite{013}
or
some types of random Hamiltonian parameters\cite{014}
and still face exactly solvable models.

On the other hand,
as can be easily seen from Eq. (\ref{2.10})
the Ising interaction between $z$ spin components
leads to interacting spinless fermions
and as a result the advantages of fermionization are less evident.
(Obviously, we can split the interaction term in the spirit of the Hartree-Fock approximation\cite{015},
however, the resulting theory will be only an approximate one.
On the other hand,
in the low-energy limit we can bosonize the fermionic Hamiltonian
obtaining an exact low-energy effective theory\cite{016,017,018}.)
We cannot examine rigorously within the Jordan-Wigner fermionization approach
the case of the next-nearest-neighbor interaction
since
\begin{eqnarray}
s_n^xs_{n+2}^x+s_n^ys_{n+2}^y
\to
c_n^{\dagger}
\left(1-2c_{n+1}^{\dagger}c_{n+1}\right)c_{n+2}
-
c_n
\left(1-2c_{n+1}^{\dagger}c_{n+1}\right)c_{n+2}^{\dagger}.
\label{2.16}
\end{eqnarray}
It is worthwhile to note here
that recently the Jordan-Wigner fermionization approach has been applied
to the spin-1/2 isotropic $XY$ model on a diamond chain\cite{019},
however,
the authors of that paper apparently missed some interaction terms in the fermionic Hamiltonian
and their statement about rigorous results for such a model is wrong.
Finally we note that an external magnetic field directed along $x$ or $y$ axes
has an enormously complicated form in the Jordan-Wigner picture.

\section{Generalization of the Jordan-Wigner Transformation
         \label{secdk3}}

The Jordan-Wigner fermionization is a powerful tool for the study of quantum spin chains.
Since the late 1980s there were several attempts
to extend this approach
to two (and three) dimensions\cite{020,021,022,023}
as well as to spin values $s>1/2$\cite{024,025,026}.
For a review on the two-dimensional Jordan-Wigner fermionization approach see also Ref.~\cite{027}.

Bearing in mind the Jordan-Wigner transformation in one dimension as a guideline
we consider in the two-dimensional case
the following relation between spin $s=1/2$ and Fermi operators
\begin{eqnarray}
d_{\vec{i}}=\exp\left(-{\rm{i}}\alpha_{\vec{i}}\right)s^-_{\vec{i}},
\;\;\;
d_{\vec{i}}^{\dagger}=\exp\left({\rm{i}}\alpha_{\vec{i}}\right)s^+_{\vec{i}},
\nonumber\\
s_{\vec{i}}^-=\exp\left({\rm{i}}\alpha_{\vec{i}}\right)d_{\vec{i}},
\;\;\;
s^+_{\vec{i}}=\exp\left(-{\rm{i}}\alpha_{\vec{i}}\right)d_{\vec{i}}^{\dagger},
\nonumber\\
\alpha_{\vec{i}}=\sum_{\vec{j}(\ne \vec{i})}B_{\vec{i}\vec{j}}d^{\dagger}_{\vec{j}}d_{\vec{j}}.
\label{3.01}
\end{eqnarray}
Here $d$, $d^{\dagger}$ are the Fermi operators,
the operators $s^{\pm}$ defined according to (\ref{3.01}) commute at different sites
if the c-number matrix $B_{\vec{i}\vec{j}}$ satisfies the relation
\begin{eqnarray}
\exp\left({\rm{i}}B_{\vec{i}\vec{j}}\right)
=
-\exp\left({\rm{i}}B_{\vec{j}\vec{i}}\right).
\label{3.02}
\end{eqnarray}

There are many choices of the matrix $B_{\vec{i}\vec{j}}$
which realize the two-dimensional Jordan-Wigner transformation.
Following Y.~R.~Wang\cite{021}
we use the Cartesian coordinates $\vec{i}=(i_x,i_y)$
to construct a complex number
$\tau_{\vec{i}}=i_x+{\rm{i}}i_y
=\vert\tau_{\vec{i}}\vert\,\exp\left({\rm{i}}\,\arg(\tau_{\vec{i}})\right)$
and then choose
\begin{eqnarray}
B_{\vec{i}\vec{j}}
=
\arg\left(\tau_{\vec{j}}-\tau_{\vec{i}}\right)
=\Im\ln\left(\tau_{\vec{j}}-\tau_{\vec{i}}\right)
=\Im\ln\left(j_x-i_x+{\rm{i}}(j_y-i_y)\right).
\label{3.03}
\end{eqnarray}
Indeed, 
for such a choice Eq. (\ref{3.02}) is satisfied,
$\exp\left({\rm{i}}B_{\vec{j}\vec{i}}\right)
=\exp\left({\rm{i}}\arg(\tau_{\vec{i}}-\tau_{\vec{j}})\right)
=\exp\left({\rm{i}}\left(\arg(\tau_{\vec{j}}-\tau_{\vec{i}})\pm\pi\right)\right)
=-\exp\left({\rm{i}}B_{\vec{i}\vec{j}}\right)$.
Another choice of the matrix $B_{\vec{i}\vec{j}}$ has the following form\cite{022}
\begin{eqnarray}
B_{\vec{i}\vec{j}}
=\pi\left(\theta\left(i_x-j_x\right)\left(1-\delta_{i_x,j_x}\right)
+\delta_{i_x,j_x}\theta\left(i_y-j_y\right)\left(1-\delta_{i_y,j_y}\right)\right);
\label{3.04}
\end{eqnarray}
here $\theta(x)$ is the Heaviside step function
(see also Ref.~\cite{027}).

After performing the Jordan-Wigner transformation (\ref{3.01})
for the two-dimensional spin-1/2 $XXZ$ Heisenberg Hamiltonian one gets
\begin{eqnarray}
H=\sum_{\langle\vec{i},\vec{j}\rangle}
\left(\frac{J_{\vec{i}\vec{j}}}{2}
\left(
d_{\vec{i}}^{\dagger}
\exp\left({\rm{i}}\left(\alpha_{\vec{j}}-\alpha_{\vec{i}}\right)\right)
d_{\vec{j}}
+
d_{\vec{i}}
\exp\left({\rm{i}}\left(\alpha_{\vec{i}}-\alpha_{\vec{j}}\right)\right)
d^{\dagger}_{\vec{j}}
\right)
\right.
\nonumber\\
\left.
+J_{\vec{i}\vec{j}}\Delta
\left(d_{\vec{i}}^{\dagger}d_{\vec{i}}-\frac{1}{2}\right)
\left(d_{\vec{j}}^{\dagger}d_{\vec{j}}-\frac{1}{2}\right)
\right)
\label{3.05}
\end{eqnarray}
with
\begin{eqnarray}
\alpha_{\vec{j}}-\alpha_{\vec{i}}
=\int_{\vec{i}}^{\vec{j}}{\rm{d}}\vec{r}\cdot\vec{A}(\vec{r}),
\nonumber\\
\vec{A}(\vec{r})
=\vec{\nabla}\alpha_{\vec{r}}
=-\sum_{\vec{r}^{\prime}(\ne\vec{r})}
\frac{\vec{n}_z\times\left(\vec{r}^{\prime}-\vec{r}\right)}{\left(\vec{r}^{\prime}-\vec{r}\right)^2}
d^{\dagger}_{\vec{r}^{\prime}}d_{\vec{r}^{\prime}}
\label{3.06}
\end{eqnarray}
(we have used Eq. (\ref{3.03}) for $\alpha_{\vec{r}}$ (\ref{3.01})).
We need further approximations
to proceed with statistical mechanics calculations for the Hamiltonian (\ref{3.05}).
Within the mean-field description one assumes
$d^{\dagger}_{\vec{r}}d_{\vec{r}}
\to \langle d^{\dagger}_{\vec{r}}d_{\vec{r}}\rangle
=\langle s^z_{\vec{r}}\rangle+1/2\to 1/2$.
We expect such an approximation to be valid in the case of zero magnetic field.
For the mean-field description in the case of nonzero magnetic field
and an analysis of the magnetization processes in the spin system see Ref.~\cite{028}.
We also note that a more sophisticated
(self-consistent site-dependent)
mean-field treatment
has been suggested as well\cite{023}.

After adopting the mean-field approach we face the problem of particles in a magnetic field
with the flux per elementary plaquette $\Phi_0=\pi$.
We may change the gauge preserving the flux per elementary plaquette
to make the Hamiltonian more convenient for further calculations.
For example,
for a square lattice we have
\begin{eqnarray}
H=
\sum_{\langle\vec{i},\vec{j}\rangle}
\left(\frac{J_{\vec{i}\vec{j}}}{2}
\left(
d_{\vec{i}}^{\dagger}d_{\vec{j}}
-
d_{\vec{i}}d^{\dagger}_{\vec{j}}
\right)
+J_{\vec{i}\vec{j}}\Delta
\left(d_{\vec{i}}^{\dagger}d_{\vec{i}}-\frac{1}{2}\right)
\left(d_{\vec{j}}^{\dagger}d_{\vec{j}}-\frac{1}{2}\right)
\right)
\nonumber\\
J_{i_x,i_y;i_x+1,i_y}=-J,
\nonumber\\
J_{i_x,i_y;i_x,i_y+1}
=J_{i_x+1,i_y;i_x+2,i_y}
=J_{i_x+1,i_y;i_x+1,i_y+1}=J.
\label{3.07}
\end{eqnarray}
In the one-dimensional case when either vertical or horizontal bonds vanish
the Hamiltonian (\ref{3.07}) transforms into Eq. (\ref{2.10}) (with $h=0$).

\begin{figure}[th]
\centerline{\psfig{file=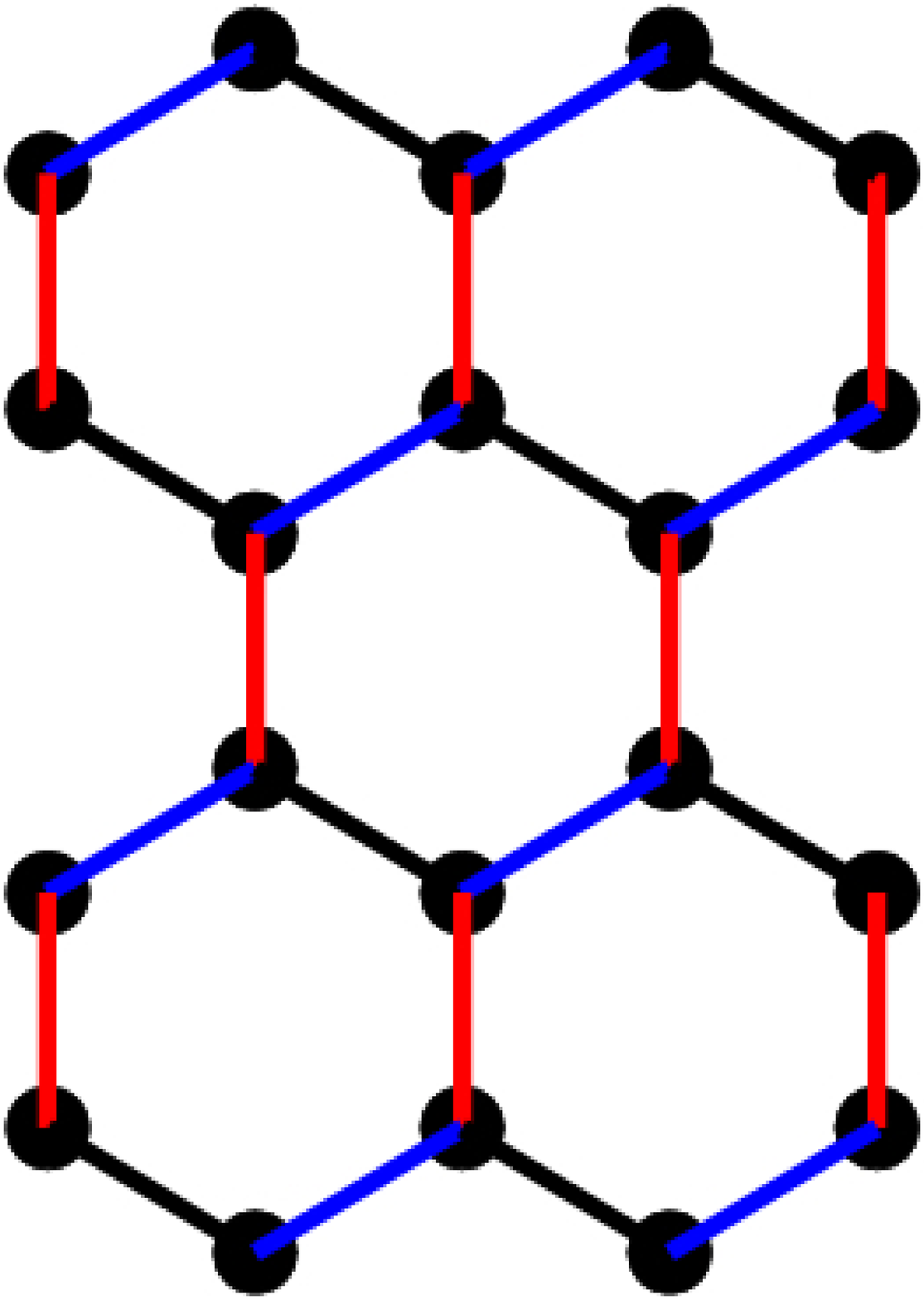,height=1.4in,angle=0}}
\vspace{0.2in}
\centerline{\psfig{file=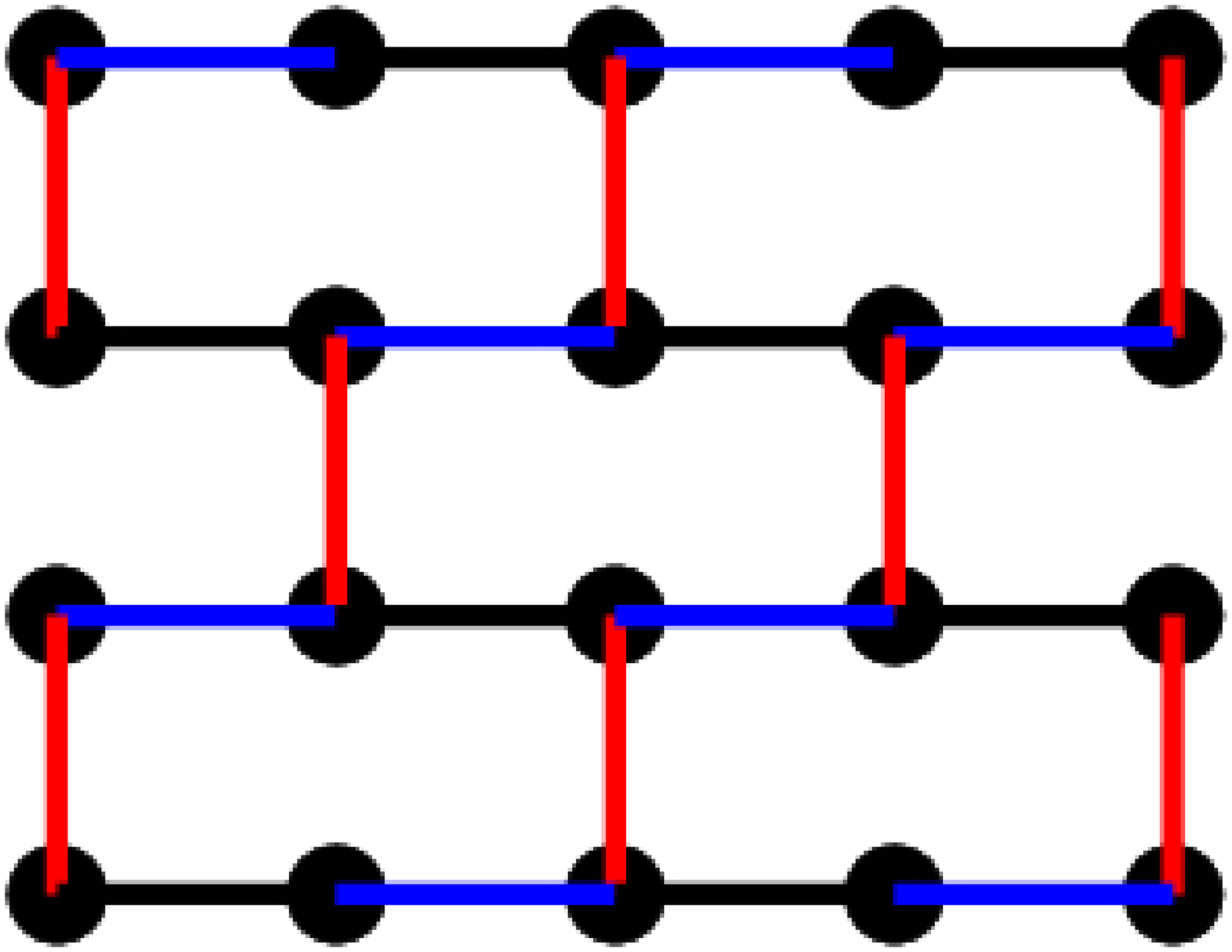,width=1.15in,angle=0}}
\vspace*{8pt}
\caption
{A honeycomb lattice (up) with its equivalent brick-wall lattice (down).
The bonds $J_1$ run from south-west to north-east,
the bonds $J_2$ run from south-east to north-west,
the bonds $J_3$ run from south to north.}
\label{fig01}
\end{figure}

Recently
A.~Kitaev has suggested a new exactly solvable two-dimensional quantum spin model\cite{029}.
This is a spin-1/2 model on a honeycomb lattice
with interactions between different components of neighboring spins along differently directed bonds.
An alternative representation of the honeycomb lattice is a brick-wall lattice
(see Fig.~\ref{fig01}).
The Hamiltonian of the model reads
\begin{eqnarray}
H=
\sum_{j+l={\rm{even}}}
\left(
J_1s_{j,l}^xs_{j+1,l}^x
+J_2s_{j-1,l}^ys_{j,l}^y
+J_3s_{j,l}^zs_{j,l+1}^z
\right);
\label{3.08}
\end{eqnarray}
$j$ and $l$ denote the column and row indices of the lattice.
We discuss in what follows a fermionic representation for the Kitaev model\cite{030}.
Let us perform the Jordan-Wigner transformation
\begin{eqnarray}
s_{j,l}^+
=a_{j,l}^{\dagger}
\exp\left({\rm{i}}\pi
\left(\sum_{i}\sum_{k<l}a_{ik}^{\dagger}a_{ik}+\sum_{i<j}a_{il}^{\dagger}a_{il}\right)
\right)
\label{3.09}
\end{eqnarray}
(compare with Eqs. (\ref{3.01}), (\ref{3.04})).
As a result we find that
\begin{eqnarray}
J_1s_{j,l}^xs_{j+1,l}^x
\to
\frac{J_1}{4}\left(
a_{j,l}^{\dagger}a_{j+1,l}^{\dagger}+a_{j,l}^{\dagger}a_{j+1,l}
-a_{j,l}a_{j+1,l}^{\dagger}-a_{j,l}a_{j+1,l}
\right),
\nonumber\\
J_2s_{j-1,l}^ys_{j,l}^y
\to
\frac{J_2}{4}\left(
-a_{j-1,l}^{\dagger}a_{j,l}^{\dagger}+a_{j-1,l}^{\dagger}a_{j,l}
-a_{j-1,l}a_{j,l}^{\dagger}+a_{j-1,l}a_{j,l}
\right),
\nonumber\\
J_3s_{j,l}^zs_{j,l+1}^z
\to
J_3\left(a_{j,l}^{\dagger}a_{j,l}-\frac{1}{2}\right)
\left(a_{j,l+1}^{\dagger}a_{j,l+1}-\frac{1}{2}\right).
\label{3.10}
\end{eqnarray}
Next we introduce the following operators
\begin{eqnarray}
c_{j,l}=a_{j,l}^{\dagger}+a_{j,l},
\;\;\;
d_{j,l}={\rm{i}}\left(a_{j,l}^{\dagger}-a_{j,l}\right),
\;\;\;
j+l={\rm{odd}};
\nonumber\\
c_{j,l}={\rm{i}}\left(a_{j,l}^{\dagger}-a_{j,l}\right),
\;\;\;
d_{j,l}=a_{j,l}^{\dagger}+a_{j,l},
\;\;\;
j+l={\rm{even}}.
\label{3.11}
\end{eqnarray}
In terms of these operators the Hamiltonian reads as follows
\begin{eqnarray}
H=\sum_{j+l={\rm{even}}}
\left(
-{\rm{i}}\frac{J_1}{4}c_{j,l}c_{j+1,l}
+{\rm{i}}\frac{J_2}{4}c_{j-1,l}c_{j,l}
+{\rm{i}}\frac{J_3}{4}D_{j,l}c_{j,l}c_{j,l+1}
\right).
\label{3.12}
\end{eqnarray}
Since $D_{j,l}={\rm{i}}d_{j,l}d_{j,l+1}$ are good quantum numbers
the Hamiltonian (\ref{3.12})
corresponds to a model of spinless fermions with local static $Z_2$ gauge fields.
Thus,
Eq. (\ref{3.12}) explains a hidden simple structure of the spin model (\ref{3.08}).

The generalizations of the Jordan-Wigner transformation for arbitrary spin values
were discussed by several authors\cite{024,025,026},
however,
these mappings have not yet provided a substantial break-through
for difficult strongly correlated problems.

\section{Spin-1/2 Isotropic $XY$ Chain in a Transverse Field:
         Dynamic Quantities
         \label{secdk4}}

We start with the simplest spin-1/2 $XY$ model,
the transverse $XX$ chain,
with the Hamiltonian (\ref{2.11}).
After performing the Jordan-Wigner transformation we arrive at
a tight-binding model for spinless fermions (\ref{2.12})
and after performing the Fourier transformation,
$c_k=\left(1/\sqrt{N}\right)\sum_{n=1}^N\exp\left({\rm{i}}kn\right)c_n$
($k=2\pi n/N$ if the number of fermions is odd
or
$k=2\pi(n+1/2)/N$ if the number of fermions is even,
$n=-N/2,-N/2+1,\ldots,N/2-1$ if $N$ is even
or
$n=-(N-1)/2,-(N-1)/2+1,\ldots,(N-1)/2$ if $N$ is odd),
the Hamiltonian (\ref{2.12}) becomes diagonal
\begin{eqnarray}
H=\sum_{k}\Lambda_k\left(c_k^{\dagger}c_k-\frac{1}{2}\right),
\;\;\;
\Lambda_k=\Omega+J\cos k.
\label{4.01}
\end{eqnarray}
As it has been mentioned above,
for the analytical calculations discussed below
we may consider only periodic boundary conditions for the fermionic Hamiltonian
(i.e. $k=2\pi n/N$ in Eq. (\ref{4.01})).

\subsection{Two-fermion excitations}

We begin with the transverse dynamic structure factor
$S_{zz}(k,\omega)$ (\ref{1.04})\cite{031,032,033}.
The calculation of the $zz$ time-dependent spin correlation function
is straightforward.
After exploiting the Jordan-Wigner transformation we have
$\langle s_n^z(t)s_{n+l}^z\rangle
-\langle s_n^z\rangle\langle s_{n+l}^z\rangle
=\langle c_n^{\dagger}(t)c_n(t)c_{n+l}^{\dagger}c_{n+l}\rangle
-\langle c_n^{\dagger}c_n\rangle\langle c_{n+l}^{\dagger}c_{n+l}\rangle$.
Here $c_n^{\dagger}(t)=\left(1/\sqrt{N}\right)
\sum_k\exp\left({\rm{i}}kn\right)c_k^{\dagger}(t)$
and
$c_k^{\dagger}(t)=c_k^{\dagger}\exp\left({\rm{i}}\Lambda_k t\right)$.
Next we have to use the Wick-Bloch-de Dominicis theorem,
$\langle c_{k_1}^{\dagger}c_{k_2}c^{\dagger}_{k_3}c_{k_4} \rangle
=\langle c_{k_1}^{\dagger}c_{k_2}\rangle\langle c^{\dagger}_{k_3}c_{k_4} \rangle
-\langle c_{k_1}^{\dagger}c^{\dagger}_{k_3}\rangle\langle c_{k_2}c_{k_4} \rangle
+\langle c_{k_1}^{\dagger}c_{k_4}\rangle\langle c_{k_2}c^{\dagger}_{k_3} \rangle$,
and to calculate the elementary contractions introducing the Fermi function
$n_k=1/\left(1+\exp\left(\beta\Lambda_k\right)\right)$,
$\langle c_{k_1}^{\dagger}c_{k_2}\rangle=\delta_{k_1k_2}n_{k_1}$,
$\langle c_{k_1}^{\dagger}c_{k_2}^{\dagger}\rangle=0$.
As a result,
the final expression for the $zz$ time-dependent spin correlation function reads
\begin{eqnarray}
\langle s_n^z(t)s_{n+l}^z\rangle
-\langle s^z\rangle^2
=
\frac{1}{N^2}\sum_{k_1,k_2}\exp\left(-{\rm{i}}\left(k_1-k_2\right)l\right)
\nonumber\\
\cdot
\exp\left({\rm{i}}\left(\Lambda_{k_1}-\Lambda_{k_2}\right)t\right)
n_{k_1}\left(1-n_{k_2}\right),
\nonumber\\
\nonumber\\
\langle s^z\rangle
=\frac{1}{N}\sum_{n=1}^N\langle s^z_n\rangle
=-\frac{1}{2N}\sum_k\tanh\frac{\beta\Lambda_k}{2}.
\label{4.02}
\end{eqnarray}
Plugging Eq. (\ref{4.02}) into Eq. (\ref{1.04})
we get the desired transverse dynamic structure factor
\begin{eqnarray}
S_{zz}(k,\omega)
=\sum_{l=1}^N\exp(-{\rm{i}}k l)
\int_{-\infty}^{\infty}{\rm{d}}t\exp({\rm{i}}\omega t)
\langle \left(s_n^z(t)-\langle s^z\rangle\right)\left(s_{n+l}^z-\langle s^z\rangle\right)\rangle
\nonumber\\
=
\int_{-\pi}^{\pi}{\rm{d}}k_1n_{k_1}\left(1-n_{k_1+k}\right)
\delta\left(\omega+\Lambda_{k_1}-\Lambda_{k_1+k}\right)
\nonumber\\
=\sum_{k^\star}
\frac{n_{k^{\star}}\left(1-n_{k+k^{\star}}\right)}
{2\left\vert J\sin\frac{k}{2}\cos\left(\frac{k}{2}+k^{\star}\right)\right\vert}
\label{4.03}
\end{eqnarray}
where $-\pi\le k^{\star}<\pi$ are the solutions of the equation
$\omega=-2J\sin(k/2)\sin(k/2+k^{\star})$.

The $zz$ dynamic structure factor (\ref{4.03}) is governed exclusively by a two-fermion
(one particle and one hole)
excitation continuum.
The properties of the two-fermion excitation continuum were discussed by G.~M\"{u}ller et al\cite{032};
we present these results briefly below.
The boundaries of the two-fermion continuum
in the plane wave-vector $k$ -- frequency $\omega$
(we assume $\omega\ge 0$, $-\pi\le k<\pi$)
are determined by the equations
\begin{eqnarray}
\omega=-\Lambda_{k_1}+\Lambda_{k_2},
\;
k=-k_1+k_2(\mod(2\pi)),
\;
\Lambda_k=\Omega+J\cos k,
\label{4.04}
\end{eqnarray}
where $-\pi\le k_1<\pi$.
Moreover,
in the ground state we have to require in addition
$n_{k_1}> 0$
and
$1-n_{k_2}> 0$,
i.e.
$\Lambda_{k_1}\le 0$
and
$\Lambda_{k_2}\ge 0$.

\begin{figure}[th]
\centerline{\psfig{file=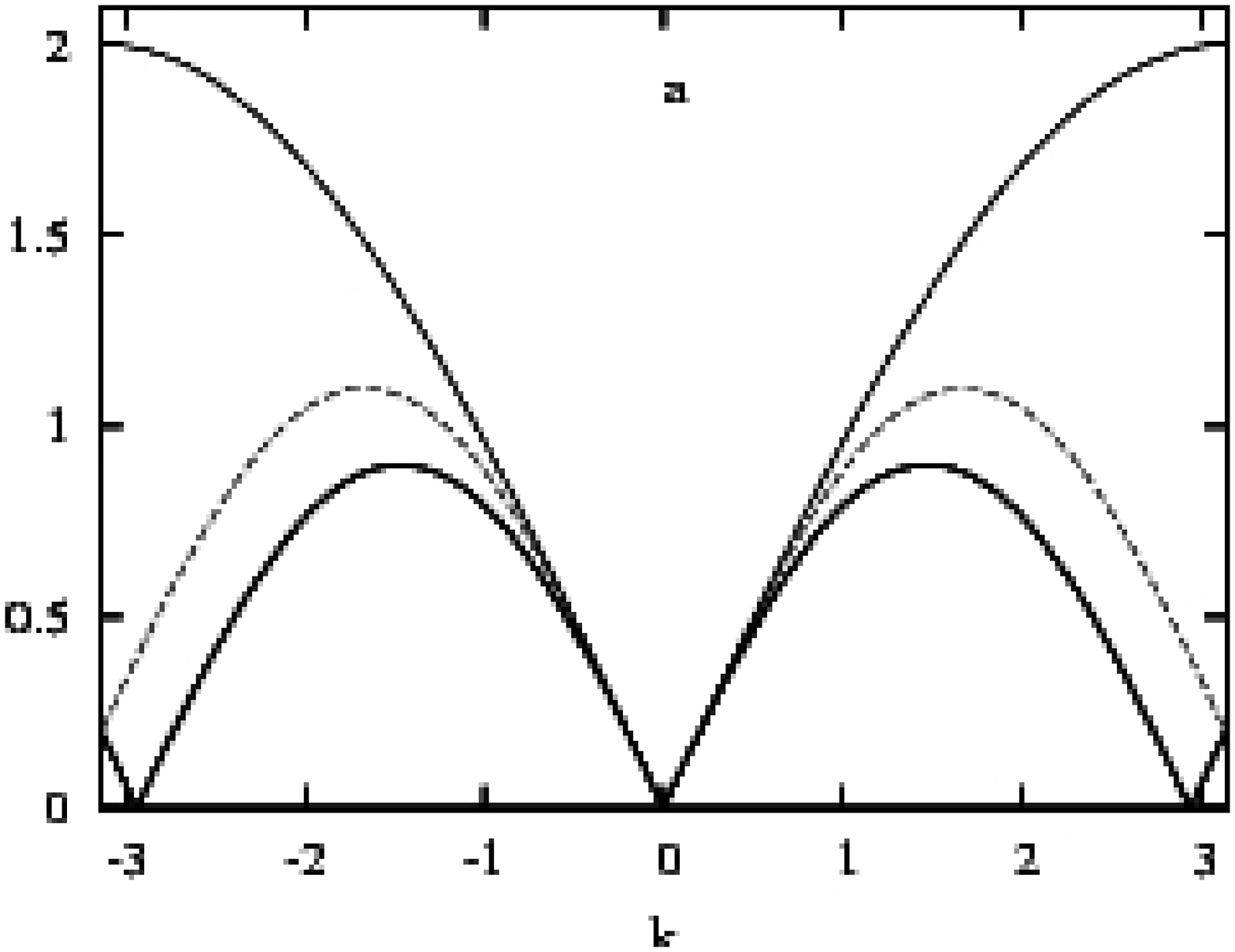,width=2.0in,angle=0}
\psfig{file=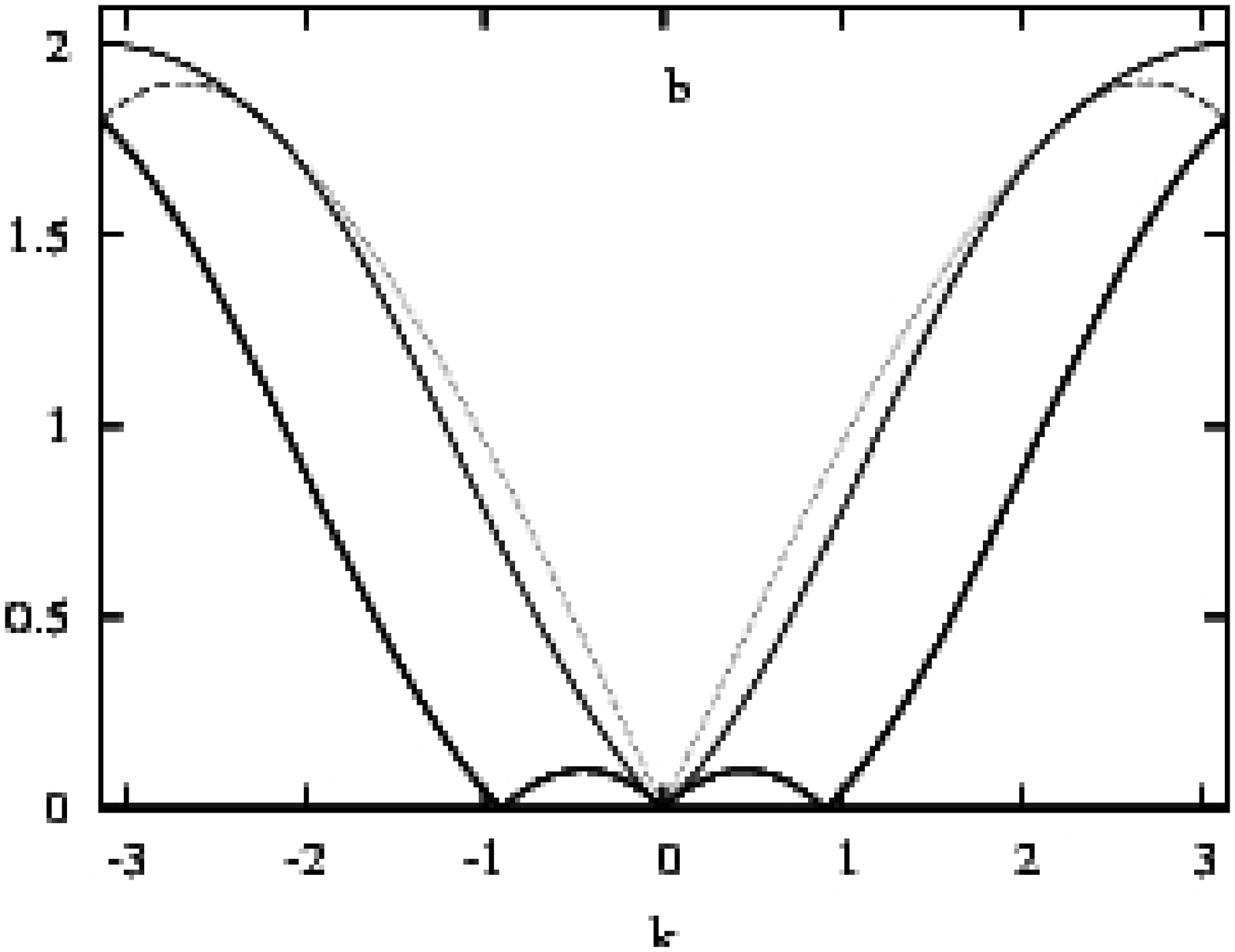,width=2.0in,angle=0}}
\vspace*{8pt}
\caption{The two-fermion excitation continuum which governs the ground-state
two-fermion dynamic quantities.
$\vert J\vert=1$,
$\vert\Omega\vert=0.1$ (a),
$\vert\Omega\vert=0.9$ (b).
We show
the lower boundaries (bold lines),
the middle boundaries (dashed lines),
the upper boundaries (thin lines)
and
the lines of potential singularities (dotted lines).}
\label{fig02}
\end{figure}

We start with the zero-temperature case.
In this case the two-fermion excitation continuum exists
as long as $\vert\Omega\vert<\vert J\vert$.
Let us introduce the parameter
$\alpha=\arccos\left(\vert\Omega\vert/\vert J\vert\right)$
and the following characteristic lines in the $k$--$\omega$ plane
\begin{eqnarray}
\frac{\omega_1(k)}{\vert J\vert}
=2\left\vert\sin\frac{k}{2}\sin\left(\frac{\vert k\vert}{2}-\alpha\right)\right\vert,
\label{4.05}
\\
\frac{\omega_2(k)}{\vert J\vert}
=2\left\vert\sin\frac{k}{2}\sin\left(\frac{\vert k\vert}{2}+\alpha\right)\right\vert,
\label{4.06}
\\
\frac{\omega_3(k)}{\vert J\vert}
=2\left\vert\sin\frac{k}{2}\right\vert.
\label{4.07}
\end{eqnarray}
The two-fermion dynamic quantities in the ground state
may have non-zero values only within a restricted region of the $k$--$\omega$ plane
with the lower boundary $\omega_l(k)=\omega_1(k)$
and the upper boundary $\omega_u(k)=\omega_2(k)$ if $\vert k\vert\le\pi-2\alpha$
or $\omega_u(k)=\omega_3(k)$ if $\pi-2\alpha\le\vert k\vert$.
Obviously,
the two-fermion dynamic quantities may have only three soft modes $k_0=\{0,\pm 2\alpha\}$.
Moreover,
there is a middle boundary of the two-fermion excitation continuum
$\omega_m(k)=\omega_2(k)$ if $\pi-2\alpha\le\vert k\vert$
along which the two-fermion dynamic quantities exhibit a jump increasing their values by 2.
Finally,
the two-fermion dynamic quantities show
one-dimensional square-root van Hove divergencies
along the curve $\omega_s(k)=\omega_3(k)$.
In Fig.~\ref{fig02}
we display the characteristic lines
(\ref{4.05}), (\ref{4.06}), (\ref{4.07})
which give the boundaries of the two-fermion continuum
and potential soft modes and singularities.

As temperature increases the lower boundary is smeared-out and finally disappears,
the upper boundary becomes $\omega_3(k)$ along which van Hove singularities occur.
In the high-temperature limit the two-fermion dynamic structure factor
becomes $\Omega$-independent.

\begin{figure}[th]
\centerline{\psfig{file=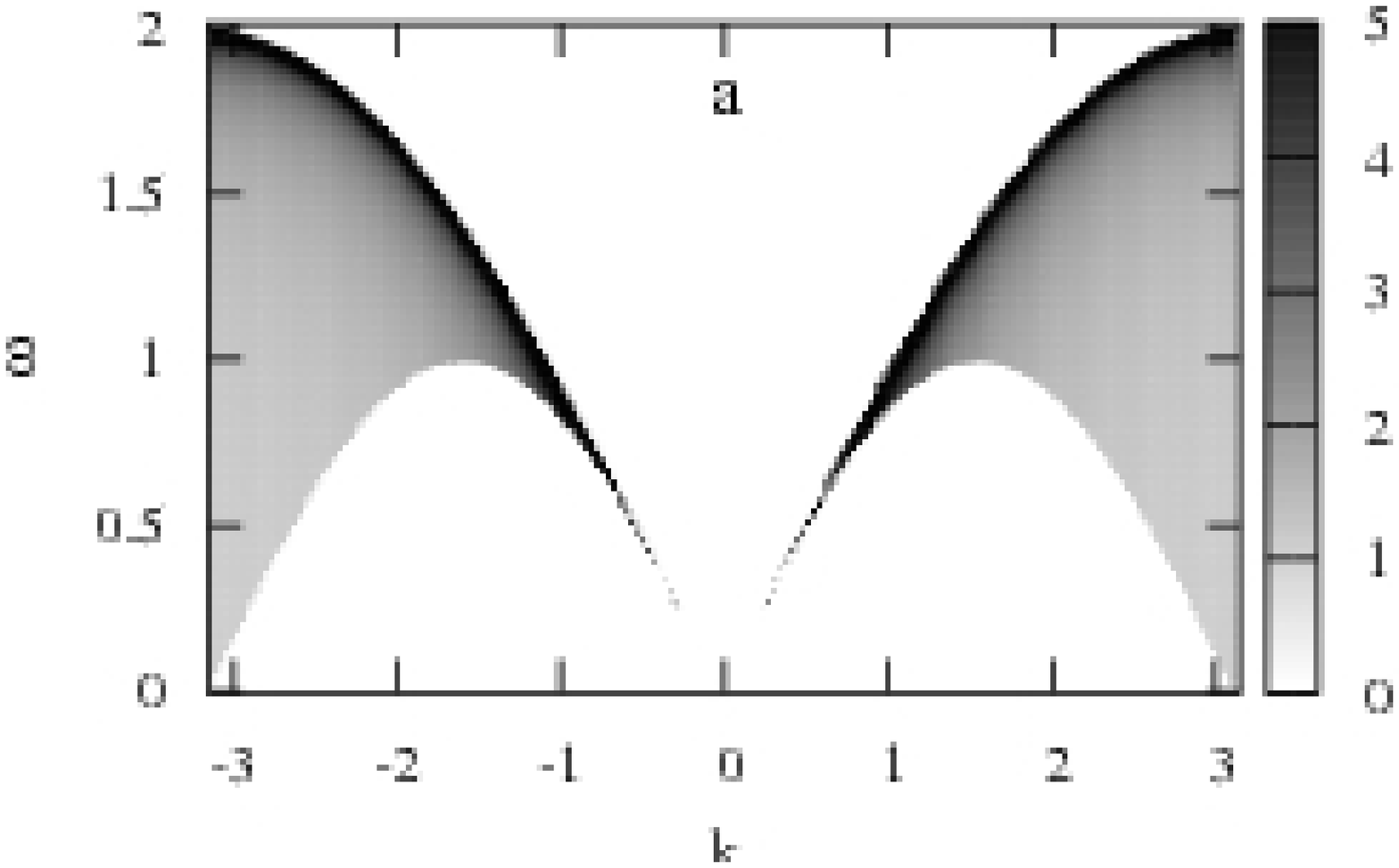,width=2.0in,angle=0}
\psfig{file=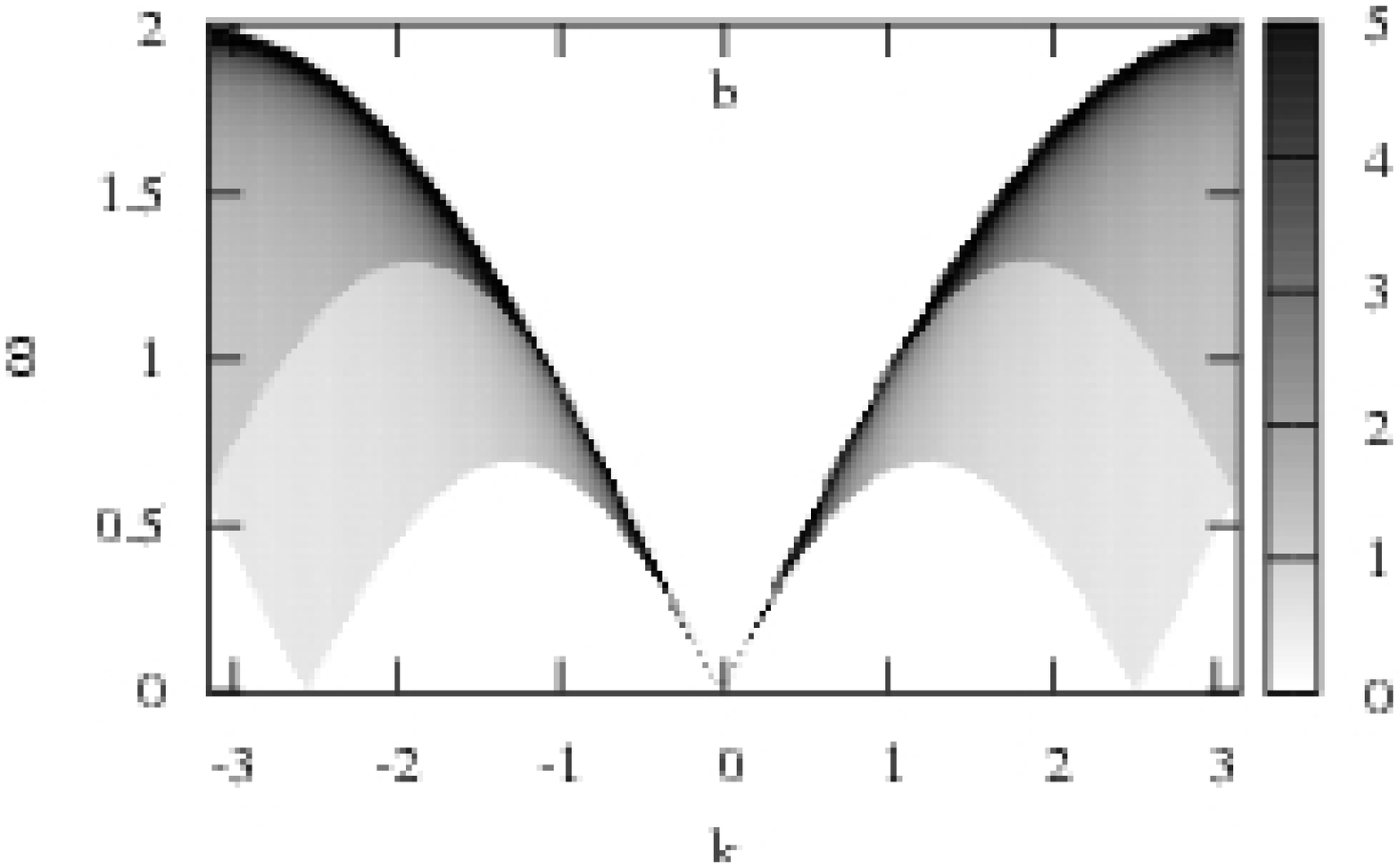,width=2.0in,angle=0}}
\centerline{\psfig{file=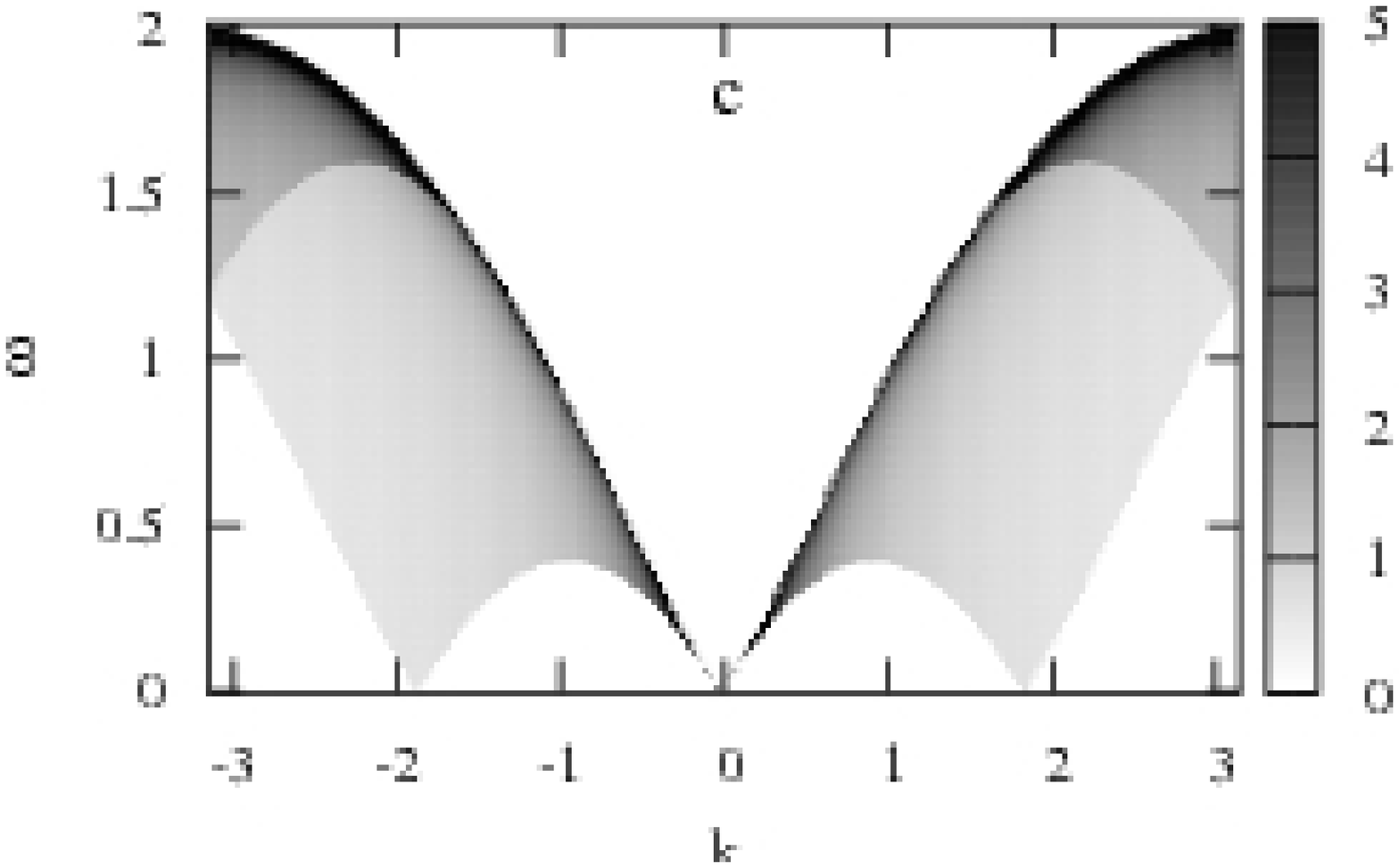,width=2.0in,angle=0}
\psfig{file=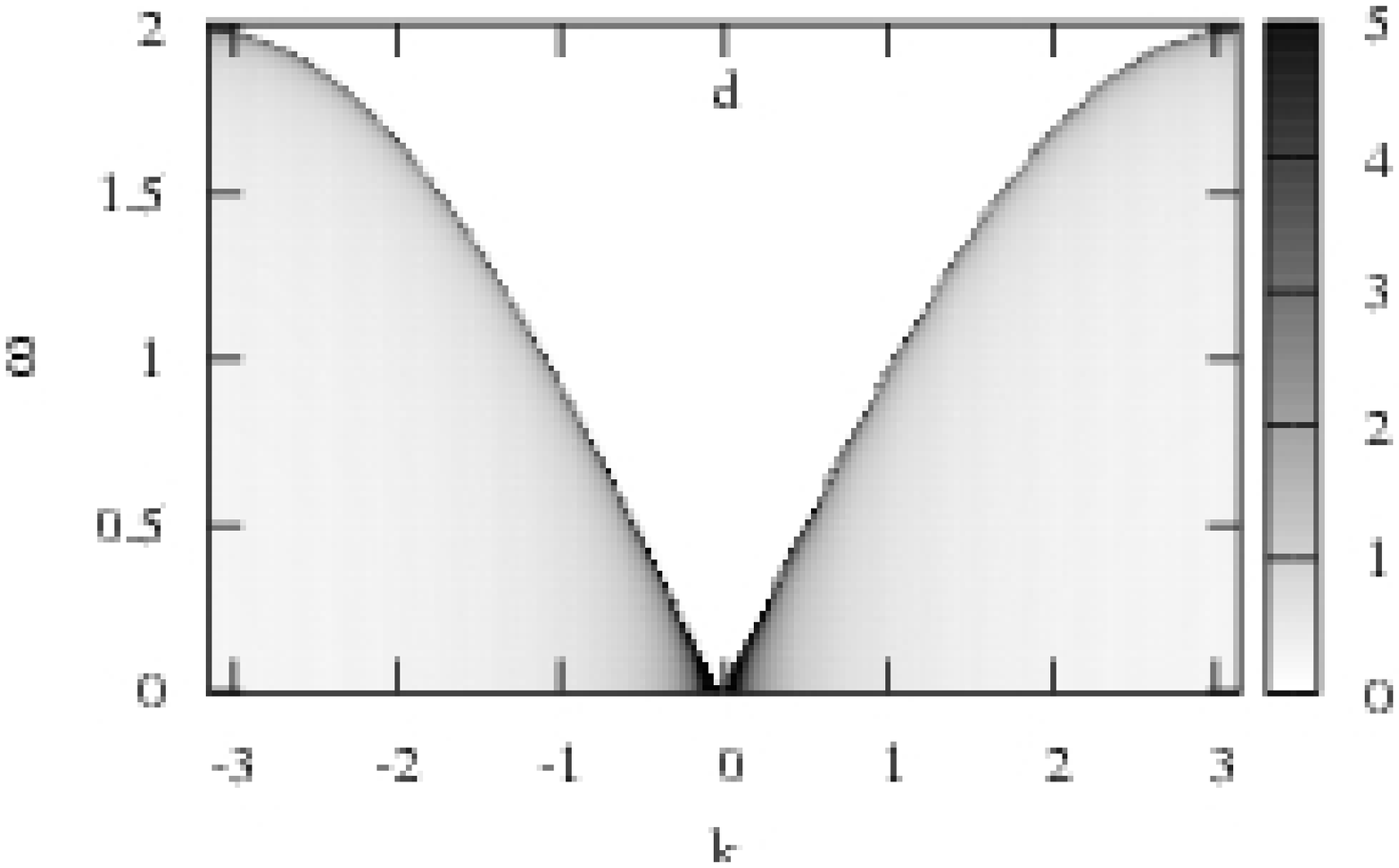,width=2.0in,angle=0}}
\vspace*{8pt}
\caption{$S_{zz}(k,\omega)$ (gray-scale plots)
for the chain (\ref{2.11}) with $J=-1$,
$\Omega=0$ (a), $\Omega=0.3$ (b), $\Omega=0.6$ (c) at $T=0$
and at $T\to\infty$ (d).}
\label{fig03}
\end{figure}

In Fig.~\ref{fig03}
we display the transverse dynamic structure factor $S_{zz}(k,\omega)$ (\ref{4.03})
at zero temperature (panels a, b, c) and in the high-temperature limit (panel d).

There are other dynamic quantities which probe the two-fermion excitation continuum.
Let us consider the dimer operator
\begin{eqnarray}
D_n=s_n^xs_{n+1}^x+s_n^ys_{n+1}^y
\to
\frac{1}{2}\left(c_n^{\dagger}c_{n+1}-c_nc_{n+1}^{\dagger}\right).
\label{4.08}
\end{eqnarray}
That operator is related to a perturbation to the Hamiltonian (\ref{2.11})
which mimics dimerization,
$\epsilon\sum_n\cos(\pi n)D_n$.
The dynamics of fluctuations of the dimer operator can be measured experimentally:
the corresponding dimer dynamic structure factor is relevant
to phonon-assisted optical absorption processes in magnetic-chain compounds\cite{034}.

\begin{figure}[th]
\centerline{\psfig{file=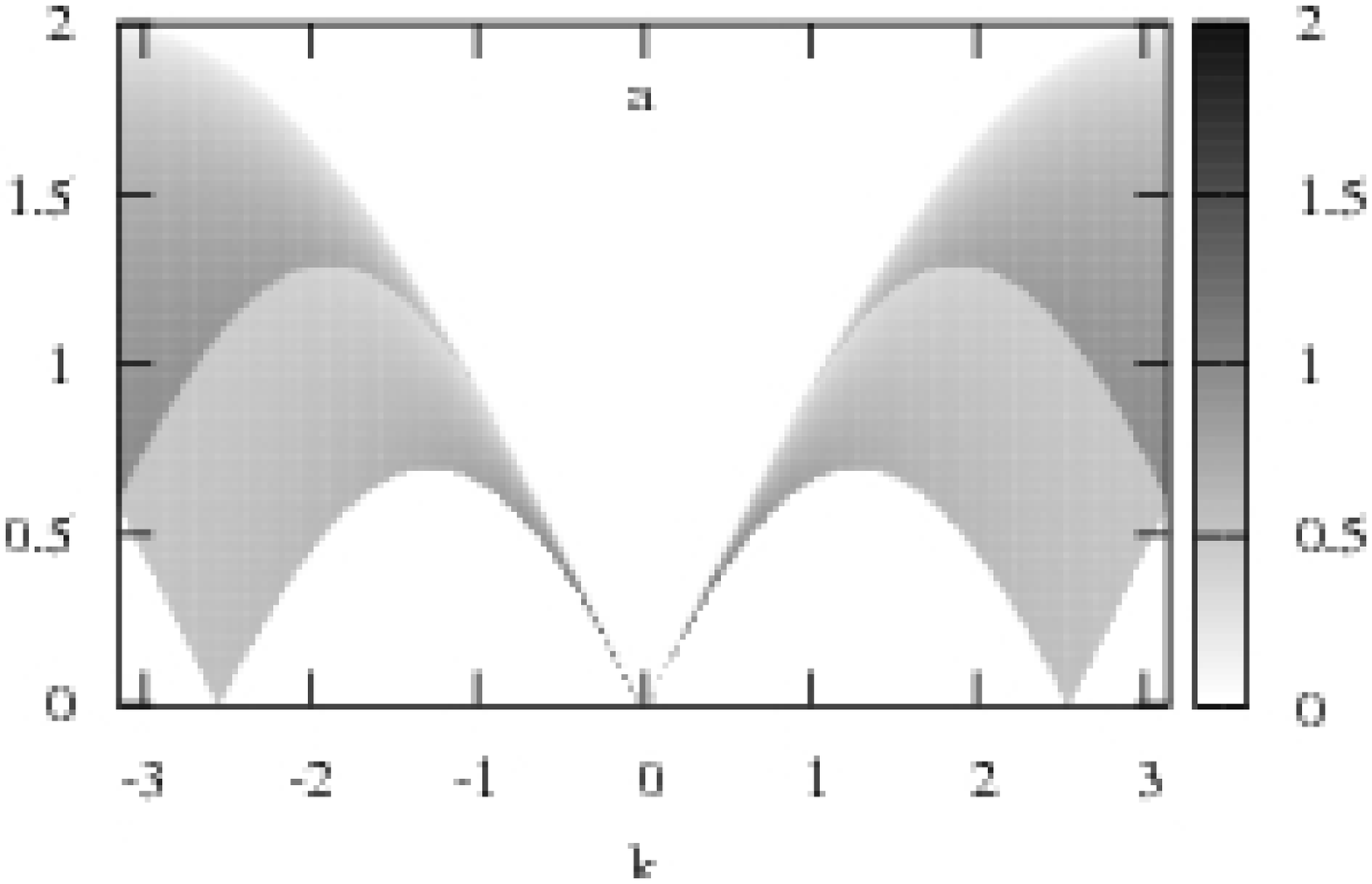,width=2.0in,angle=0}
\psfig{file=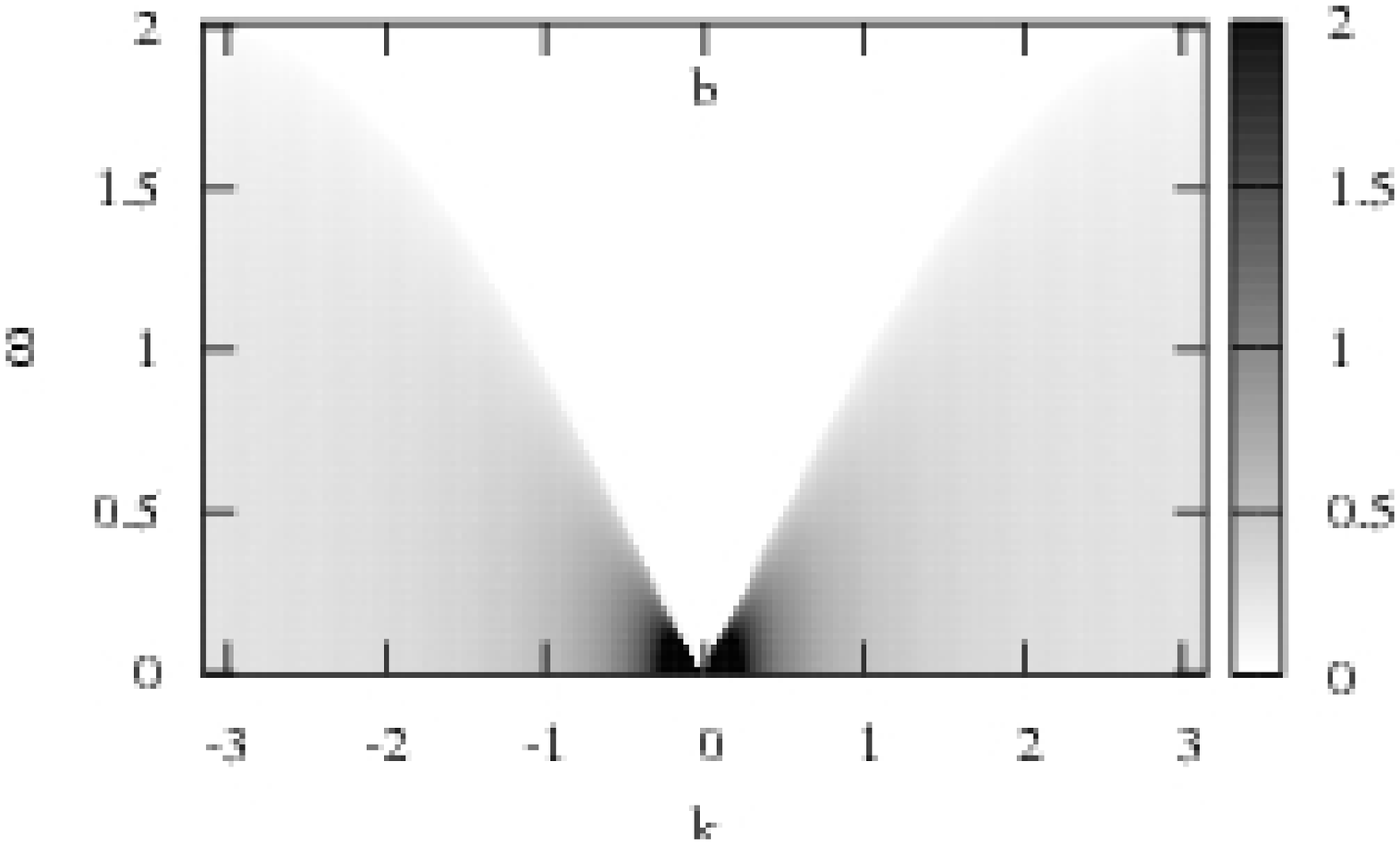,width=2.0in,angle=0}}
\centerline{\psfig{file=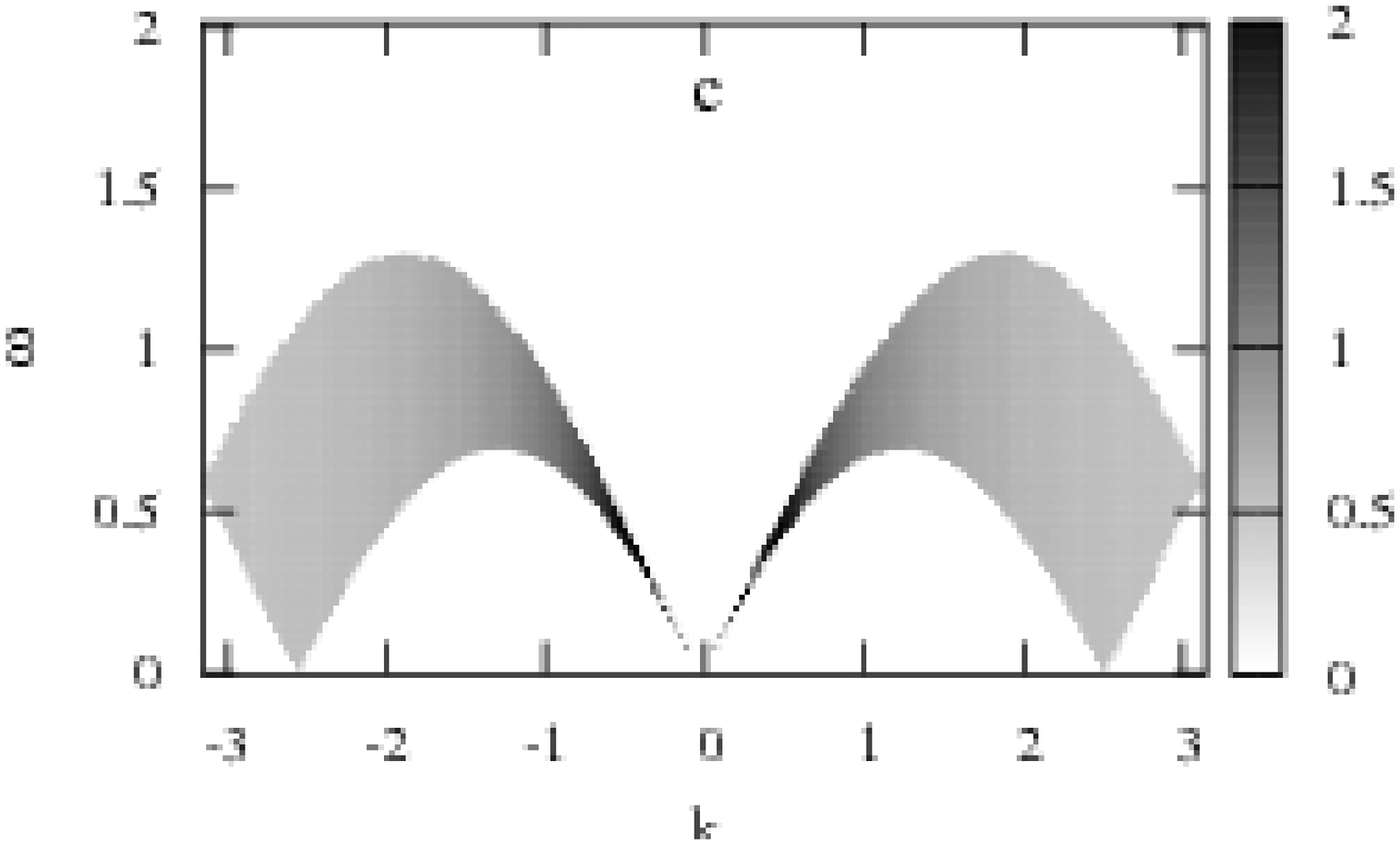,width=2.0in,angle=0}
\psfig{file=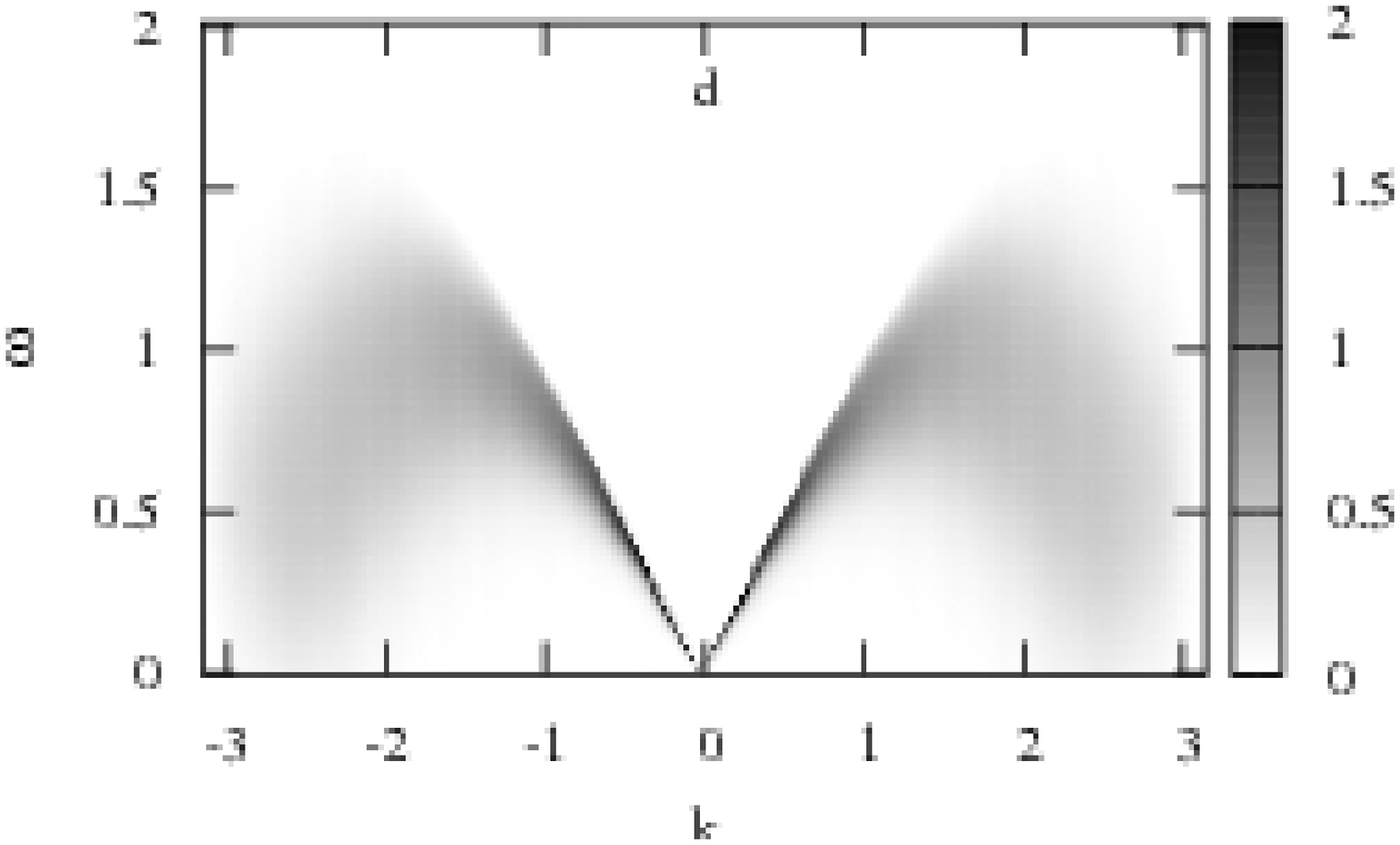,width=2.0in,angle=0}}
\vspace*{8pt}
\caption{$S_{DD}(k,\omega)$ (a, b) 
and $S_{zD}(k,\omega)$ 
(multiplied by $\exp\left(-{\rm{i}}k/2\right)$)
(c, d)
(gray-scale plots)
for the chain (\ref{2.11}) with $J=-1$.
$S_{DD}(k,\omega)$:
$\Omega=0.3$, $T=0$ (a),
$T\to\infty$ (b).
$S_{zD}(k,\omega)$:
$\Omega=0.3$, $T=0$ (c),
$\Omega=0.3$, $T=0.1$ (d).}
\label{fig04}
\end{figure}

The calculation of the time-dependent dimer-dimer correlation function
repeats all steps discussed above while deriving (\ref{4.02})
and ends up with
\begin{eqnarray}
\langle D_n(t) D_{n+l}\rangle-\langle D\rangle^2
=\frac{1}{N^2}\sum_{k_1,k_2}
\cos^2\frac{k_1+k_2}{2}
\exp\left(-{\rm{i}}\left(k_1-k_2\right)l\right)
\nonumber\\
\cdot
\exp\left({\rm{i}}\left(\Lambda_{k_1}-\Lambda_{k_2}\right)t\right)
n_{k_1}\left(1-n_{k_2}\right),
\nonumber\\
\langle D\rangle
=\frac{1}{N}\sum_{n=1}^N\langle D_n\rangle
=-\frac{1}{2N}\sum_k\cos k\tanh\frac{\beta\Lambda_k}{2}.
\label{4.09}
\end{eqnarray}
Inserting Eq. (\ref{4.09}) into Eq. (\ref{1.04})
we get the dimer dynamic structure factor
\begin{eqnarray}
S_{DD}(k,\omega)
=\sum_{l=1}^N\exp(-{\rm{i}}k l)
\int_{-\infty}^{\infty}{\rm{d}}t\exp({\rm{i}}\omega t)
\langle \left(D_n(t)-\langle D\rangle\right)\left(D_{n+l}-\langle D\rangle\right)\rangle
\nonumber\\
=
\int_{-\pi}^{\pi}{\rm{d}}k_1
\cos^2\left(k_1+\frac{k}{2}\right)
n_{k_1}\left(1-n_{k_1+k}\right)
\delta\left(\omega+\Lambda_{k_1}-\Lambda_{k_1+k}\right)
\nonumber\\
=\sum_{k^\star}
\frac{\cos^2\left(\frac{k}{2}+k^{\star}\right)n_{k^{\star}}\left(1-n_{k+k^{\star}}\right)}
{2\left\vert J\sin\frac{k}{2}\cos\left(\frac{k}{2}+k^{\star}\right)\right\vert}.
\label{4.10}
\end{eqnarray}
We can also calculate the $zD$ and $Dz$ dynamic structure factors
\begin{eqnarray}
S_{zD}(k,\omega)
=
\exp\left({\rm{i}}\frac{k}{2}\right)
\int_{-\pi}^{\pi}{\rm{d}}k_1
\cos\left(k_1+\frac{k}{2}\right)
\nonumber\\
\cdot
n_{k_1}\left(1-n_{k_1+k}\right)
\delta\left(\omega+\Lambda_{k_1}-\Lambda_{k_1+k}\right),
\nonumber\\
S_{Dz}(k,\omega)
=\left(S_{zD}(k,\omega)\right)^{*}.
\label{4.11}
\end{eqnarray}
In Fig.~\ref{fig04}
we display the dynamic structure factors
$S_{DD}(k,\omega)$ (\ref{4.10})
and
$S_{zD}(k,\omega)$ (\ref{4.11})
at zero temperature (panels a, c),
at low temperature (panel d)
and in the high-temperature limit (panel b).

Comparing Eqs. (\ref{4.03}), (\ref{4.10}), (\ref{4.11})
we immediately recognize that all these dynamic quantities
are governed exclusively by the two-fermion excitation continuum
(for other two-fermion dynamic quantities see below
and also Refs.~\cite{035,005})
and therefore
all of them exhibit generic properties
inherent in the two-fermion dynamic quantities.
However,
they also exhibit some specific properties
originating from additional factors in the integrands
in Eqs. (\ref{4.03}), (\ref{4.10}), (\ref{4.11})
(e.g. singularities may be suppressed etc.,
compare Fig.~\ref{fig03}b and Fig.~\ref{fig04}a).
In general, the dynamic structure factor governed by the two-fermion excitation continuum
can be written in the following form
\begin{eqnarray}
S_{AB}(k,\omega)
=
\int_{-\pi}^{\pi}{\rm{d}}k_1{\rm{d}}k_2
C_{AB}\left(k_1,k_2\right)
\nonumber\\
\cdot
n_{k_1}\left(1-n_{k_2}\right)
\delta\left(\omega+\Lambda_{k_1}-\Lambda_{k_2}\right)
\delta_{k+k_1-k_2,0},
\nonumber\\
C_{zz}(k_1,k_2)=1,
\nonumber\\
C_{DD}(k_1,k_2)=\cos^2\frac{k_1+k_2}{2},
\nonumber\\
C_{zD}(k_1,k_2)=\frac{1}{2}
\left(\exp\left(-{\rm{i}}k_1\right)+\exp\left({\rm{i}}k_2\right)\right),
\nonumber\\
C_{Dz}(k_1,k_2)=\left(C_{zD}(k_1,k_2)\right)^{*}.
\label{4.12}
\end{eqnarray}
All these quantities show generic properties
(spectral boundaries, soft modes, singularity structure)
and specific properties
controlled by $C_{AB}(k_1,k_2)$.

\subsection{Four-fermion excitations}

We proceed by considering more complicate dynamic quantities.
Namely, consider a trimer operator\cite{036}
\begin{eqnarray}
T_n=s_n^xs_{n+2}^x+s_n^ys_{n+2}^y
\nonumber\\
\to
\frac{1}{2}\left(
c_n^{\dagger}c_{n+2}-c_nc_{n+2}^{\dagger}
-2c_n^{\dagger}c_{n+1}^{\dagger}c_{n+1}c_{n+2}
+2c_nc_{n+1}^{\dagger}c_{n+1}c_{n+2}^{\dagger}
\right).
\label{4.13}
\end{eqnarray}
That operator enters as a perturbation to the Hamiltonian (\ref{2.11})
which mimics trimerization,
$\epsilon\sum_n\cos(2\pi n/3)T_n$.
The dynamics of fluctuations of the trimer operator, 
although it can be analyzed rigorously,
is less evident from the experimental point of view.
Its importance, however, is justified as a quantity
of intermediate complexity between the $zz$ and the $xx$ and $xy$ dynamic quantities.

The calculation of the time-dependent trimer-trimer correlation function
contains no new conceptual ideas but is somewhat tedious.
The final result for the time-dependent trimer correlation function reads
\begin{eqnarray}
\langle T_n(t)T_{n+l}\rangle-\langle T\rangle^2
=\frac{1}{N^2}\sum_{k_1,k_2}
C_{TT}^{(2)}(k_1,k_2)
\exp\left(-{\rm{i}}\left(k_1-k_2\right)l\right)
\nonumber\\
\cdot
\exp\left({\rm{i}}\left(\Lambda_{k_1}-\Lambda_{k_2}\right)t\right)
n_{k_1}\left(1-n_{k_2}\right)
\nonumber\\
+\frac{1}{N^4}\sum_{k_1,k_2,k_3,k_4}
C_{TT}^{(4)}(k_1,k_2,k_3,k_4)
\exp\left(-{\rm{i}}\left(k_1+k_2-k_3-k_4\right)l\right)
\nonumber\\
\cdot
\exp\left({\rm{i}}\left(\Lambda_{k_1}+\Lambda_{k_2}-\Lambda_{k_3}-\Lambda_{k_4}\right)t\right)
n_{k_1}n_{k_2}\left(1-n_{k_3}\right)\left(1-n_{k_4}\right),
\nonumber\\
\langle T\rangle
=\frac{1}{N}\sum_{n=1}^N\langle T_n\rangle
=c_2+2c_1^2-2c_0c_2
\label{4.14}
\end{eqnarray}
with
\begin{eqnarray}
C_{TT}^{(2)}(k_1,k_2)
=\left(1-2c_0\right)^2\cos^2\left(k_1+k_2\right)
\nonumber\\
+4c_1\left(1-2c_0\right)
\left(\cos^2\left(k_1+\frac{k_2}{2}\right)
+\cos^2\left(\frac{k_1}{2}+k_2\right)\right)
\nonumber\\
+4c_1^2\left(\cos^2k_1+\cos^2k_2\right)
\nonumber\\
+8\left(-c_2+c_1^2+2c_0c_2\right)
\cos^2\frac{k_1+k_2}{2}
+8c_1^2\cos^2\frac{k_1-k_2}{2}
\nonumber\\
+4c_1\left(1-2c_0-4c_2\right)\left(\cos^2\frac{k_1}{2}+\cos^2\frac{k_2}{2}\right)
\nonumber\\
+4c_2-8c_1-8c_1^2+4c_2^2+16c_0c_1-8c_0c_2+16c_1c_2,
\label{4.15}
\\
C_{TT}^{(4)}(k_1,k_2,k_3,k_4)
\nonumber\\
=16\sin^2\frac{k_1-k_2}{2}\sin^2\frac{k_3-k_4}{2}
\cos^2\frac{k_1+k_2+k_3+k_4}{2}\ge 0
\label{4.16}
\end{eqnarray}
and $c_p=\left(1/N\right)\sum_k\cos\left(pk\right)n_k$.
Obviously,
the calculation of such an average as
$\langle c_{k_1}^{\dagger}c_{k_2}^{\dagger}c_{k_3}c_{k_4}
c_{k_5}^{\dagger}c_{k_6}^{\dagger}c_{k_7}c_{k_8}\rangle$
according to the Wick-Bloch-de Dominicis theorem is rather complicated.
Substituting (\ref{4.14}) into Eq. (\ref{1.04})
we obtain the following result for the trimer dynamic structure factor
\begin{eqnarray}
S_{TT}(k,\omega)
=
\sum_{l=1}^N\exp(-{\rm{i}}k l)
\int_{-\infty}^{\infty}{\rm{d}}t\exp({\rm{i}}\omega t)
\langle \left(T_n(t)-\langle T\rangle\right)\left(T_{n+l}-\langle T\rangle\right)\rangle
\nonumber\\
=
S_{TT}^{(2)}(k,\omega)+S_{TT}^{(4)}(k,\omega)
\label{4.17}
\end{eqnarray}
with
\begin{eqnarray}
S_{TT}^{(2)}(k,\omega)
=
\int_{-\pi}^{\pi}{\rm{d}}k_1
C_{TT}^{(2)}(k_1,k_1+k)
n_{k_1}\left(1-n_{k_1+k}\right)
\delta\left(\omega+\Lambda_{k_1}-\Lambda_{k_1+k}\right),
\label{4.18}
\\
S_{TT}^{(4)}(k,\omega)
=
\frac{1}{4\pi^2}
\int_{-\pi}^{\pi}{\rm{d}}k_1{\rm{d}}k_2{\rm{d}}k_3
C_{TT}^{(4)}(k_1,k_2,k_3,k_1+k_2-k_3+k)
\nonumber\\
\cdot
n_{k_1}n_{k_2}\left(1-n_{k_3}\right)\left(1-n_{k_1+k_2-k_3+k}\right)
\delta\left(\omega+\Lambda_{k_1}+\Lambda_{k_2}-\Lambda_{k_3}-\Lambda_{k_1+k_2-k_3+k}\right).
\label{4.19}
\end{eqnarray}
For further details see Ref.~\cite{036}.

As can be seen from Eqs. (\ref{4.17}), (\ref{4.18}), (\ref{4.19})
the trimer dynamic structure factor is governed
both by the two-fermion (one particle and one hole) excitation continuum
discussed above,
the term $S_{TT}^{(2)}(k,\omega)$,
and by the four-fermion (two particles and two holes) excitation continuum,
the term $S_{TT}^{(4)}(k,\omega)$.
The four-fermion excitation continuum is determined by the conditions
\begin{eqnarray}
\omega=-\Lambda_{k_1}-\Lambda_{k_2}+\Lambda_{k_3}+\Lambda_{k_4},
\;
k=-k_1-k_2+k_3+k_4(\mod(2\pi)),
\label{4.20}
\end{eqnarray}
where $-\pi\le k_1,k_2,k_3< \pi$
and $\Lambda_k=\Omega+J\cos k$.
Moreover,
in the ground state we have to require in addition
$n_{k_1}> 0$,
$n_{k_2}> 0$,
$1-n_{k_3}> 0$,
$1-n_{k_4}> 0$,
i.e.
$\Lambda_{k_1}\le 0$,
$\Lambda_{k_2}\le 0$,
$\Lambda_{k_3}\ge 0$,
$\Lambda_{k_4}\ge 0$.

\begin{figure}[th]
\centerline{\psfig{file=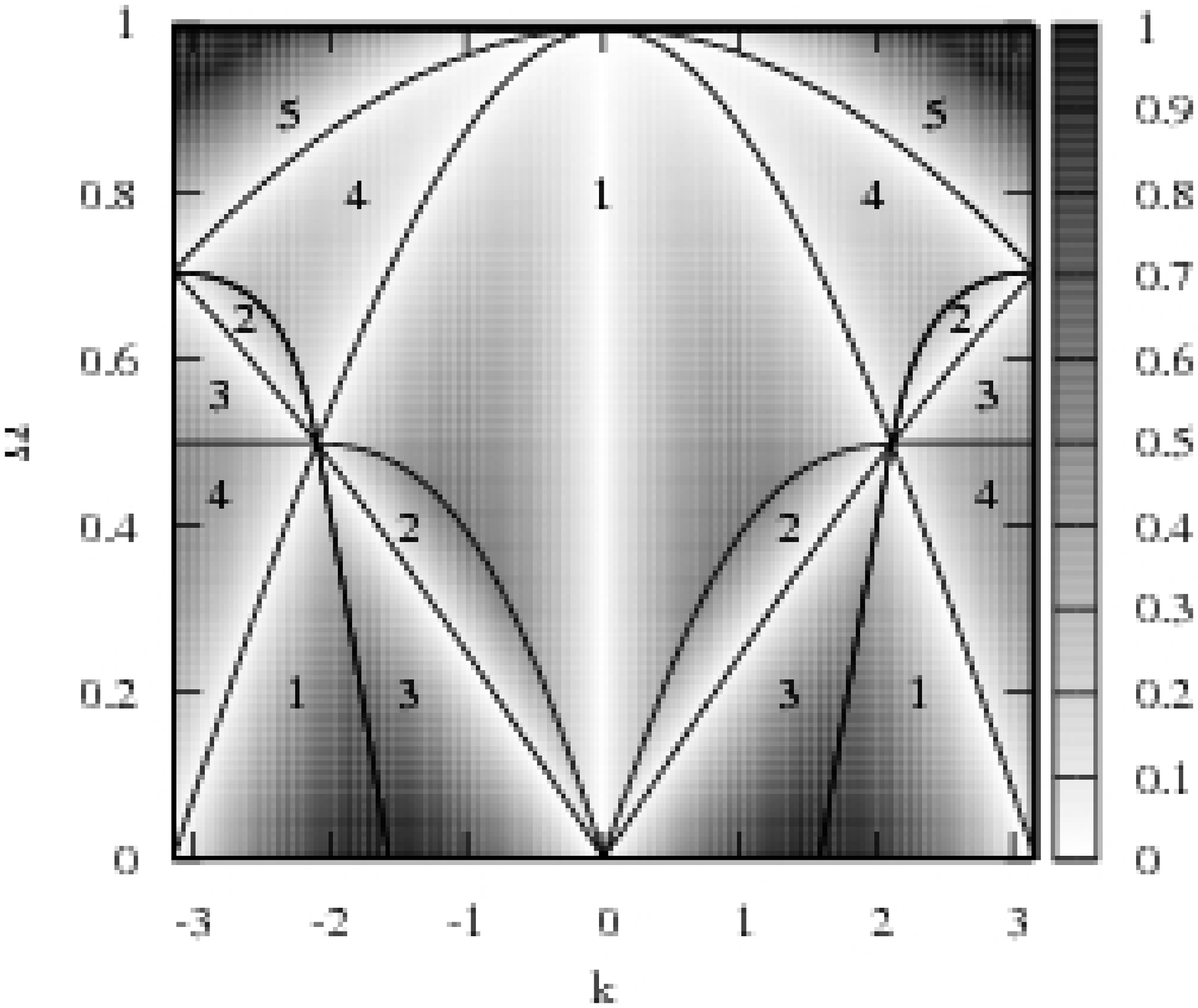,width=2.0in,angle=0}
\psfig{file=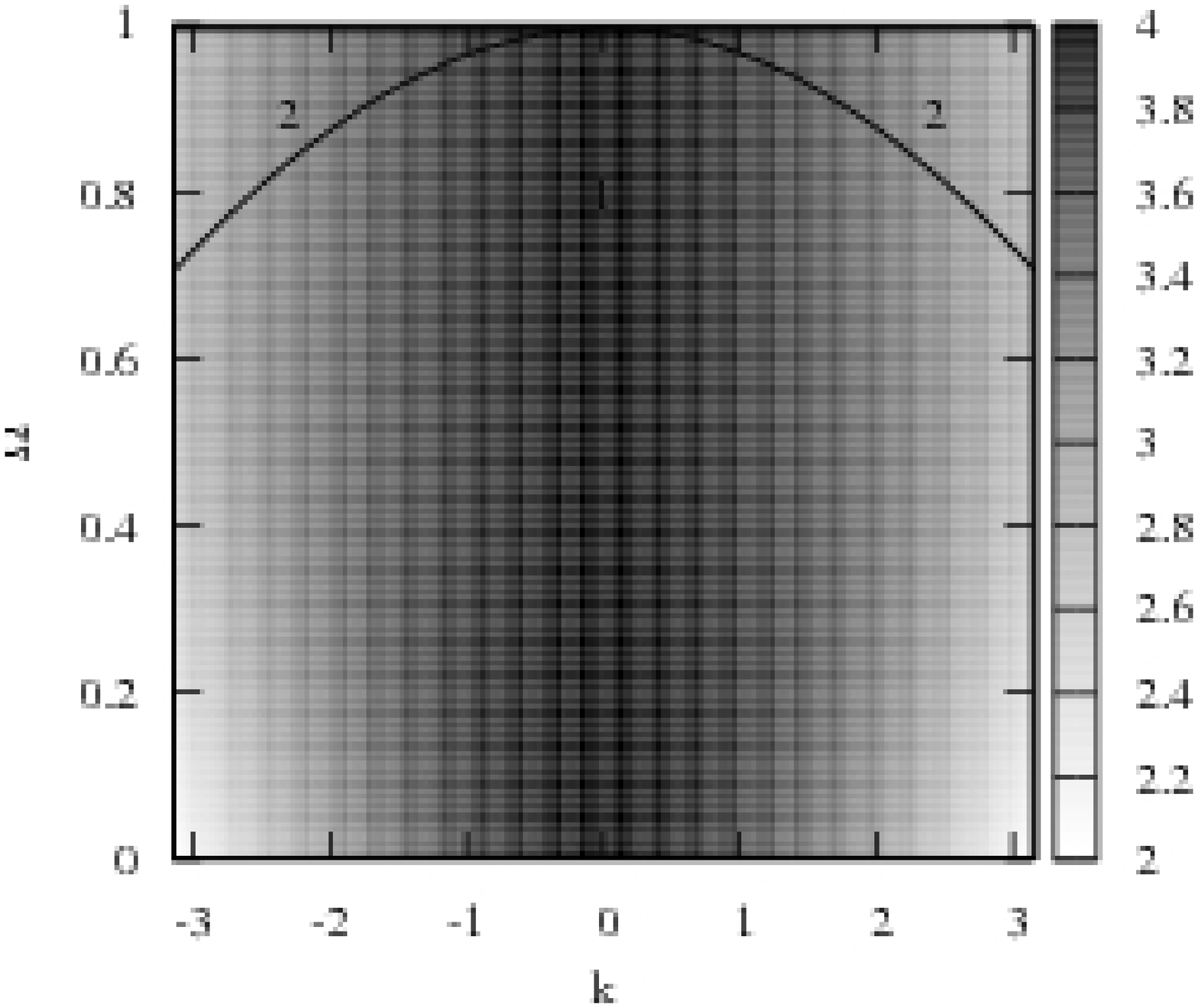,width=2.0in,angle=0}}
\vspace*{8pt}
\caption{The lower boundary $\omega_l(k)$ (left panel)
and the upper boundary $\omega_u(k)$ (right panel)
of the four-fermion excitation continuum
in the plane wave-vector $k$ -- transverse field $\Omega$,
$\vert J\vert=1$.}
\label{fig05}
\end{figure}

\begin{figure}[th]
\centerline{\psfig{file=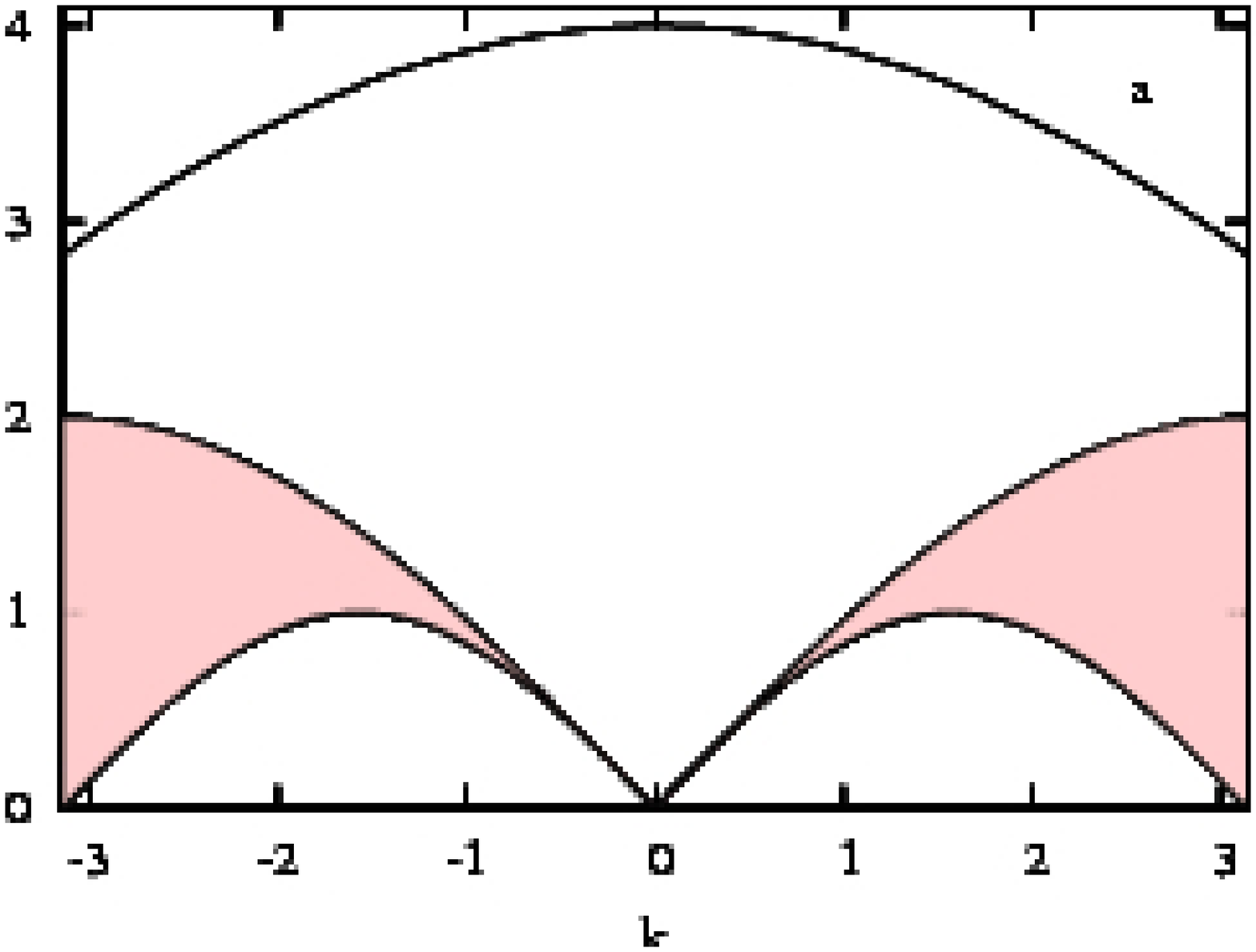,width=2.0in,angle=0}
\psfig{file=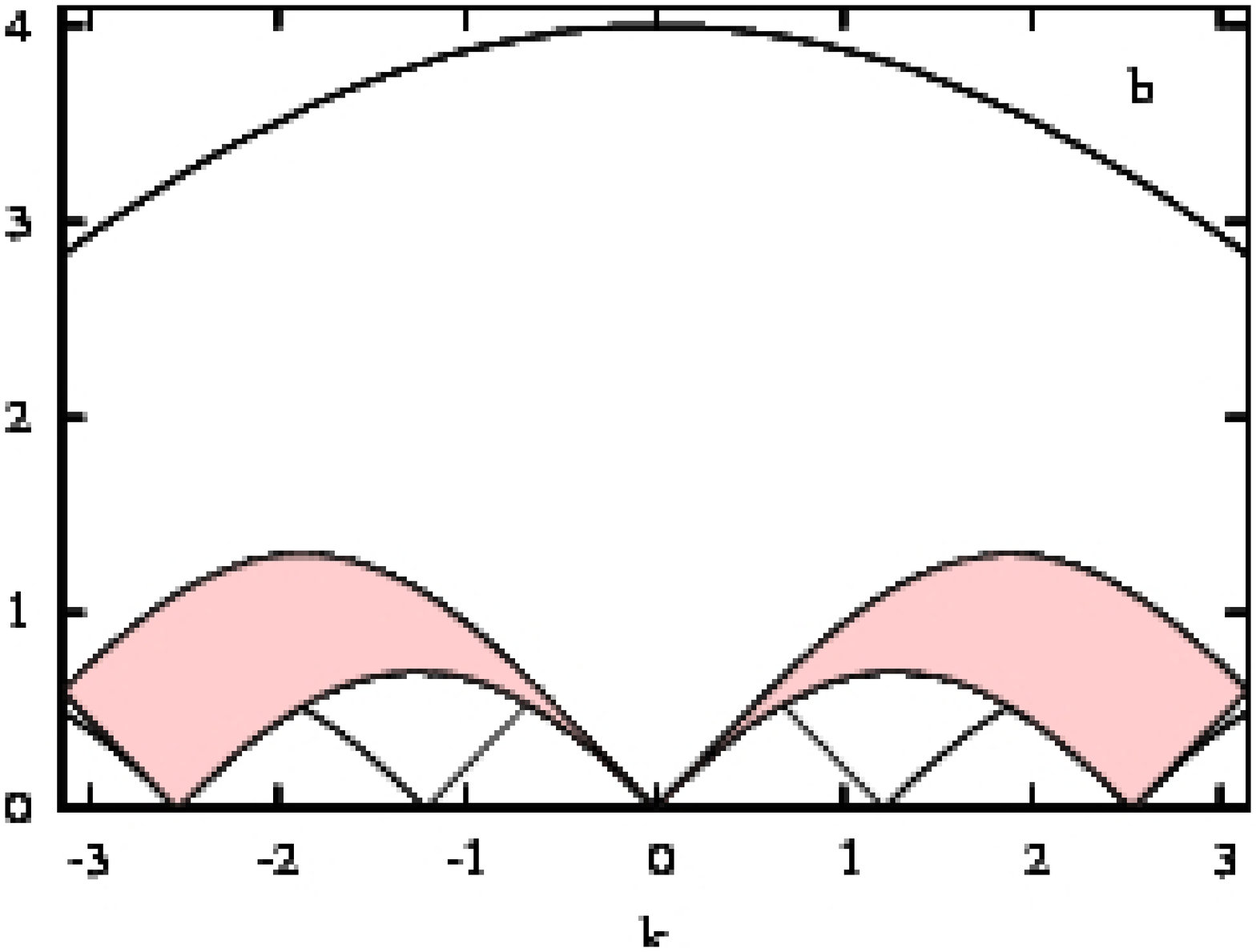,width=2.0in,angle=0}}
\vspace*{8pt}
\caption{Lower boundaries and upper boundaries
of the two-fermion and four-fermion excitation continua
for $\vert J\vert=1$ and $\Omega=0$ (a) and $\Omega=0.3$ (b)
at $T=0$.
The two-fermion continuum is shown shaded.}
\label{fig06}
\end{figure}

We start with the zero-temperature case.
The lower boundary is given by one of the following curves
\begin{eqnarray}
\frac{\omega^{(1)}_l(k)}{\vert J\vert}
=2\sin\frac{\vert k\vert}{2}\sin\left(\alpha-\frac{\vert k\vert}{2}\right),
\nonumber\\
\frac{\omega^{(2)}_l(k)}{\vert J\vert}
=4\cos\frac{k}{4}\cos\left(\alpha+\frac{\vert k\vert}{4}\right),
\nonumber\\
\frac{\omega^{(3)}_l(k)}{\vert J\vert}
=-2\sin\left(\alpha+\frac{\vert k\vert}{2}\right)\sin\left(2\alpha+\frac{\vert k\vert}{2}\right),
\nonumber\\
\frac{\omega^{(4)}_l(k)}{\vert J\vert}
=-2\sin\left(\alpha-\frac{\vert k\vert}{2}\right)\sin\left(2\alpha-\frac{\vert k\vert}{2}\right),
\nonumber\\
\frac{\omega^{(5)}_l(k)}{\vert J\vert}
=-4\sin\frac{\vert k\vert}{4}\sin\left(\alpha-\frac{\vert k\vert}{4}\right)
\label{4.21}
\end{eqnarray}
depending on the value of $\Omega$, $\vert\Omega\vert\le\vert J\vert$
and the value of $k$, $\pi\le k<\pi$
as is shown in the left panel in Fig.~\ref{fig05}.
The boundary between the region $i$
(where $\omega_l^{(i)}(k)$ is the lower boundary)
and the region $j$
(where $\omega_l^{(j)}(k)$ is the lower boundary)
(see the left panel in Fig.~\ref{fig05})
is given by the formula $\vert k\vert=l_{ij}(\alpha)$
where
$l_{12}(\alpha)=4\arctan\left(\left(\tan\alpha-\sqrt{\tan^2\alpha-3}\right)/3\right)$,
$\vert k\vert\le2\pi/3$;
$l_{13}(\alpha)=\pi-\alpha$,
$\pi/2\le\vert k\vert\le 2\pi/3$;
$l_{14}(\alpha)=2\alpha$;
$l_{23}(\alpha)=2\pi-4\alpha$;
$l_{34}(\alpha)=\vert k\vert+\cos\alpha-1/2$,
$2\pi/3\le\vert k\vert\le\pi$;
$l_{45}(\alpha)=4\alpha$
(for further details see Ref.~\cite{036}).
The four-fermion contribution to dynamic quantities may exhibit soft modes
$\vert k_0\vert=\{0,2\pi-4\alpha,2\alpha,4\alpha\}$.
Next we pass to the upper boundary of the four-fermion excitation continuum
which is given by one of the following curves
\begin{eqnarray}
\frac{\omega^{(1)}_u(k)}{\vert J\vert}
=4\cos\frac{k}{4},
\label{4.22}
\\
\frac{\omega^{(2)}_u(k)}{\vert J\vert}
=4\cos\frac{k}{4}\cos\left(\alpha-\frac{\vert k\vert}{4}\right)
\label{4.23}
\end{eqnarray}
depending on the value of $\Omega$, $\vert\Omega\vert\le\vert J\vert$
and the value of $k$, $\pi\le k<\pi$
as is shown in the right panel in Fig.~\ref{fig05}.
The boundary between the regions $1$ and $2$
is given by the curve $\vert k\vert=4\alpha$.
In Fig.~\ref{fig06}
we compare the ground-state two-fermion and four-fermion excitation continua
for two values of the transverse field $\Omega$.
The four-fermion excitation continuum always contains the two-fermion excitation continuum.
The lower boundaries may coincide
(e.g. in the zero-field case the lower boundary is $\vert J\vert \sin\vert k\vert$ for both continua,
panel a in Fig.~\ref{fig06}a)
whereas the upper boundaries are different.

\begin{figure}[th]
\centerline{\psfig{file=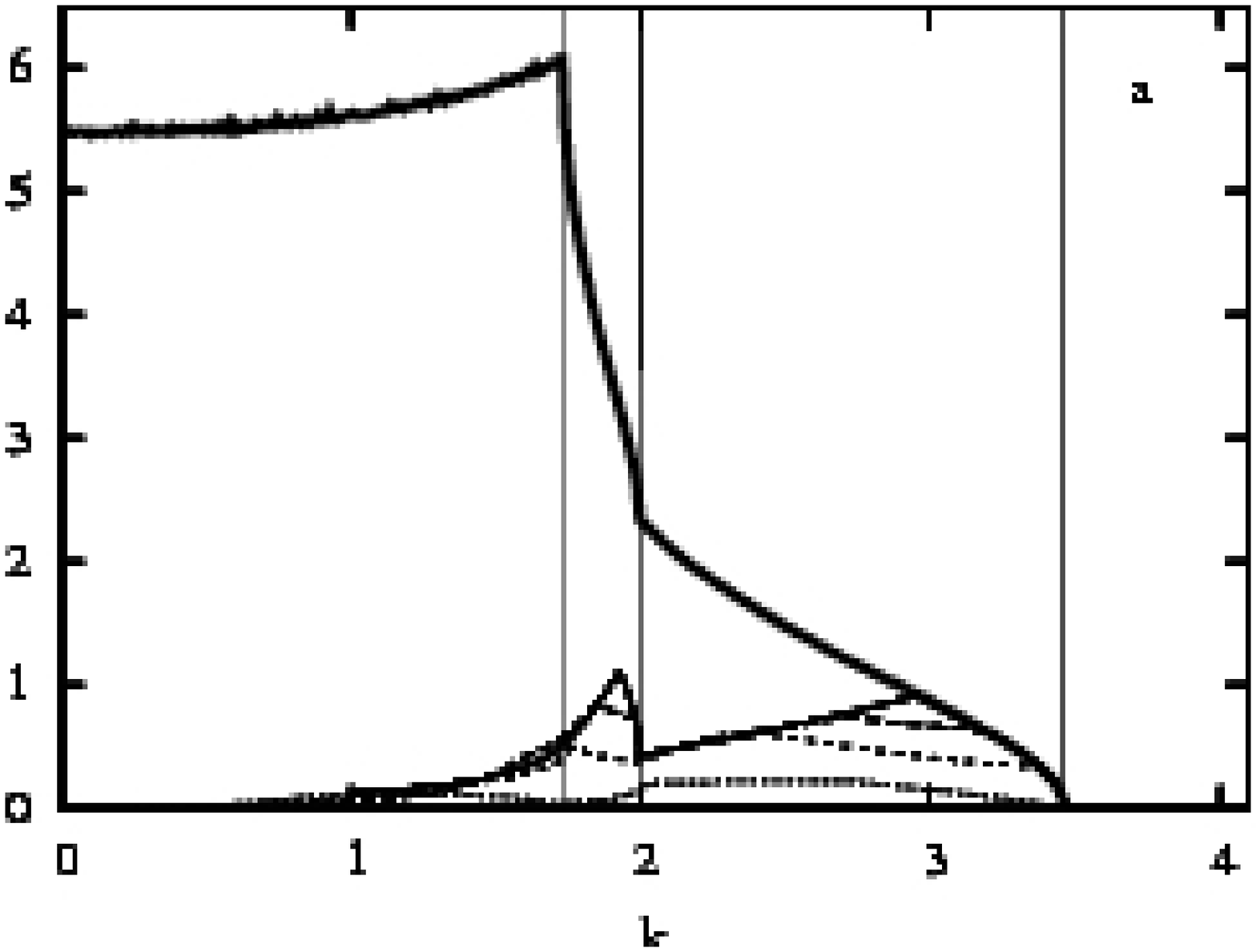,width=2.0in,angle=0}
\psfig{file=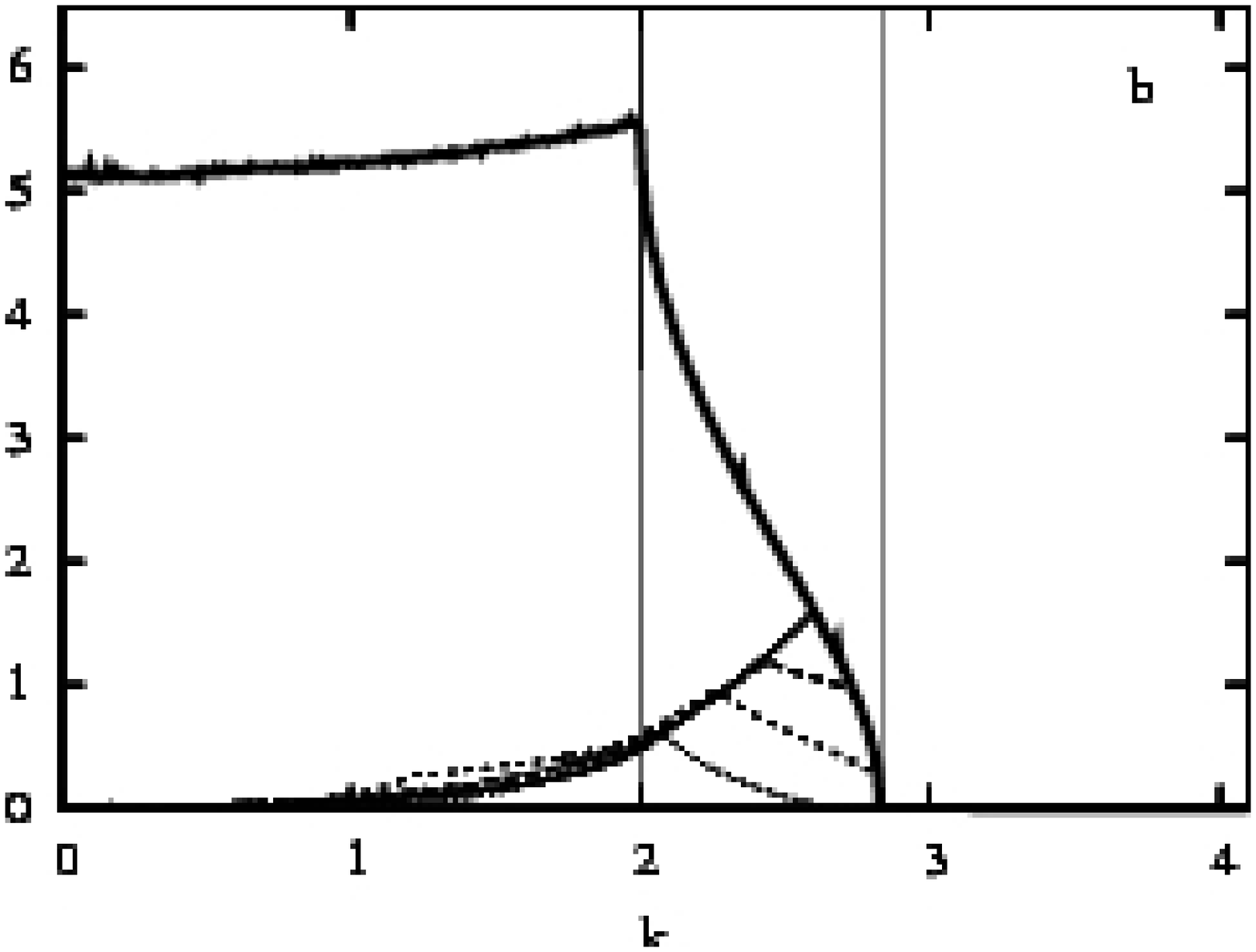,width=2.0in,angle=0}}
\vspace*{8pt}
\caption{$S(k,\omega)$ (\ref{4.24}) vs $\omega$
at $k=2\pi/3$ (a) and $k=\pi$ (b)
for $S(k_1,k_2,k_3,k)=1$ (bold curves)
and
$S(k_1,k_2,k_3,k)
=n_{k_1}n_{k_2}\left(1-n_{k_3}\right)\left(1-n_{k_1+k_2-k_3+k}\right)$,
$T=0$ 
for $\Omega=0$ (solid curves),
$\Omega=0.3$ (long-dashed curves),
$\Omega=0.6$ (short-dashed curves),
$\Omega=0.9$ (dotted curves).
Vertical lines denote the values of $\omega_{s}^{(j)}(k)$, $j=1,2,3$ (\ref{4.25}).}
\label{fig07}
\end{figure}

Next we turn to the van Hove singularities inherent in the four-fermion dynamic quantities.
Evidently,
the quantity
\begin{eqnarray}
S(k,\omega)
=\int_{-\pi}^{\pi}{\rm{d}}k_1
\int_{-\pi}^{\pi}{\rm{d}}k_2
\int_{-\pi}^{\pi}{\rm{d}}k_3
S(k_1,k_2,k_3,k)
\nonumber\\
\cdot
\delta\left(\omega-\vert J\vert\cos k_1-\vert J\vert\cos k_2
+\vert J\vert\cos k_3+\vert J\vert\cos \left(k+k_1+k_2-k_3\right)\right)
\label{4.24}
\end{eqnarray}
may exhibit van Hove singularities characteristic to the three-dimensional density of states.
The lines of potential singularities are as follows
\begin{eqnarray}
\frac{\omega^{(1)}_s(k)}{\vert J\vert}
=2\sin\frac{\vert k\vert}{2},
\nonumber\\
\frac{\omega^{(2)}_s(k)}{\vert J\vert}
=4\sin\frac{\vert k\vert}{4},
\nonumber\\
\frac{\omega^{(3)}_s(k)}{\vert J\vert}
=4\cos\frac{k}{4}.
\label{4.25}
\end{eqnarray}
The four-fermion dynamic quantities may exhibit cusp singularities
(akin to density-of-states effects in thee dimensions)
along these curves.
We illustrate potential singularities in the frequency profiles of $S(k,\omega)$ (\ref{4.24})
at different $k$
in Fig.~\ref{fig07}.

For nonzero temperatures the lower boundary is smeared out and finally disappears.
The upper boundary is given by Eq. (\ref{4.22}).
In the high-temperature limit the properties of the four-fermion excitation continuum
become $\Omega$-independent.

\begin{figure}[th]
\centerline{\psfig{file=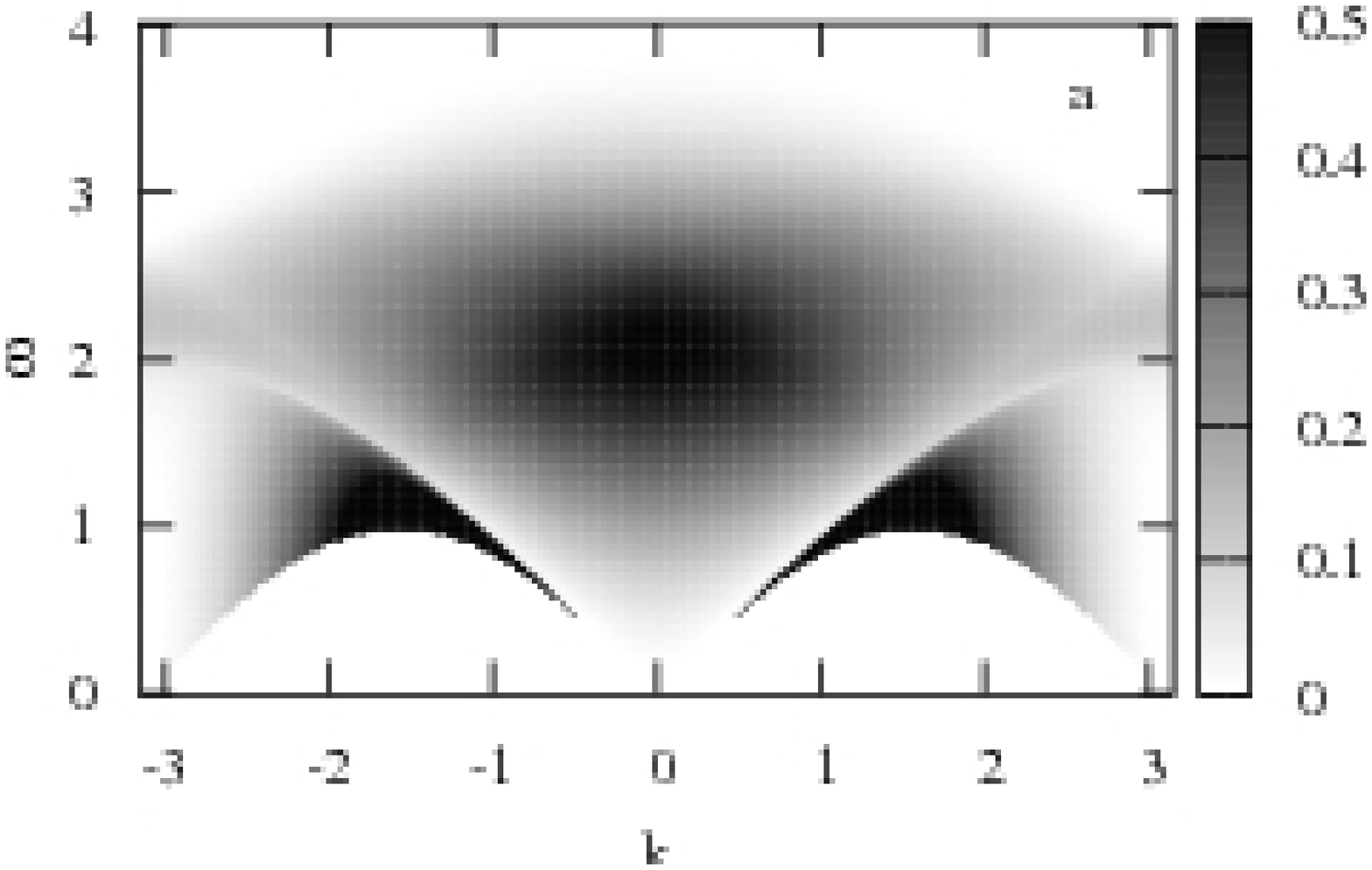,width=2.0in,angle=0}
\psfig{file=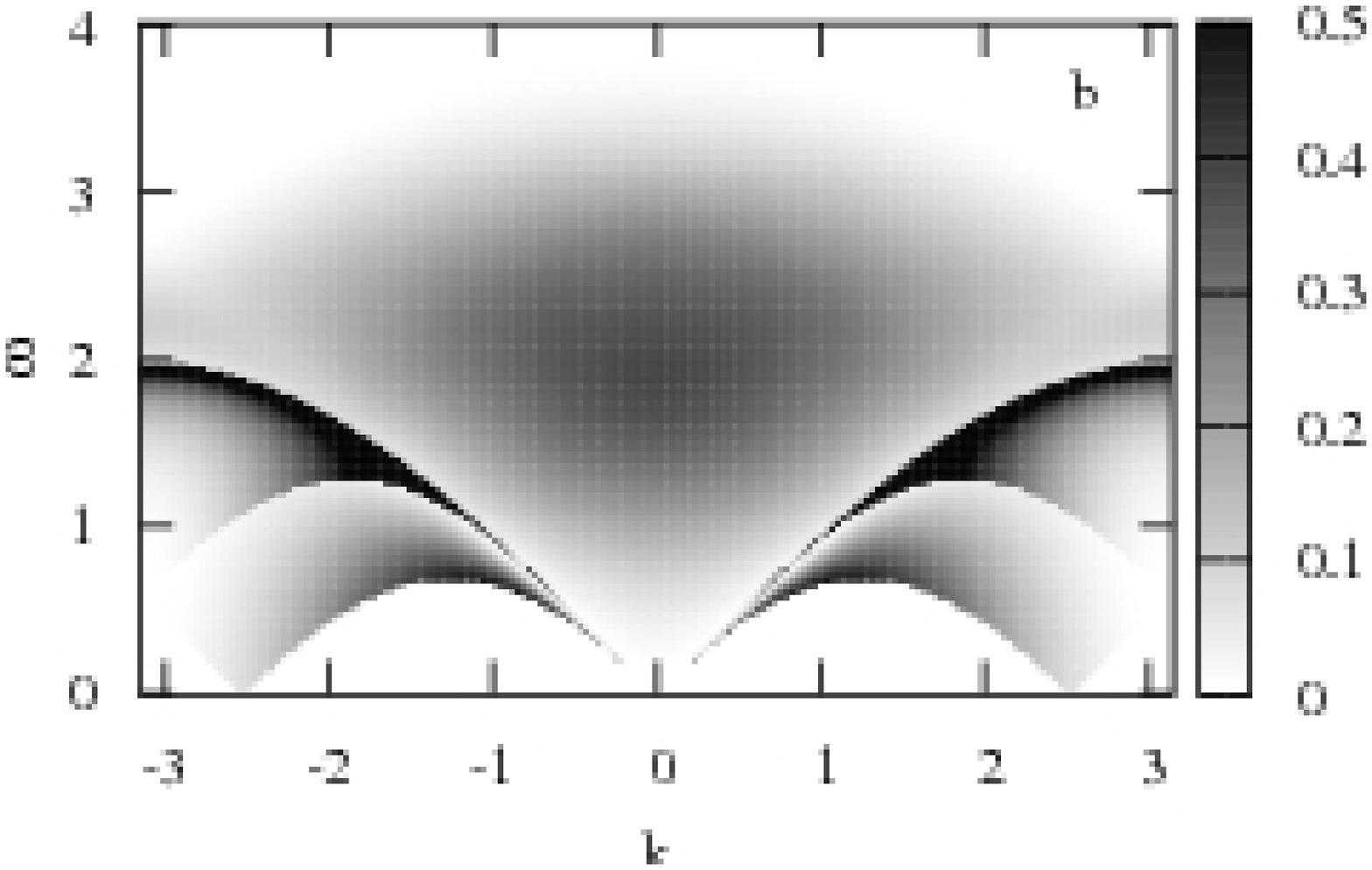,width=2.0in,angle=0}}
\centerline{\psfig{file=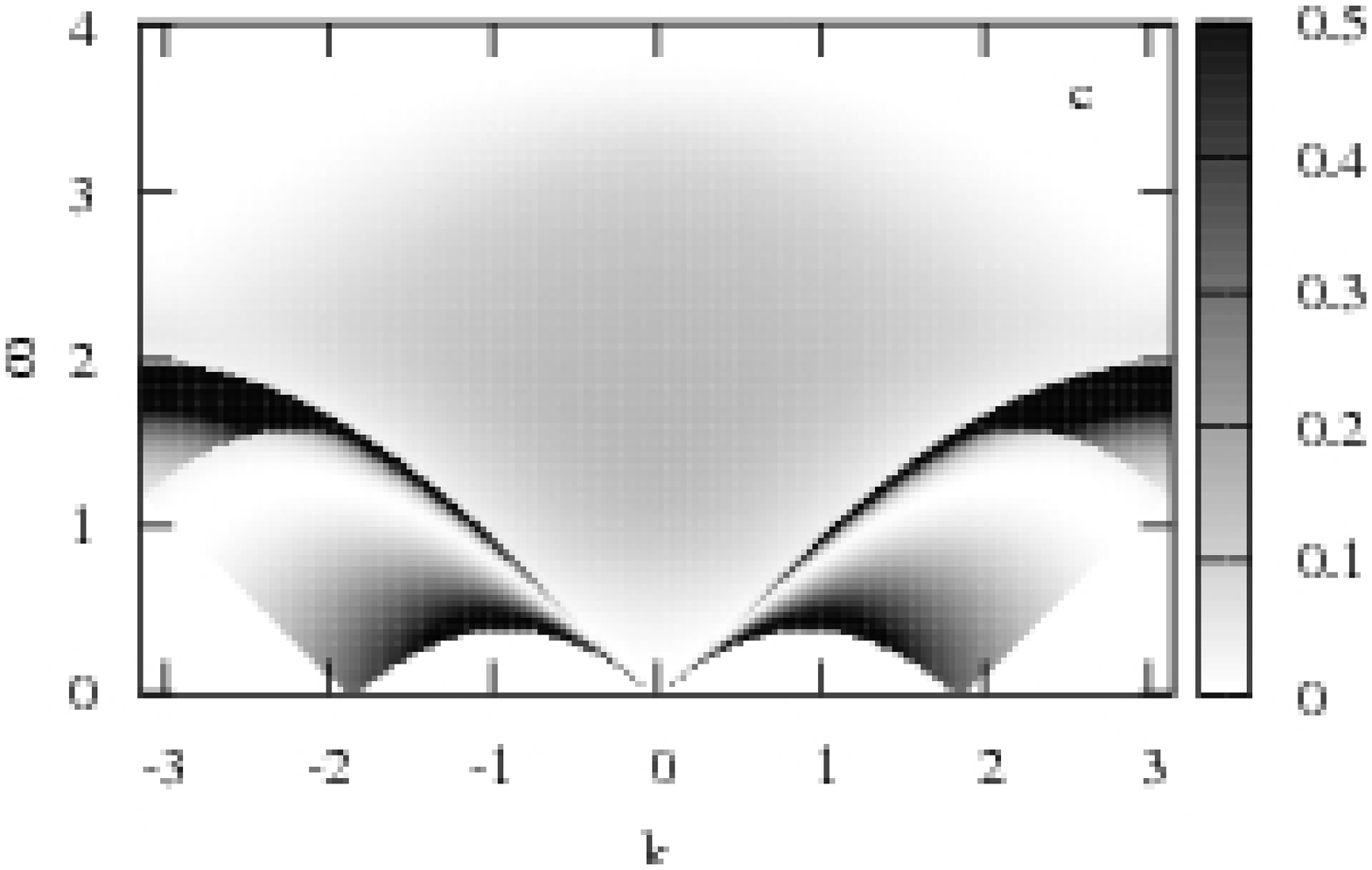,width=2.0in,angle=0}
\psfig{file=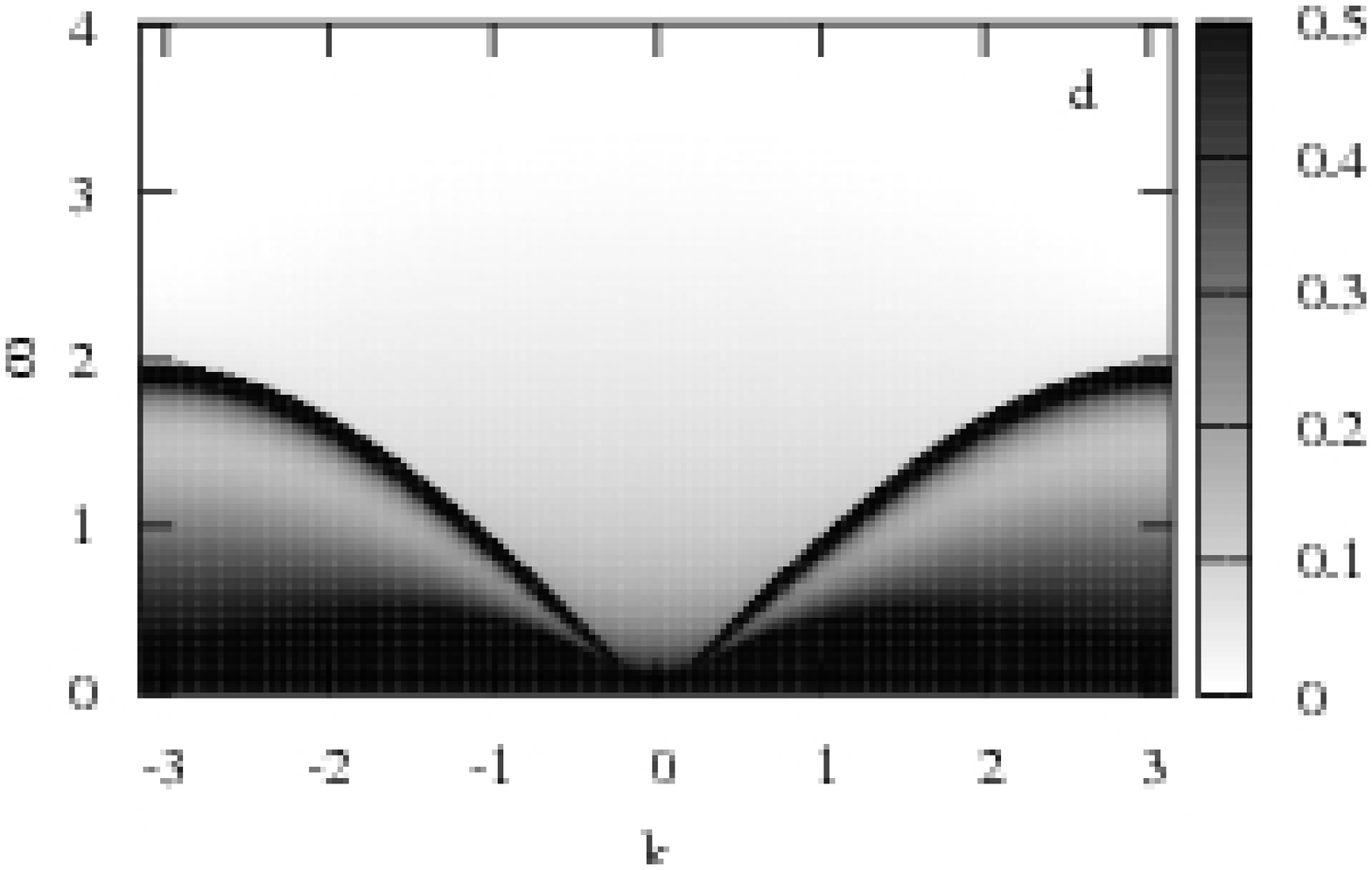,width=2.0in,angle=0}}
\vspace*{8pt}
\caption{$S_{TT}(k,\omega)$ (\ref{4.17})
(gray-scale plots) of the chain (\ref{2.11}) with
$J=-1$,
$\Omega=0$ (a), $\Omega=0.3$ (b), $\Omega=0.6$ (c) at $T=0$
and at $T\to\infty$ (d).}
\label{fig08}
\end{figure}

After discussing some generic properties of the four-fermion dynamic quantities
(inherent in any four-fermion dynamic quantity)
we illustrate some specific properties
conditioned by the function
$C_{TT}^{(4)}(k_1,k_2,k_3,k_4)$ (\ref{4.16}).
In Fig.~\ref{fig08}
we display the trimer dynamic structure factor (\ref{4.17}).
The contributions of the two-fermion excitation continuum and the four-fermion excitation continuum
to this quantity can be easily distinguished.

We may formally introduce the polymer operator
\begin{eqnarray}
{\cal{P}}_n^{(l)}
=s_n^xs_{n+l}^x+s_n^ys_{n+l}^y
\label{4.26}
\end{eqnarray}
(evidently
${\cal{P}}_n^{(1)}=D_n$
and
${\cal{P}}_n^{(2)}=T_n$).
Now the dynamic polymer structure factor
$S_{{\cal{P}}{\cal{P}}}(k,\omega)$ will involve $2m$-fermion excitations
with $m=1,2,\ldots,l$.
These quantities are of moderate complexity in comparison with
$S_{xx}(k,\omega)$
and
$S_{xy}(k,\omega)$
which are enormously complex (see below).
We also note that in the limit $l\to\infty$
\begin{eqnarray}
\langle {\cal{P}}_n^{(l)}(t){\cal{P}}_{n+m}^{(l)}\rangle
\stackrel{l\to \infty}{\longrightarrow}
2\langle s_n^x(t)s_{n+m}^x\rangle^2
+2\langle s_n^x(t)s_{n+m}^y\rangle^2.
\label{4.27}
\end{eqnarray}
The last term in Eq. (\ref{4.27}) is nonzero only if $\Omega\ne 0$.

In passing,
we note that the multimagnon continua of quantum spin chains
have been discussed recently on general ground by T.~Barnes\cite{037}.
Obviously, there is an essential difference in comparison with our case,
since the Jordan-Wigner fermions obey the Fermi statistics
and this point has important consequences
for the four-fermion excitation continuum considered in some detail above.

\subsection{Many-fermion excitations}

We pass to dynamic structure factors which are governed by many-fermion excitations.
Let us recall that according to the Jordan-Wigner transformation
we have
\begin{eqnarray}
s_n^x=\frac{1}{2}\left(1-2c^{\dagger}_1c_1\right)
\ldots\left(1-2c^{\dagger}_{n-1}c_{n-1}\right)\left(c_n^{\dagger}+c_n\right)
\nonumber\\
=\frac{1}{2}\varphi_1^+\varphi_1^-\ldots\varphi_{n-1}^+\varphi_{n-1}^-\varphi_n^+,
\nonumber\\
s_n^y=\frac{1}{2{\rm{i}}}\varphi_1^+\varphi_1^-\ldots\varphi_{n-1}^+\varphi_{n-1}^-\varphi_n^-,
\nonumber\\
s_n^z=-\frac{1}{2}\varphi_n^+\varphi_n^-
\label{4.28}
\end{eqnarray}
where we have introduced the operators
$\varphi_m^{\pm}=c_m^{\dagger}\pm c_m$.
Obviously the operators $\varphi_m^{\pm}$ are linear combinations
of the operators $c_k$, $c_k^{\dagger}$ in terms of which the Hamiltonian is diagonal
(see Eq. (\ref{4.01})).

Consider now the $xx$ time-dependent spin correlation function
\begin{eqnarray}
4\langle s_j^x(t)s^x_{j+n}\rangle
=
\langle\varphi_1^+(t)\varphi_1^-(t)\ldots\varphi_{j-1}^+(t)\varphi_{j-1}^-(t)\varphi_j^+(t)
\nonumber\\
\cdot
\varphi_1^+\varphi_1^-\ldots\varphi_{j-1}^+\varphi_{j-1}^-\varphi_j^+\varphi_j^-
\varphi_{j+1}^+\varphi_{j+1}^-\ldots\varphi_{j+n-1}^+\varphi_{j+n-1}^-\varphi_{j+n}^+\rangle.
\label{4.29}
\end{eqnarray}
It contains a product of $2(2j+n-1)$ $\varphi^{\pm}$ operators
(in contrast to
$4\langle s_j^z(t)s^z_{j+n}\rangle
=\langle\varphi_j^+(t)\varphi_j^-(t)\varphi_{j+n}^+\varphi_{j+n}^-\rangle$
which contains the product of only four $\varphi^{\pm}$ operators).
Therefore the calculation of $xx$ and $xy$ dynamic quantities
(which are governed by many-fermion excitations)
is essentially more complicated.

Exact analytical results for $xx$ and $xy$ dynamic quantities are rather scarce.
At the high-temperature limit $T\to\infty$ we know\cite{038,039} that
\begin{eqnarray}
4\langle s_j^x(t)s^x_{j+n}\rangle
=
\delta_{n,0}\cos\left(\Omega t\right)\exp\left(-\frac{1}{4}J^2t^2\right),
\nonumber\\
4\langle s_j^x(t)s^y_{j+n}\rangle
=
-\delta_{n,0}\sin\left(\Omega t\right)\exp\left(-\frac{1}{4}J^2t^2\right).
\label{4.30}
\end{eqnarray}

At zero temperature the $xx$ and $xy$ time-dependent correlation functions
are extremely simple only when $\vert\Omega\vert>\vert J\vert$\cite{040}.
Consider for example the case $\Omega>\vert J\vert$
when the ground state is completely polarized
$\vert {\rm{GS}}_s\rangle=\prod_{n=1}^N\vert\downarrow_n\rangle$
(in spin language)
or completely empty
$c_k\vert {\rm{GS}}_c\rangle=0$
(in fermionic language).
Owing to the simplicity of the ground state
$s^+_m\vert {\rm{GS}}_s\rangle
=c_m^{\dagger}\vert {\rm{GS}}_c\rangle
=\left(1/\sqrt{N}\right)\sum_k\exp\left({\rm{i}}km\right)
c_k^{\dagger}\vert {\rm{GS}}_c\rangle$,
$s_m^-\vert {\rm{GS}}_s\rangle=0$.
Therefore
\begin{eqnarray}
4\langle s_j^x(t)s^x_{j+n}\rangle
=\langle {\rm{GS}}_s\vert s_j^-(t)s_{j+n}^+ \vert{\rm{GS}}_s\rangle
\nonumber\\
=\frac{1}{N}\sum_k
\exp\left({\rm{i}}kn-{\rm{i}}\left(\Omega+J\cos k\right)t\right),
\nonumber\\
4\langle s_j^x(t)s^y_{j+n}\rangle
=-{\rm{i}}\langle {\rm{GS}}_s\vert s_j^-(t)s_{j+n}^+ \vert{\rm{GS}}_s\rangle
=-4{\rm{i}}\langle s_j^x(t)s^x_{j+n}\rangle
\label{4.31}
\end{eqnarray}
and as a result
\begin{eqnarray}
S_{xx}(k,\omega)={\rm{i}}S_{xy}(k,\omega)
=\frac{\pi}{2}\delta\left(\omega-\Omega-J\cos k\right).
\label{4.32}
\end{eqnarray}

Many results at $T=0$ refer to the asymptotic behavior
of the $xx$ or $xy$ time-dependent spin correlation functions\cite{041,042}.
From the paper by A.~R.~Its et al
we know the long-time asymptotic behavior at nonzero temperatures
\begin{eqnarray}
\langle s_j^+(t)s_{j+n}^-\rangle
\sim
\left\{
\begin{array}{ll}
\exp\left(f(n,0)\right),                   & \frac{n}{Jt}>1, \\
t^{2\left(\nu_-^2+\nu_+^2\right)}\exp\left(f(n,t)\right), & \frac{n}{Jt}<1,
\end{array}
\right.
\nonumber\\
f(n,t)=\frac{1}{2\pi}
\int_{-\pi}^{\pi}{\rm{d}}p\vert n+Jt\,\sin p\vert
\ln\left\vert\tanh\frac{\beta\left(\Omega-J\cos p\right)}{2}\right\vert,
\nonumber\\
\nu_{\pm}
=\frac{1}{2\pi}
\ln\left\vert\tanh\frac{\beta\left(\Omega\mp J\sqrt{1-\left(\frac{n}{Jt}\right)^2}\right)}{2}\right\vert.
\label{4.33}
\end{eqnarray}

On the other hand,
we can obtain the $xx$ and $xy$ dynamic quantities numerically\cite{044,045,046,snm,047,048,049}.
Consider the slightly more complicated inhomogeneous spin-1/2 anisotropic $XY$ chain
in a transverse field with the Hamiltonian
\begin{eqnarray}
H=\sum_{j=1}^N\Omega_js_j^z
+\sum_{j=1}^{N-1}
\left(J_j^{xx}s_j^xs_{j+1}^x+J_j^{xy}s_j^xs_{j+1}^y+
J_j^{yx}s_j^ys_{j+1}^x+J_j^{yy}s_j^ys_{j+1}^y
\right)
\nonumber\\
\to
-\frac{1}{2}\sum_{j=1}^N\Omega_j
+\sum_{i,j=1}^N
\left(
c_i^{\dagger}A_{ij}c_j+\frac{1}{2}
\left(c_i^{\dagger}B_{ij}c_j^{\dagger}
-c_iB_{ij}^{*}c_j\right)
\right)
\label{4.34}
\end{eqnarray}
where
\begin{eqnarray}
A_{ij}
=\Omega_i\delta_{ij}
+J_i^{+-}\delta_{j,i+1}
+J_{i-1}^{-+}\delta_{j,i-1}=A_{ji}^{*},
\nonumber\\
B_{ij}
=J_i^{++}\delta_{j,i+1}
-J_{i-1}^{++}\delta_{j,i-1}
=-B_{ji},
\nonumber\\
J_j^{+-}=\frac{1}{4}
\left( J_j^{xx}+J_j^{yy}+{\rm{i}}\left(J_j^{xy}-J_j^{yx}\right)\right)
=\left(J_j^{-+}\right)^{*},
\nonumber\\
J_j^{++}=\frac{1}{4}
\left( J_j^{xx}-J_j^{yy}-{\rm{i}}\left(J_j^{xy}+J_j^{yx}\right)\right)
=\left(J_j^{--}\right)^{*}.
\label{4.35}
\end{eqnarray}
To diagonalize a form bilinear in Fermi operators like (\ref{4.34})
we perform
the linear canonical transformation
\begin{eqnarray}
\eta_k
=\sum_{n=1}^N\left(g_{kn}c_n+h_{kn}c_n^{\dagger}\right),
\;\;\;
\eta_k^{\dagger}
=\sum_{n=1}^N\left(g^{*}_{kn}c_n^{\dagger}+h^{*}_{kn}c_n\right).
\label{4.36}
\end{eqnarray}
The resulting Hamiltonian reads as follows
\begin{eqnarray}
H
=\sum_{k=1}^N\Lambda_k
\left(\eta_k^{\dagger}\eta_k-\frac{1}{2}\right),
\nonumber\\
\left\{\eta_{k^{\prime}},\eta^{\dagger}_{k^{\prime\prime}}\right\}
=\delta_{k^{\prime}k^{\prime\prime}},
\;\;\;
\left\{\eta_{k^{\prime}},\eta_{k^{\prime\prime}}\right\}
=\left\{\eta^{\dagger}_{k^{\prime}},\eta^{\dagger}_{k^{\prime\prime}}\right\}=0
\label{4.37}
\end{eqnarray}
if the coefficients $g_{kn}$, $h_{kn}$
satisfy the set of equations
\begin{eqnarray}
\left(
\begin{array}{cc}
{\bf{g}}_k & {\bf{h}}_k
\end{array}
\right)
{\bf{M}}=\Lambda_k
\left(
\begin{array}{cc}
{\bf{g}}_k & {\bf{h}}_k
\end{array}
\right),
\nonumber\\
{\bf{g}}_k
=
\left(
\begin{array}{ccc}
g_{k1} & \ldots & g_{kN}
\end{array}
\right),
\;\;\;
{\bf{h}}_k
=
\left(
\begin{array}{ccc}
h_{k1} & \ldots & h_{kN}
\end{array}
\right),
\;\;\;
{\bf{M}}
=\left(
\begin{array}{cc}
{\bf{A}}      & {\bf{B}} \\
-{\bf{B}}^{*} & -{\bf{A}}^{*}
\end{array}
\right).
\label{4.38}
\end{eqnarray}
Further it may be convenient to introduce the linear combinations
$\Phi_{kn}=g_{kn}+h_{kn}$
and
$\Psi_{kn}=g_{kn}-h_{kn}$
which enter the relations
\begin{eqnarray}
\varphi_j^{+}
=c_j^{\dagger}+c_j
=\sum_{p=1}^N
\left(
\Phi_{pj}\eta_p^{\dagger}+\Phi_{pj}^{*}\eta_p
\right),
\nonumber\\
\varphi_j^{-}
=c_j^{\dagger}-c_j
=\sum_{p=1}^N
\left(
\Psi_{pj}\eta_p^{\dagger}-\Psi_{pj}^{*}\eta_p
\right).
\label{4.39}
\end{eqnarray}
We calculate the time-dependent spin correlation functions
using the Wick-Bloch-de Dominicis theorem.
For example,
\begin{eqnarray}
4\langle s_n^z(t) s_{n+m}^z \rangle
=\langle\varphi_n^+(t)\varphi_n^-(t)\varphi_{n+m}^+\varphi_{n+m}^-\rangle
\nonumber\\
=\langle\varphi_n^+\varphi_n^-\rangle\langle\varphi_{n+m}^+\varphi_{n+m}^-\rangle
\nonumber\\
-\langle\varphi_n^+(t)\varphi_{n+m}^+\rangle\langle\varphi_{n}^-(t)\varphi_{n+m}^-\rangle
+\langle\varphi_n^+(t)\varphi_{n+m}^-\rangle\langle\varphi_{n}^-(t)\varphi_{n+m}^+\rangle.
\label{4.40}
\end{eqnarray}
The r.h.s. of Eq. (\ref{4.40})
may be compactly written
as the Pfaffian
of the $4\times 4$ antisymmetric matrix
\begin{eqnarray}
4\langle s_n^z(t) s_{n+m}^z \rangle
\nonumber\\
={\mbox{Pf}}
\left(
\begin{array}{cccc}
0 &
\langle \varphi^+_n\varphi_n^- \rangle &
\langle \varphi^+_n(t)\varphi_{n+m}^+ \rangle &
\langle \varphi^+_n(t)\varphi_{n+m}^- \rangle \\
-\langle \varphi^+_n\varphi_n^- \rangle &
0 &
\langle \varphi^-_n(t)\varphi_{n+m}^+ \rangle &
\langle \varphi^-_n(t)\varphi_{n+m}^- \rangle \\
-\langle \varphi^+_n(t)\varphi_{n+m}^+ \rangle &
-\langle \varphi^-_n(t)\varphi_{n+m}^+ \rangle &
0 &
\langle \varphi^+_{n+m}\varphi_{n+m}^- \rangle \\
-\langle \varphi^+_n(t)\varphi_{n+m}^- \rangle &
-\langle \varphi^-_n(t)\varphi_{n+m}^- \rangle &
-\langle \varphi^+_{n+m}\varphi_{n+m}^- \rangle &
0
\end{array}
\right).
\label{4.41}
\end{eqnarray}
Similarly (see Eq. (\ref{4.29})),
for the more complicated $xx$ time-dependent spin correlation function we have
\begin{eqnarray}
4\langle s_n^x(t) s_{n+m}^x \rangle
\nonumber\\
={\mbox{Pf}}
\left(
\begin{array}{ccccc}
0                                          &
\langle \varphi^+_1\varphi_{1}^- \rangle &
\langle \varphi^+_1\varphi_{2}^+ \rangle &
\ldots                                     &
\langle \varphi^+_1(t)\varphi_{n+m}^+ \rangle \\
-\langle \varphi^+_1\varphi_{1}^- \rangle    &
0                                              &
\langle \varphi^-_{1}\varphi_{2}^+ \rangle &
\ldots                                         &
\langle \varphi^-_{1}(t)\varphi_{n+m}^+ \rangle \\
\vdots &
\vdots &
\vdots &
\cdots &
\vdots \\
-\langle \varphi^+_1(t)\varphi_{n+m}^+ \rangle &
-\langle \varphi^-_1(t)\varphi_{n+m}^+ \rangle &
-\langle \varphi^+_2(t)\varphi_{n+m}^+ \rangle &
\ldots                                          &
0
\end{array}
\right),
\label{4.42}
\end{eqnarray}
i.e. $\langle s_j^x(t)s_{j+n}^x\rangle$
can be written as a Pfaffian of a 
$2(2j+n-1)\times2(2j+n-1)$ antisymmetric matrix.
The elementary contractions
involved in (\ref{4.41}), (\ref{4.42})
read
\begin{eqnarray}
\langle\varphi_j^+(t)\varphi_{m}^+\rangle
=\sum_{p=1}^N
\left(\Phi_{pj}\Phi^{*}_{pm}F(\Lambda_p)+\Phi^{*}_{pj}\Phi_{pm}F(-\Lambda_p)\right),
\nonumber\\
\langle\varphi_j^+(t)\varphi_{m}^-\rangle
=\sum_{p=1}^N
\left(-\Phi_{pj}\Psi^{*}_{pm}F(\Lambda_p)+\Phi^{*}_{pj}\Psi_{pm}F(-\Lambda_p)\right),
\nonumber\\
\langle\varphi_j^-(t)\varphi_{m}^+\rangle
=\sum_{p=1}^N
\left(\Psi_{pj}\Phi^{*}_{pm}F(\Lambda_p)-\Psi^{*}_{pj}\Phi_{pm}F(-\Lambda_p)\right),
\nonumber\\
\langle\varphi_j^-(t)\varphi_{m}^-\rangle
=-\sum_{p=1}^N
\left(\Psi_{pj}\Psi^{*}_{pm}F(\Lambda_p)+\Psi^{*}_{pj}\Psi_{pm}F(-\Lambda_p)\right),
\nonumber\\
F(\Lambda_p)=\frac{\exp\left({\rm{i}}\Lambda_p t\right)}{1+\exp\left(\beta\Lambda_p\right)}.
\label{4.43}
\end{eqnarray}

It is worthwhile to recall some properties of the Pfaffians
which are used for calculating them.
In the first numerical studies
the authors used the relation
\begin{eqnarray}
\left({\mbox{Pf}}{\bf{A}}\right)^2
={\mbox{det}}{\bf{A}}
\label{4.44}
\end{eqnarray}
and computed numerically the determinants
which gave the values of Pfaffians according to (\ref{4.44}).
On the other hand,
the Pfaffian may be computed directly\cite{045}
noting that
\begin{eqnarray}
{\mbox{Pf}}\left({\bf{U}}^T{\bf{A}}{\bf{U}}\right)
={\mbox{det}}{\bf{U}}\;{\mbox{Pf}}{\bf{A}}
\label{4.45}
\end{eqnarray}
and that
\begin{eqnarray}
{\mbox{Pf}}
\left(
\begin{array}{cccccc}
0      &
R_{12} &
0      &
0      &
\ldots &
0      \\
-R_{12} &
0       &
0       &
0       &
\ldots  &
0       \\
0         &
0         &
0         &
R_{34}    &
\ldots    &
0         \\
0       &
0       &
-R_{34} &
0       &
\ldots  &
0       \\
\vdots &
\vdots &
\vdots &
\vdots &
\ldots &
\vdots \\
0 &
0 &
0 &
0 &
\ldots &
0
\end{array}
\right)
=R_{12}R_{34}\ldots .
\label{4.46}
\end{eqnarray}

\begin{figure}[th]
\centerline{\psfig{file=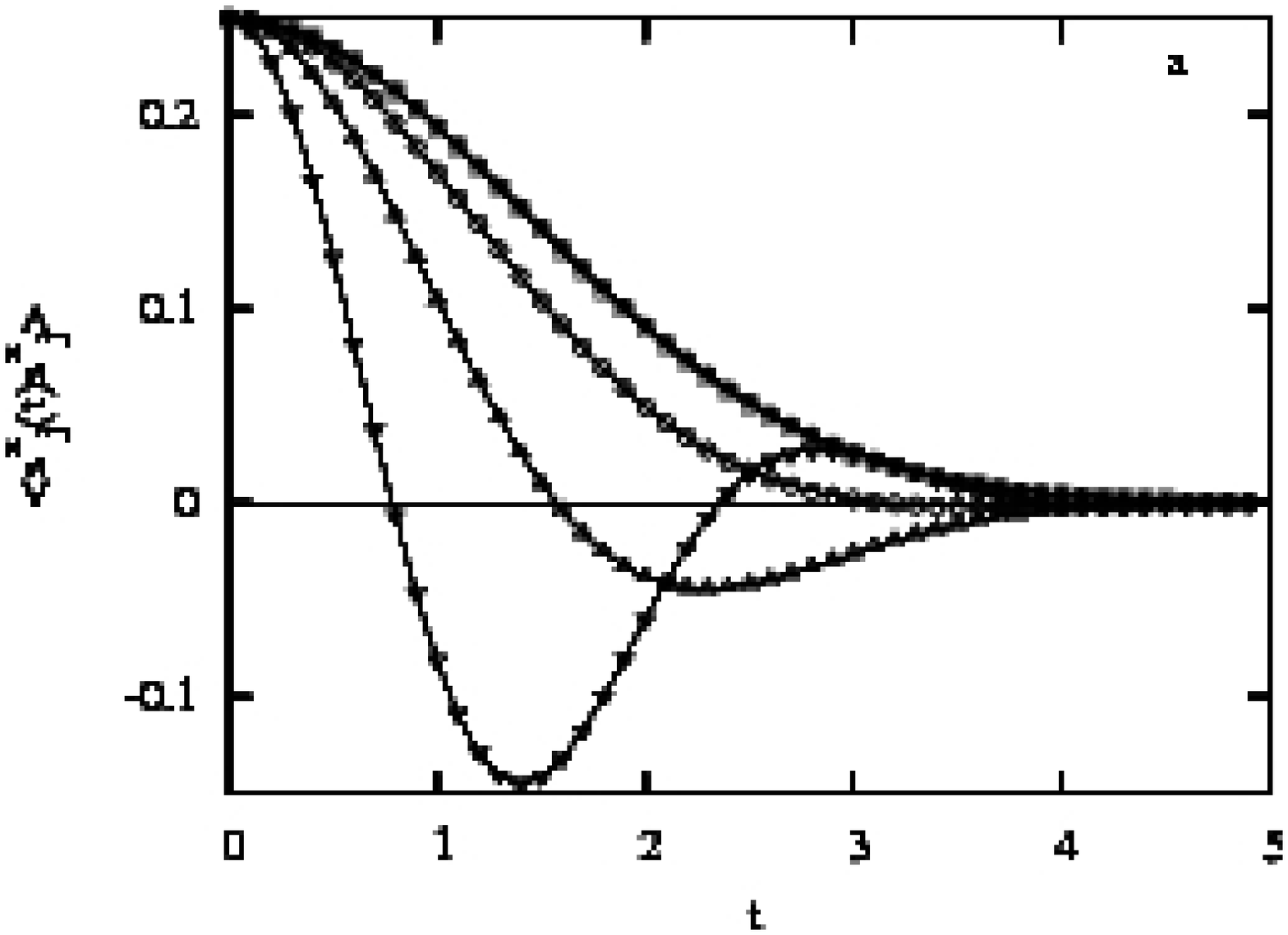,width=2.0in,angle=0}
\psfig{file=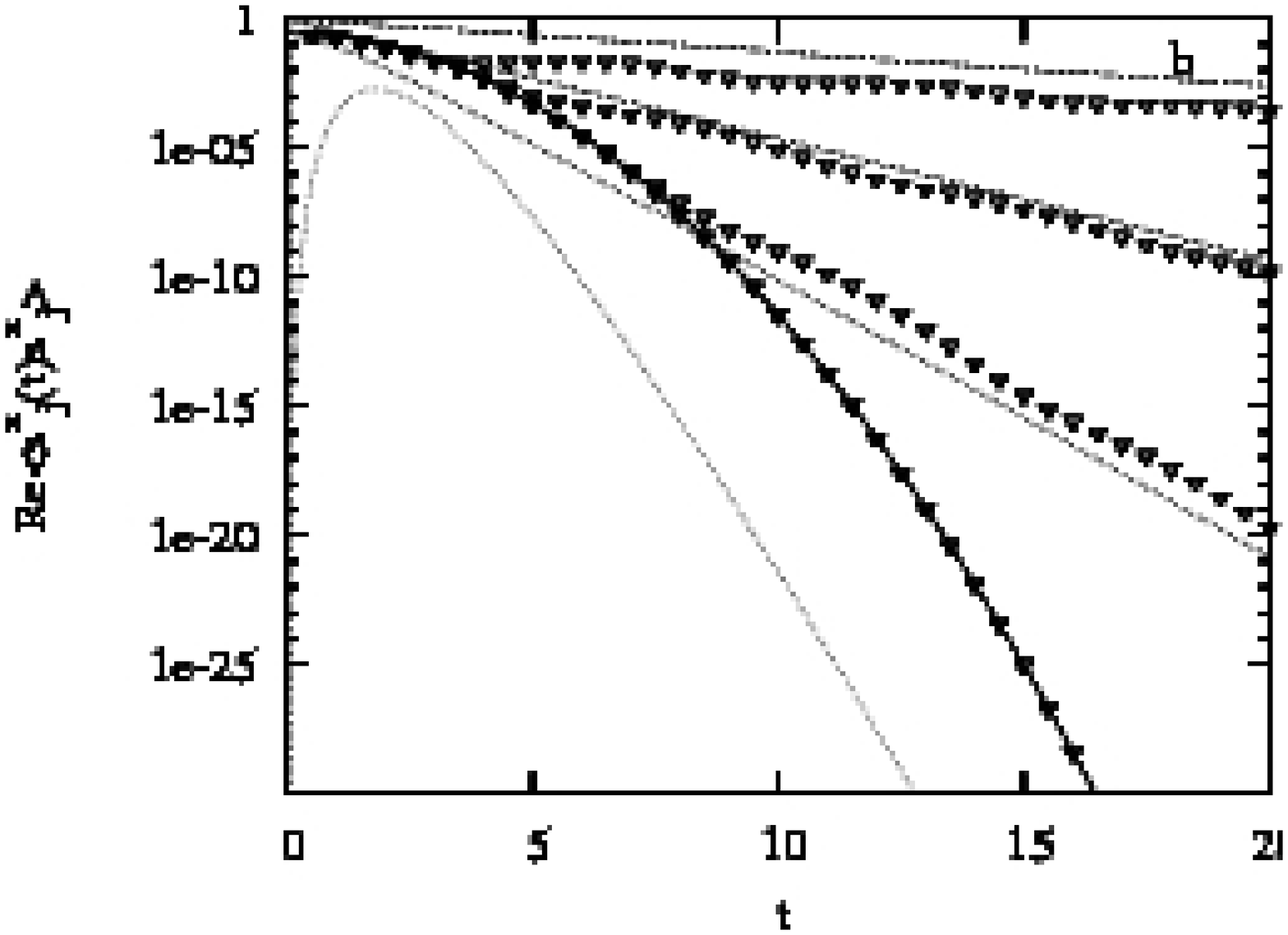,width=2.0in,angle=0}}
\vspace*{8pt}
\caption{Panel a:
Time-dependence of the autocorrelation function $\langle s_j^x(t)s_j^x\rangle$, $j=51$
at infinite temperature obtained numerically (symbols)
and analytically (see Eq. (\ref{4.30})) (solid curves).
$\Omega=2,\,1$ (downward and upward triangles),
$\Omega=0.5$ (open circles),
$\Omega=0.1$ (squares),
$\Omega=0$ (filled circles).
Panel b:
Time-dependence of the real part of the autocorrelation function $\langle s_j^x(t)s_j^x\rangle$, $j=51$
at $\Omega=0$ for various temperatures
obtained numerically (symbols) in comparison with asymptotics (\ref{4.33}).
$\beta=5,\,1,\,0.1,\,0.00001$ (from top to bottom).
The exact analytical result for $\beta=0$ is also shown (the lowest curve).
Evidently only the slopes of the asymptotics should be compared with the numerical results.}
\label{fig09}
\end{figure}

\begin{figure}[th]
\centerline{\psfig{file=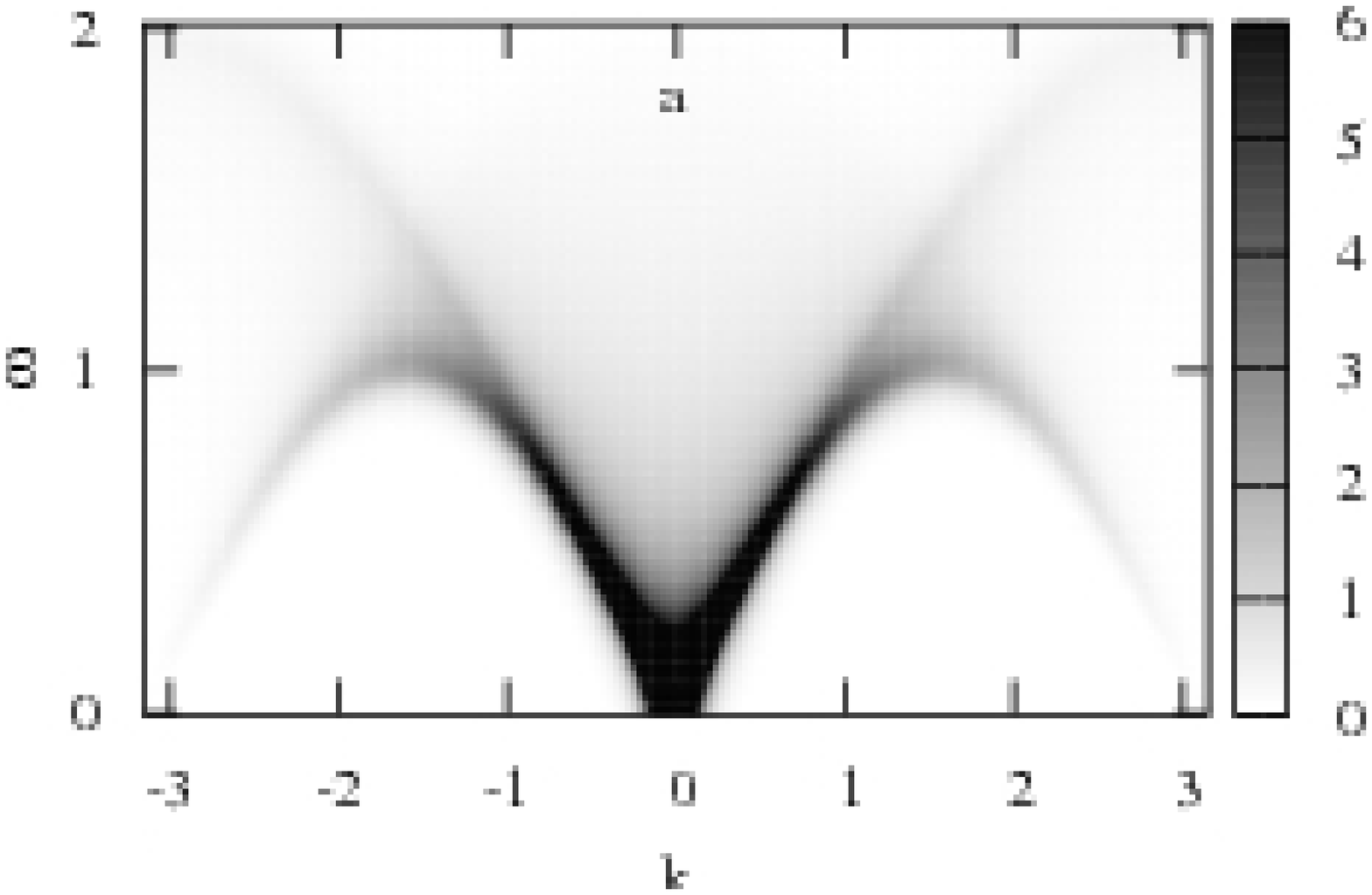,width=2.0in,angle=0}
\psfig{file=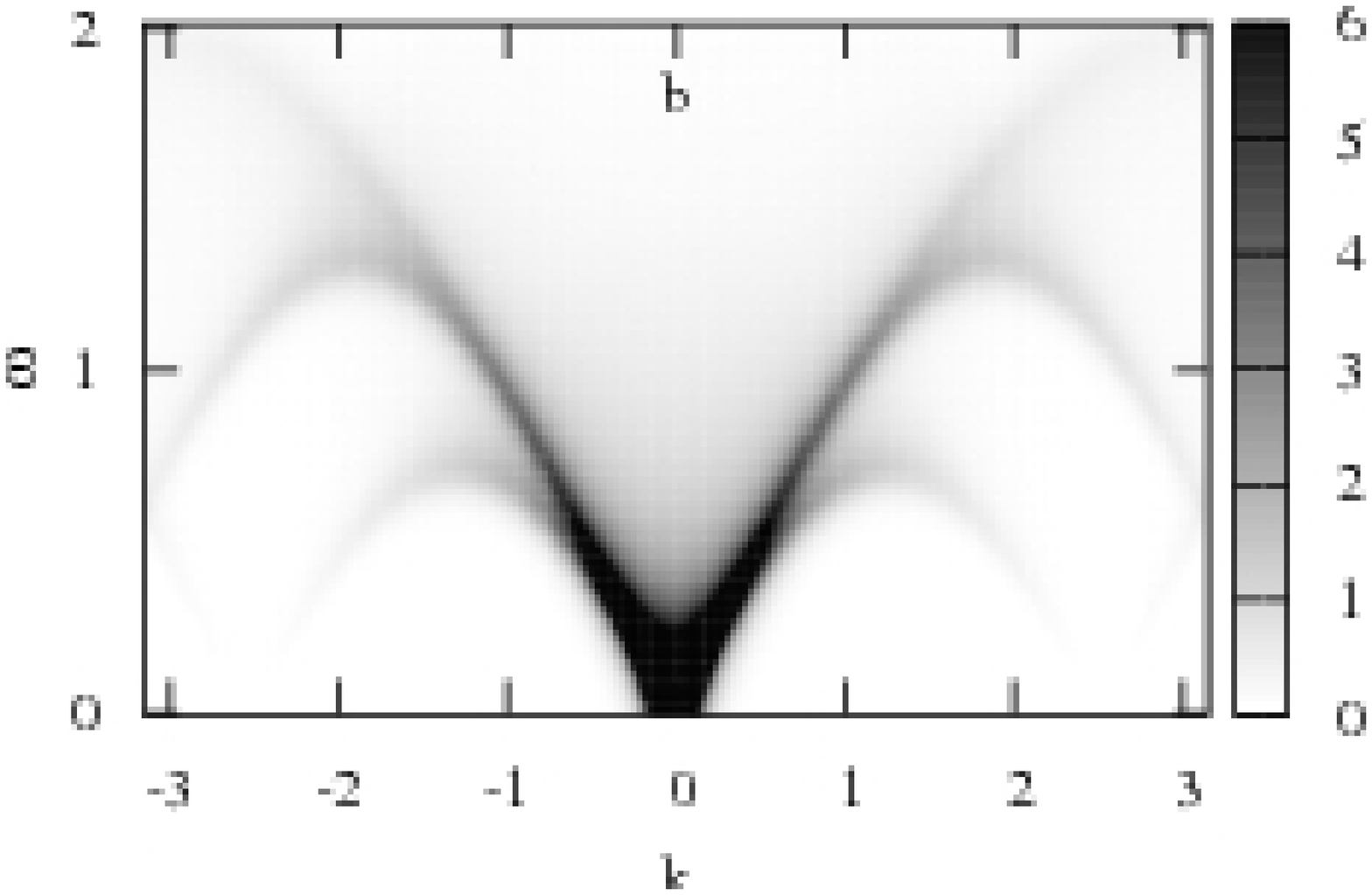,width=2.0in,angle=0}}
\centerline{\psfig{file=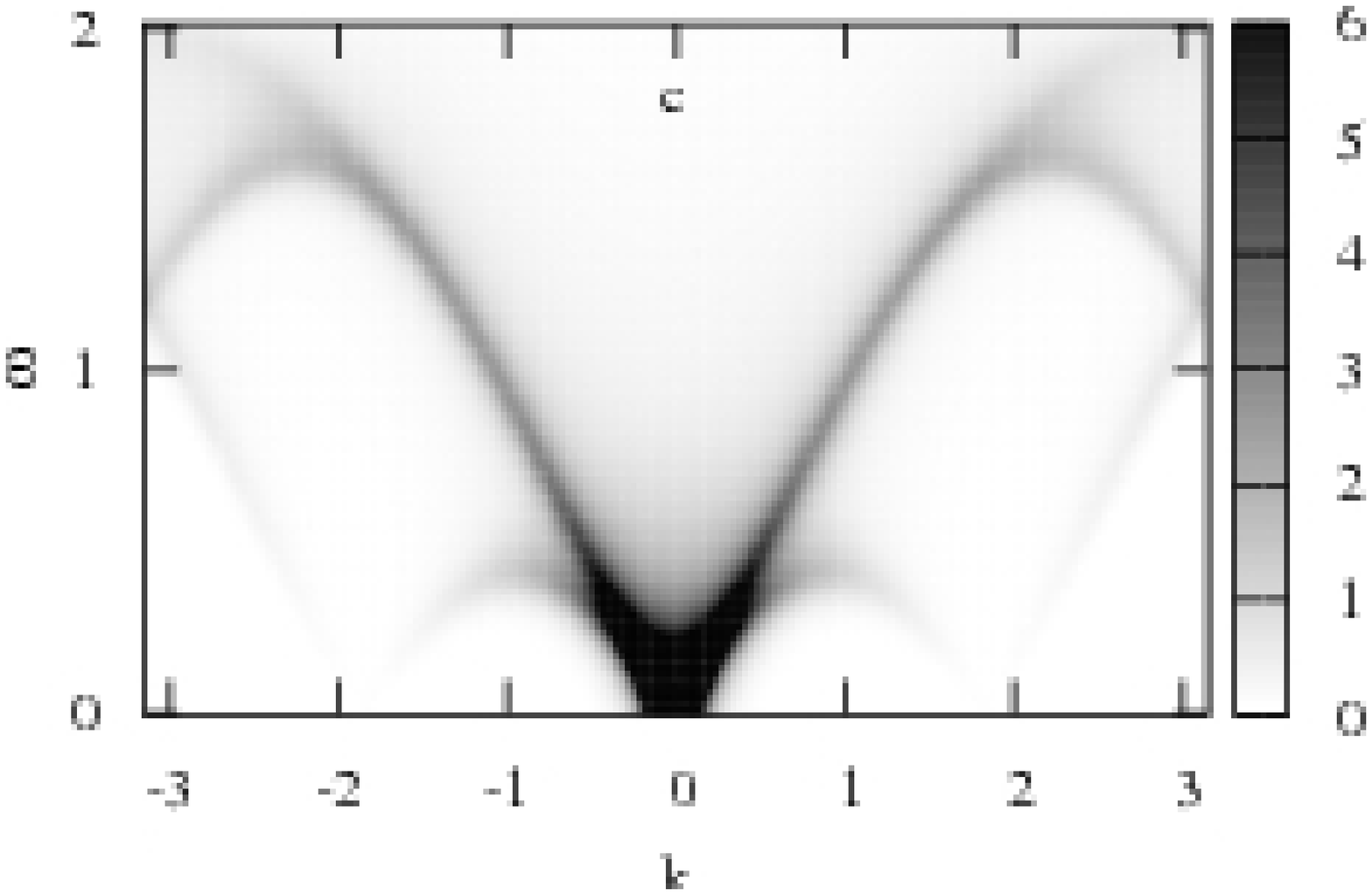,width=2.0in,angle=0}
\psfig{file=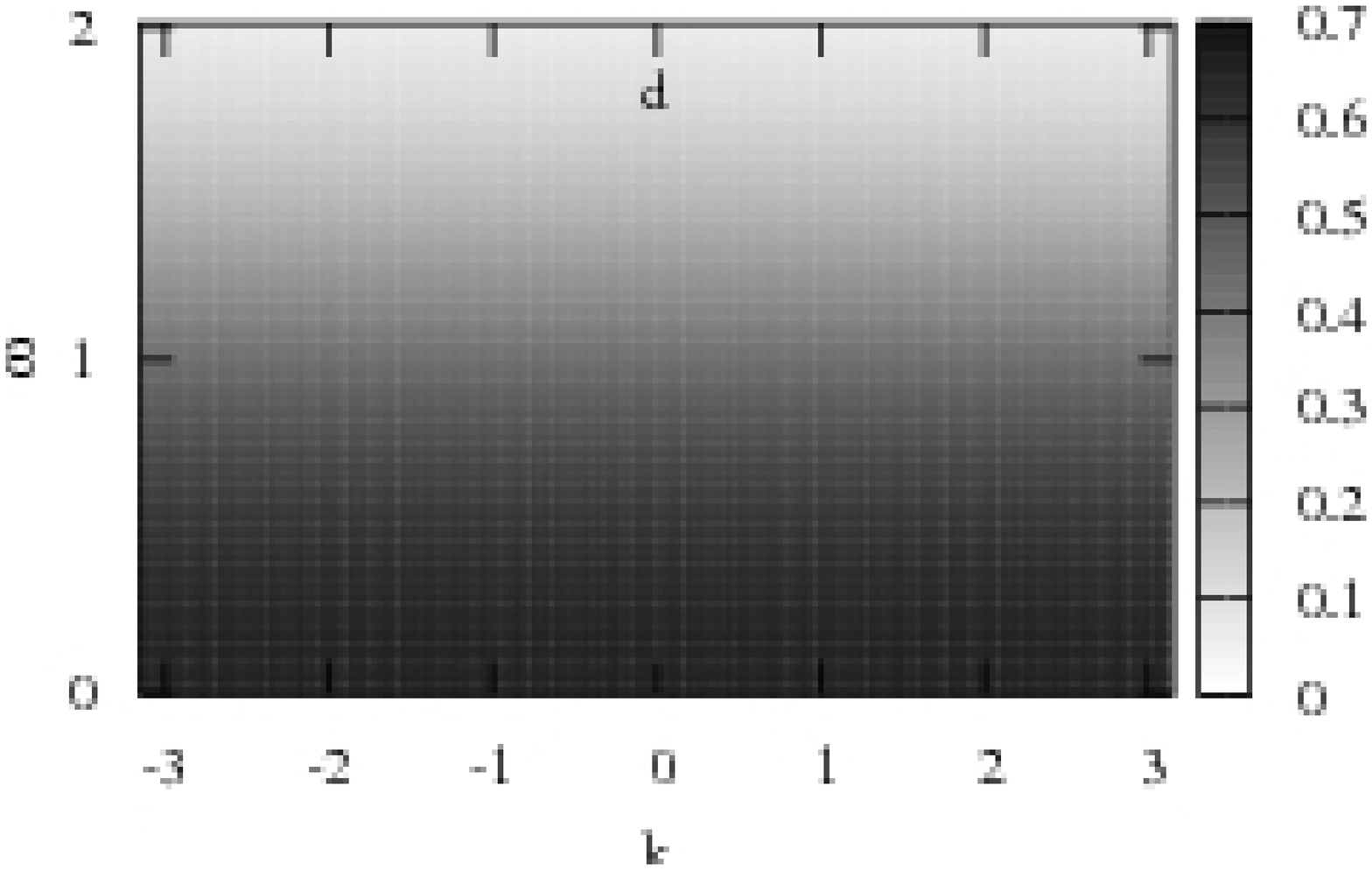,width=2.0in,angle=0}}
\vspace*{8pt}
\caption{$S_{xx}(k,\omega)$ (gray-scale plots)
for the chain (\ref{2.11}) with $J=-1$,
for $\Omega=0.0001$ (a),
$\Omega=0.3$ (b),
$\Omega=0.6$ (c)
at low temperature $\beta=20$
and
for $\Omega=0.6$
in the high-temperature limit $\beta=0$ (d).}
\label{fig10}
\end{figure}

\begin{figure}[th]
\centerline{\psfig{file=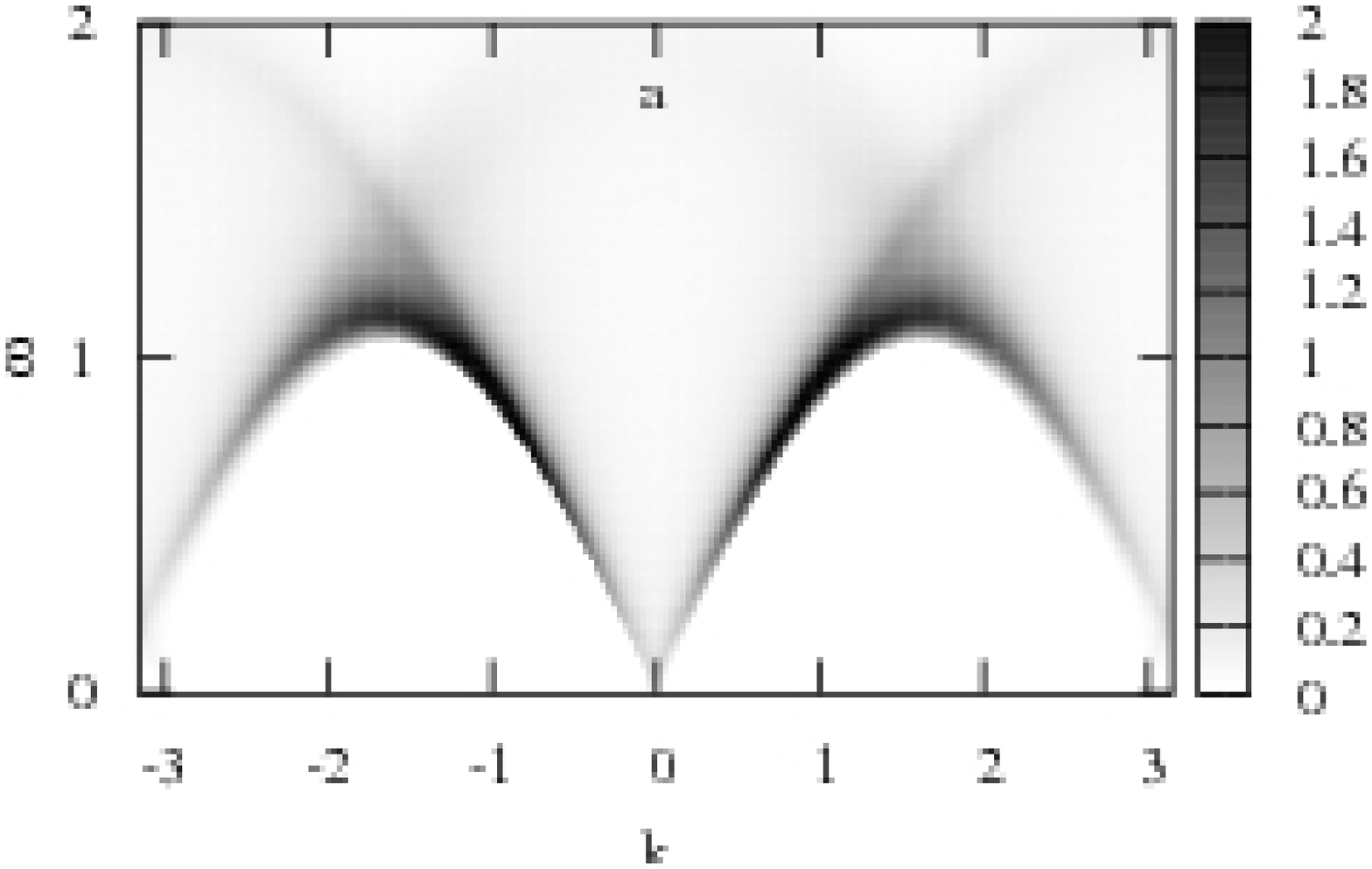,width=2.0in,angle=0}
\psfig{file=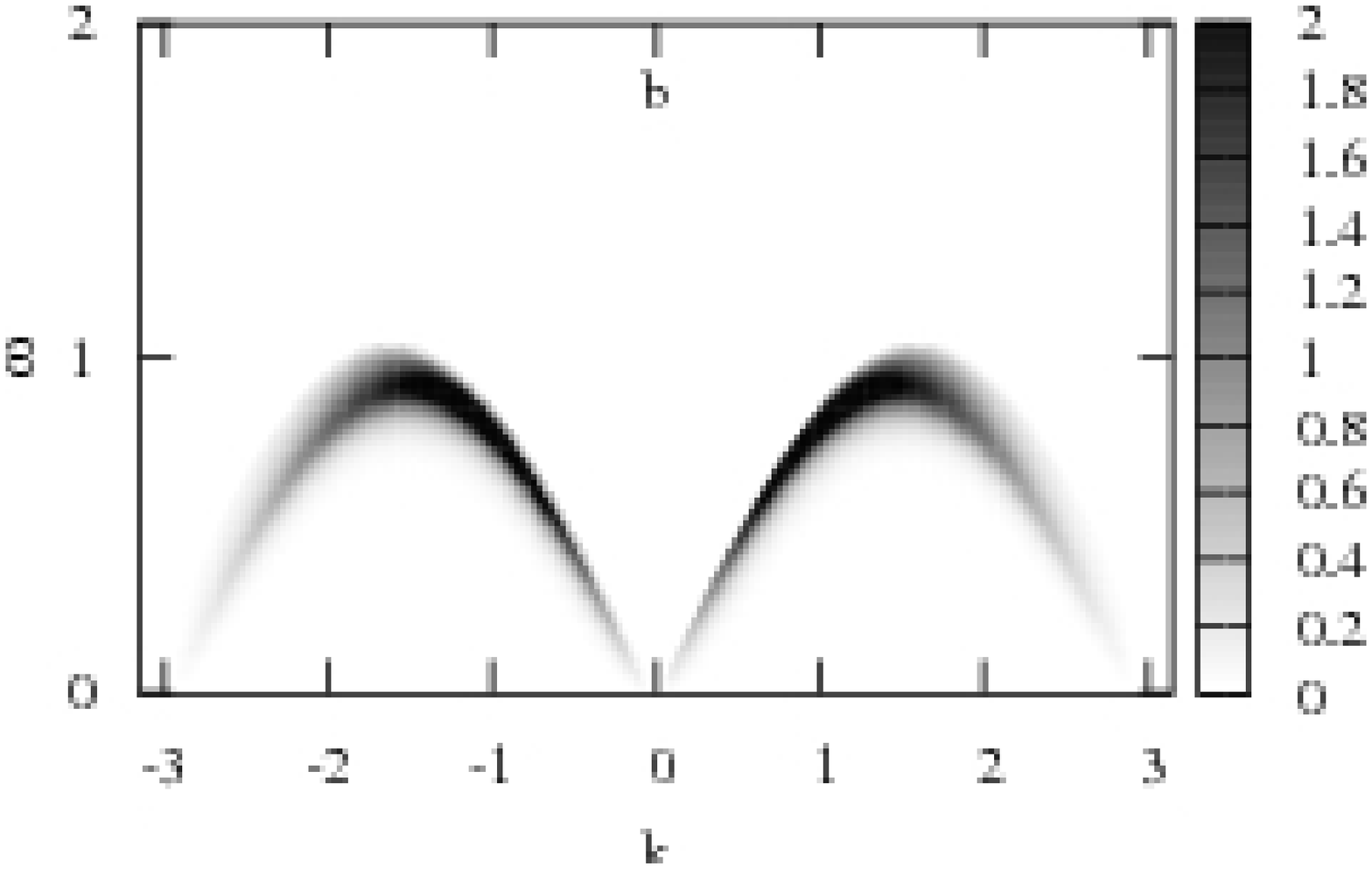,width=2.0in,angle=0}}
\centerline{\psfig{file=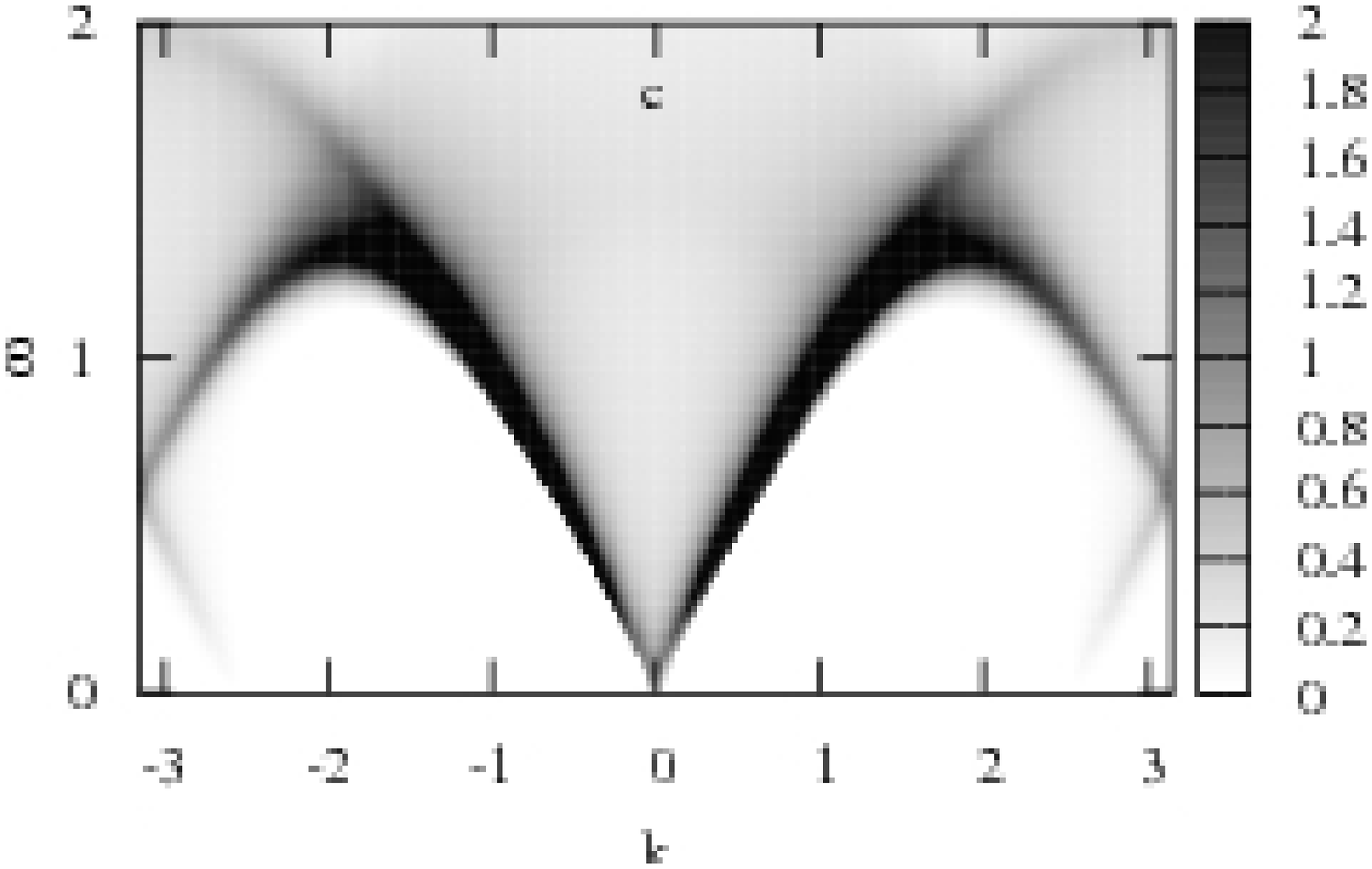,width=2.0in,angle=0}
\psfig{file=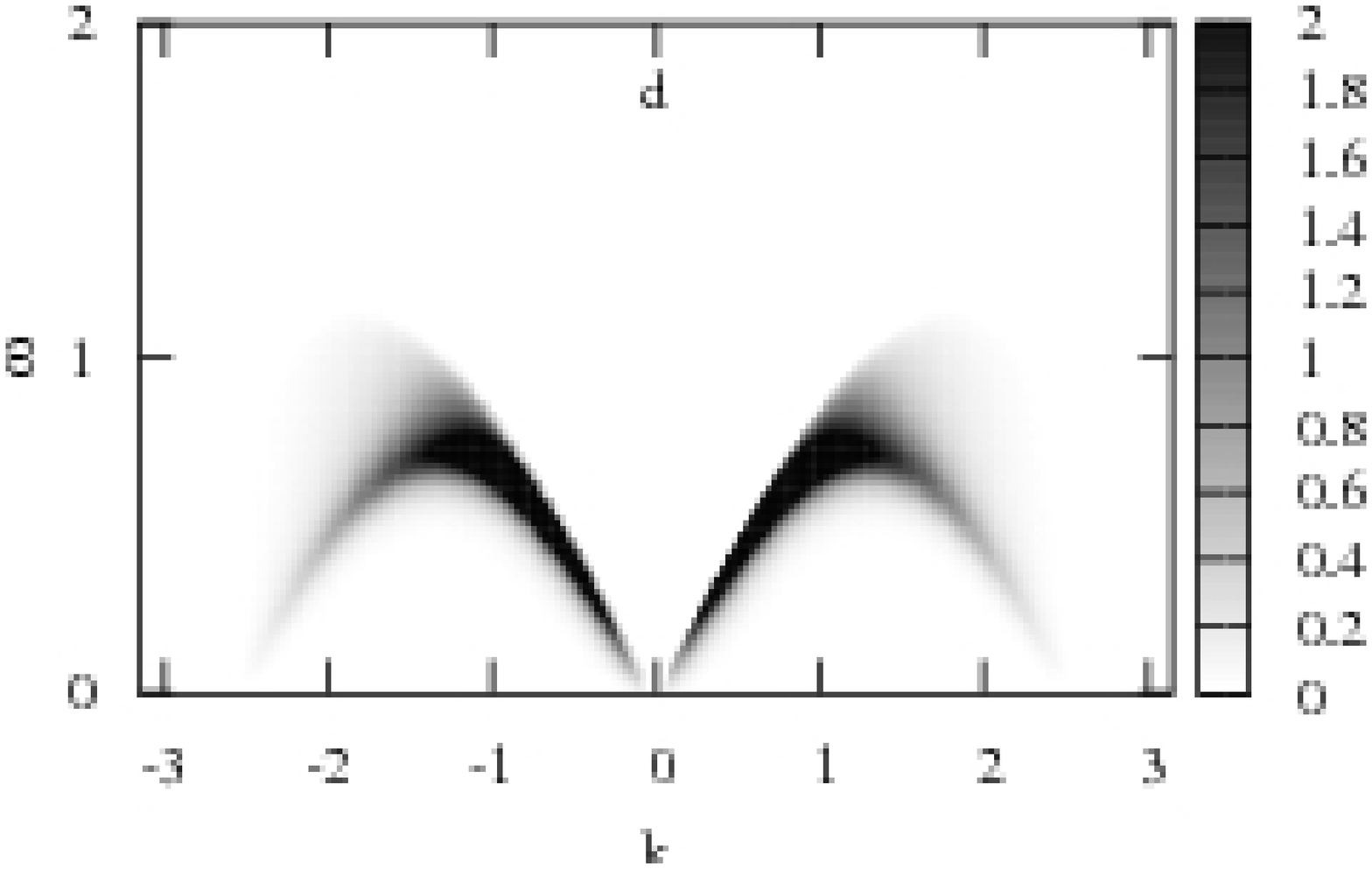,width=2.0in,angle=0}}
\vspace*{8pt}
\caption{${\rm{i}}S_{xy}(k,\omega)$ (gray-scale plots)
for the chain (\ref{2.11}) with $J=-1$
for $\Omega=0.1$ (a, b),
$\Omega=0.3$ (c, d)
at low temperature $\beta=20$.
Positive parts are shown in panels a and c,
negative parts are shown in panels b and d.}
\label{fig11}
\end{figure}

We use the approach described to calculate the $xx$ and $xy$ dynamic structure factors
for the spin-1/2 transverse $XX$ chain numerically\cite{050}.
To estimate the quality of the numerical procedure
we compare our numerical findings
with exact analytical results in the high-temperature limit
and
with exact asymptotics for finite temperatures
in Fig.~\ref{fig09}.
Knowing the time-dependent correlation functions
we obtain the corresponding dynamic structure factors according to
\begin{eqnarray}
S_{xx}(k,\omega)
=\sum_{n=0,\pm 1,\ldots}\exp\left(-{\rm{i}}k n\right)
2\Re \left(
\int_{0}^{\infty}{\rm{d}}t\exp\left({\rm{i}}\left(\omega+{\rm{i}}\epsilon\right)t\right)
\langle s_j^x(t)s_{j+n}^x\rangle
\right),
\nonumber\\
S_{xy}(k,\omega)
=\sum_{n=0,\pm 1,\ldots}\exp\left(-{\rm{i}}k n\right)
2{\rm{i}}\Im \left(
\int_{0}^{\infty}{\rm{d}}t\exp\left({\rm{i}}\left(\omega+{\rm{i}}\epsilon\right)t\right)
\langle s_j^x(t)s_{j+n}^y\rangle
\right)
\label{4.47}
\end{eqnarray}
with $\epsilon\to +0$.
In practice we consider chains of $N=400$ sites,
take $j=41,\,51,\,61$,
$n$ up to 50 or up to 100,
and set $\epsilon=0 \ldots 0.001 \ldots 0.1$
(see Ref.~\cite{050}).
The results of our calculations for
$S_{xx}(k,\omega)$
and
$S_{xy}(k,\omega)$
are illustrated in Figs.~\ref{fig10},~\ref{fig11}.

Let us recall,
the transverse dynamic structure factor $S_{zz}(k,\omega)$ probes two-particle excitations,
i.e. it is governed by the excitations which are composed of two Jordan-Wigner spinless fermions.
The two-fermion excitation continuum has a sharp upper frequency
cutoff at which $S_{zz}(k,\omega)$ may diverge.
At $T=0$ it has also a sharp lower frequency cutoff which touches $\omega=0$
at $k_0$ (soft modes).
$S_{zz}(k,\omega)$ is almost structureless
(apart from upper boundary singularities)
and exists only for $\vert\Omega\vert<\vert J\vert$.
In the high-temperature limit $T\to\infty$
$S_{zz}(k,\omega)$ becomes $\Omega$-independent.
All these features are nicely seen in Fig.~\ref{fig03}.

In contrast,
the dynamic structure factors $S_{xx}(k,\omega)$ and $S_{xy}(k,\omega)$
are many-particle quantities in terms of the Jordan-Wigner spinless fermions.
The frequency range of these quantities is not {\it a priori} restricted,
however,
in the low-temperature limit $T\to 0$
$S_{xx}(k,\omega)$ and $S_{xy}(k,\omega)$
are rather small (but nonzero) outside the two-fermion excitation continuum.
These quantities show washed-out excitation branches
roughly following the boundary of the two-fermion excitation continuum.
[Although the results presented in Figs.~\ref{fig10},~\ref{fig11}
refer to the case $J<0$
(the ferromagnetic sign of the $XX$ exchange interaction)
the results for $J>0$
(the antiferromagnetic sign of the $XX$ exchange interaction)
follow by symmetry.
In fact,
while changing the sign of $XX$ exchange interaction,
$+J\to-J$,
we get $S_{xx}(k,\omega)$, $S_{xy}(k,\omega)$ given by  Eq. (\ref{4.47})
in which the wave-vector is changed $k\to k\mp\pi$.]
From the exact calculation in the strong-field zero-temperature limit (\ref{4.32})
we know that $S_{xx}(k,\omega)$, $S_{xy}(k,\omega)$ are proportional to
$\delta\left(\omega-\Lambda_k\right)$,
$\Lambda_k=\Omega+J\cos k$.
In the high-temperature limit $T\to\infty$
$S_{xx}(k,\omega)$ and $S_{xy}(k,\omega)$
become $k$-independent, but depend on $\Omega$.
All the features described can be seen in Figs.~\ref{fig10},~\ref{fig11}.

\begin{figure}[th]
\centerline{\psfig{file=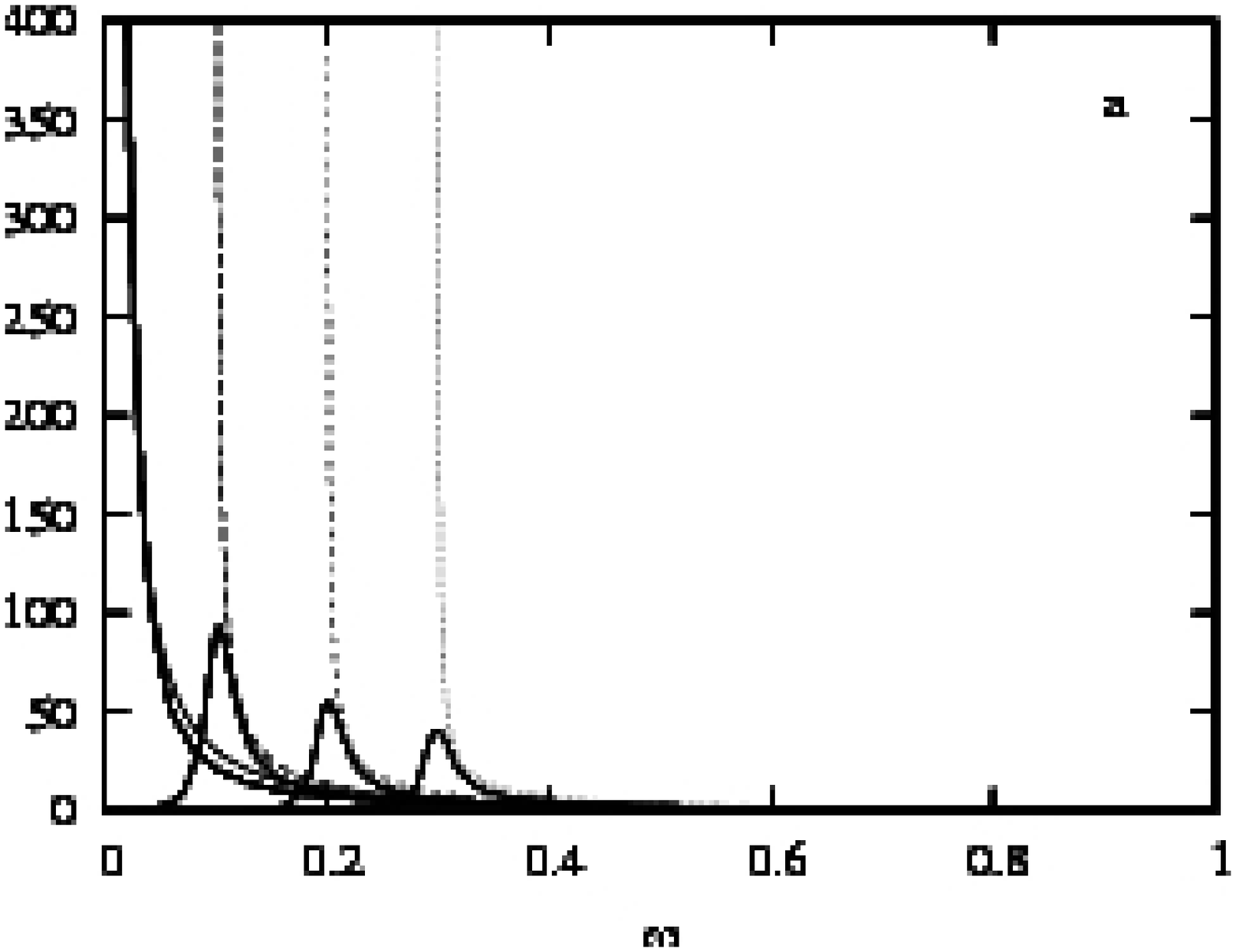,width=2.0in,angle=0}
\psfig{file=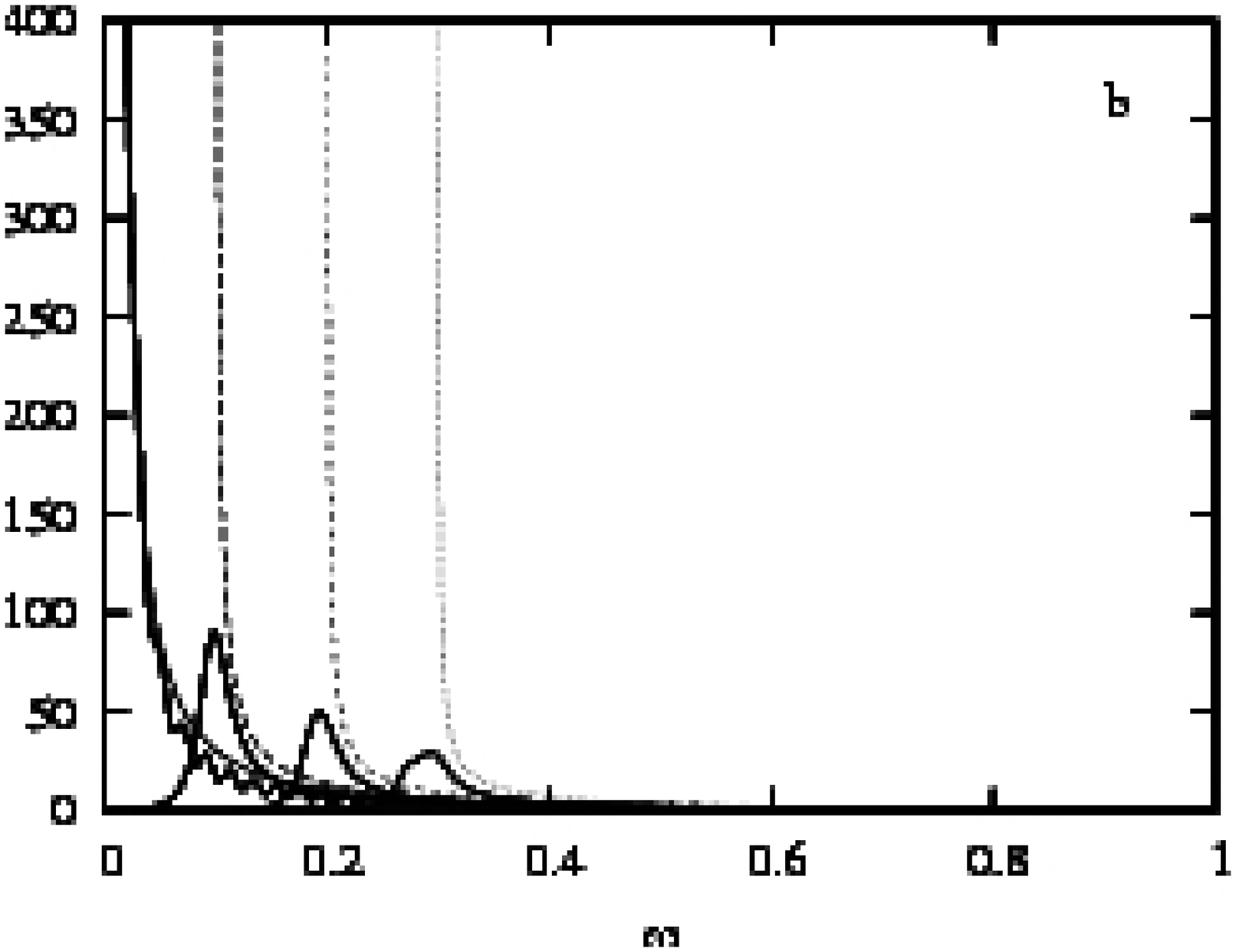,width=2.0in,angle=0}} \vspace*{8pt}
\caption{$S_{xx}(k,\omega)$ for the chain (\ref{2.11}) with $J=-1$;
frequency profiles at $k=0,\,0.1,\,0.2,\,0.3$ (from left to right)
at $\Omega=0$ (panel a) and $\Omega=0.3$ (panel b).
Bosonization results which follow from Eq. (\ref{4.48})
are shown by thin lines
($v=1$ for $\Omega=0$ and $v=0.9539\ldots$ for $\Omega=0.3$);
numerical results at low temperature $\beta=100$
are shown by solid lines.}
\label{fig12}
\end{figure}

It is instructive to compare our precise numerical findings in the low-temperature limit
with the results for the ground-state dynamic structure factors
obtained by bosonization\cite{016,017,018}.
In the case $\Omega=0$ within the framework of the bosonization approach
we have
\begin{eqnarray}
S_{\alpha\alpha}(k,\omega)
\sim
\frac{\theta\left(\omega-\vert v k\vert\right)}
{\left(\omega^2-\left(vk\right)^2\right)^{1-\frac{\eta_{\alpha}}{2}}},
\;\;\;
\alpha=x,z.
\label{4.48}
\end{eqnarray}
Here $v=J$ is the velocity and
$\eta_x=1/2$, $\eta_z=2$ are the exponents which describe correctly the singularity
at the lower continuum boundary
$\omega_l(k)=\vert J\sin k\vert\to \vert Jk\vert$ as $k\to 0$ or $k\to \pm \pi$.
In the case of nonzero transverse fields $\Omega\ne 0$
the values of the Fermi momentum and Fermi velocity are changed.
In Fig.~\ref{fig12}
we compare the predictions of the bosonization approach (\ref{4.48})
with the numerical results at low temperatures.

To summarize,
in this section we have discussed dynamic properties
of the spin-1/2 transverse $XX$ chain within the Jordan-Wigner fermionization approach.
Within this scheme the spin Hamiltonian
corresponds to the Hamiltonian of noninteracting spinless fermions.
The transverse dynamic structure factor corresponds
to the fermionic density dynamic structure factor
and probes the two-fermion excitation continuum.
There are more dynamic structure factors
which probe the two-fermion excitation continuum,
e.g., the dimer dynamic structure factor.
All two-fermion dynamic quantities have common features
(spectral boundaries, potential soft modes and singularities)
and specific features.
There are also dynamic quantities which probe the four-fermion excitation continuum;
as an example of such a quantity we have discussed the trimer dynamic structure factor.
Remarkably,
the dynamic structure factors which are associated with the dynamics of fluctuations
of the $x$ or $y$ spin components
(in contrast to the transverse dynamic structure factor
which is associated with dynamics of fluctuations
of the $z$ spin component)
are enormously complex within the Jordan-Wigner description
since they probe many-fermion excitations.
Nevertheless
the two-fermion excitation continuum
is still important for these dynamic quantities at low temperatures.
As we have observed in our numerics,
most of the spectral weight is concentrated
along the boundaries of the two-fermion excitation continuum
(it was also noted earlier for the $XXZ$ Heisenberg chain\cite{051}).
This is not the case in the high-temperature limit
when these dynamic structure factors show Gaussian ridges.
In the next two sections
we shall follow to what extent our observations survive for more complicated 
spin-1/2 $XY$ chains.

\section{Dimerized Spin-1/2 Isotropic $XY$ Chain in a Transverse Field
         \label{secdk5}}

Now we pass to the dimerized spin-1/2 $XX$ chain in a transverse field.
The Hamiltonian of the model reads
\begin{eqnarray}
H=\sum_{n}
J\left(1-(-1)^n\delta\right)
\left(s_n^xs_{n+1}^x+s_{n}^ys_{n+1}^y\right)
+\Omega\sum_{n}s_n^z
\nonumber\\
\to
\sum_{n}
\frac{J}{2}\left(1-(-1)^n\delta\right)
\left(c_n^{\dagger}c_{n+1}-c_{n}c_{n+1}^{\dagger}\right)
+\Omega\sum_{n}\left(c_{n}^{\dagger}c_{n}-\frac{1}{2}\right),
\label{5.01}
\end{eqnarray}
where $\delta$ is the dimerization parameter
($0<\delta<1$).
After performing consequently the Fourier transformation,
$c_n^{\dagger}=\left(1/\sqrt{N}\right)\sum_{k}\exp\left({\rm{i}}k n\right)c_k^{\dagger}$,
$k=2\pi p/N$, $p=-N/2,\ldots,N/2-1$ ($N$ is even),
and the Bogolyubov transformation,
$c_k=u_{k+\pi}\eta_k+{\rm{i}}v_k\eta_{k+\pi}$,
$u_k=\left(1/\sqrt{2}\right)\sqrt{1+\vert\cos k\vert/\epsilon_k}$,
$v_k={\rm{sgn}}\left(\sin(2k)\right)\left(1/\sqrt{2}\right)\sqrt{1-\vert\cos k\vert/\epsilon_k}$,
$\epsilon_k=\sqrt{\cos^2k+\delta^2\sin^2k}$,
the Hamiltonian becomes diagonal,
$H=\sum_k\Lambda_k\left(\eta_k^{\dagger}\eta_k-1/2\right)$
with the elementary excitation energy
$ \Lambda_k=\Omega+\lambda_k$,
$\lambda_k={\rm{sgn}}(\cos k)J\epsilon_k$
(for further details see Refs.~\cite{033,052}).

The calculation of the transverse dynamic structure factor
follows the scheme described in Sec.~\ref{secdk4}
and ends up with the following result
\begin{eqnarray}
S_{zz}(k,\omega)
\nonumber\\
=\int_{-\pi}^{\pi}
{\rm{d}}k_1
\left(
\left(u_{k_1}u_{k_1+k}+v_{k_1}v_{k_1+k}\right)^2n_{k_1}\left(1-n_{k_1+k}\right)
\delta\left(\omega+\lambda_{k_1}-\lambda_{k_1+k}\right)
\right.
\nonumber\\
\left.
+\left(u_{k_1}v_{k_1+k}-v_{k_1}u_{k_1+k}\right)^2n_{k_1}\left(1-n_{k_1+k+\pi}\right)
\delta\left(\omega+\lambda_{k_1}-\lambda_{k_1+k+\pi}\right)
\right)
\label{5.02}
\end{eqnarray}
(see also Refs.~\cite{033,053,054,052}).

Again for the $xx$ and $xy$ dynamic structure factors exact analytical results
are rather scarce.
In the high-temperature limit only the autocorrelation functions survive\cite{039}
\begin{eqnarray}
\langle s_j^x(t)s_j^x\rangle=\frac{1}{4}\Re Z_j(t),
\;\;\;
\langle s_j^x(t)s_j^y\rangle=\frac{1}{4}\Im Z_j(t),
\nonumber\\
Z_j(t)=\frac{\Theta_1\left(J_+t,\frac{J_-}{J_+}\right)}{\Theta_1\left(0,\frac{J_-}{J_+}\right)}
\frac{H_1\left(J_+t,\frac{J_-}{J_+}\right)}{H_1\left(0,\frac{J_-}{J_+}\right)}
\nonumber\\
\cdot
\exp\left(
-{\rm{i}}\Omega t
-\left(1-\frac{{\cal{E}}\left(\frac{J_-}{J_+}\right)}{{\cal{K}}\left(\frac{J_-}{J_+}\right)}\right)J_+^2t^2
\right).
\label{5.03}
\end{eqnarray}
Here the Jacobian theta and eta functions are defined as follows
\begin{eqnarray}
\Theta_1\left(u,k\right)
=\sum_{n=-\infty}^{\infty} q^{n^2}\exp\left(2n{\rm{i}}z\right),
\nonumber\\
H_1\left(u,k\right)
=\sum_{n=-\infty}^{\infty} q^{\left(n+\frac{1}{2}\right)^2}\exp\left((2n+1){\rm{i}}z\right),
\nonumber\\
q=\exp\left(-\pi\frac{{\cal{K}}\left(\sqrt{1-k^2}\right)}{{\cal{K}}(k)}\right),
\;\;\;
z=\frac{\pi u}{2{\cal{K}}(k)}
\label{5.04}
\end{eqnarray}
and the complete elliptic integrals of the 1st and the 2nd kinds are given by
\begin{eqnarray}
{\cal{K}}(k)=\int_{0}^{\frac{\pi}{2}}\frac{{\rm{d}}\theta}{\sqrt{1-k^2\sin^2\theta}},
\;\;\;
{\cal{E}}(k)=\int_{0}^{\frac{\pi}{2}}{\rm{d}}\theta\sqrt{1-k^2\sin^2\theta}.
\label{5.05}
\end{eqnarray}
Moreover, $J_{\pm}^2=J^2\left(1\pm\delta\right)^2/4$.

In the strong-field limit $\vert \Omega\vert >\vert J\vert$
at $T=0$
we can repeat the calculation of the previous section
to find,
for example,
for the $xx$ time-dependent correlation function
and the corresponding dynamic structure factor
the following result
\begin{eqnarray}
4\langle s_j^x(t)s_{j+n}^x\rangle
=\frac{1}{N}\sum_{k}\exp\left({\rm{i}}kn\right)
\left(
u_k^2\exp\left(-{\rm{i}}\Lambda_k t\right)
+
v_k^2\exp\left(-{\rm{i}}\Lambda_{k+\pi} t\right)
\right.
\nonumber\\
\left.
-{\rm{i}}\left(-1\right)^{j+n}u_kv_k
\left(\exp\left(-{\rm{i}}\Lambda_k t\right)-\exp\left(-{\rm{i}}\Lambda_{k+\pi} t\right)\right)
\right),
\label{5.06}
\\
S_{xx}(k,\omega)
=\frac{\pi}{2}
\left(u_k^2\delta\left(\omega-\Lambda_k\right)
+v_k^2\delta\left(\omega-\Lambda_{k+\pi}\right)
\right).
\label{5.07}
\end{eqnarray}

For arbitrary values of temperature and transverse field
the $xx$ and $xy$ dynamic structure can be obtained numerically\cite{052}.

Let us discuss the dynamic quantities of the dimerized transverse $XX$ chain.
We begin with the transverse dynamic structure factor which can be written
as (compare with Eq. (\ref{4.12}))
\begin{eqnarray}
S_{zz}(k,\omega)
=\int_{-\pi}^{\pi}{\rm{d}}k_1{\rm{d}}k_2
C^{(1)}(k_1,k_2)
\nonumber\\
\cdot
n_{k_1}\left(1-n_{k_2}\right)
\delta\left(\omega+\lambda_{k_1}-\lambda_{k_2}\right)
\delta_{k+k_1-k_2,0}
\nonumber\\
+\int_{-\pi}^{\pi}{\rm{d}}k_1{\rm{d}}k_2
C^{(2)}(k_1,k_2)
\nonumber\\
\cdot
n_{k_1}\left(1-n_{k_2}\right)
\delta\left(\omega+\lambda_{k_1}-\lambda_{k_2}\right)
\delta_{k+k_1-k_2+\pi,0},
\nonumber\\
C^{(1)}(k_1,k_2)
=\left(u_{k_1}u_{k_2}+v_{k_1}v_{k_2}\right)^2
\nonumber\\
C^{(2)}(k_1,k_2)
=\left(u_{k_1}v_{k_2}-v_{k_1}u_{k_2}\right)^2.
\label{5.08}
\end{eqnarray}
As can be seen from Eq. (\ref{5.08}) the transverse dynamic structure probes
two-fermion excitations.
$S_{zz}(k,\omega)$ may have nonzero value within a restricted region of the $k$--$\omega$ plane
when there is such a wave-vector $k_1$, $-\pi\le k_1<\pi$ that
$\omega=-\lambda_{k_1}+\lambda_{k_1+k}$
or
$\omega=-\lambda_{k_1}+\lambda_{k_1+k+\pi}$.
Moreover, at zero temperature there are additional restrictions arising from the Fermi functions.
The lines of potential singularities
follow from the analysis of the equations
${\rm{d}}\lambda_{k_1}/{\rm{d}}k_1-{\rm{d}}\lambda_{k_1+k}/{\rm{d}}k_1=0$
and
${\rm{d}}\lambda_{k_1}/{\rm{d}}k_1-{\rm{d}}\lambda_{k_1+k+\pi}/{\rm{d}}k_1=0$.
The characteristic lines in the $k$--$\omega$ plane
which determine the behavior of the transverse dynamic structure factor
were reported for the first time by J.~H.~Taylor and G.~M\"{u}ller\cite{033}.

\begin{figure}[th]
\centerline{\psfig{file=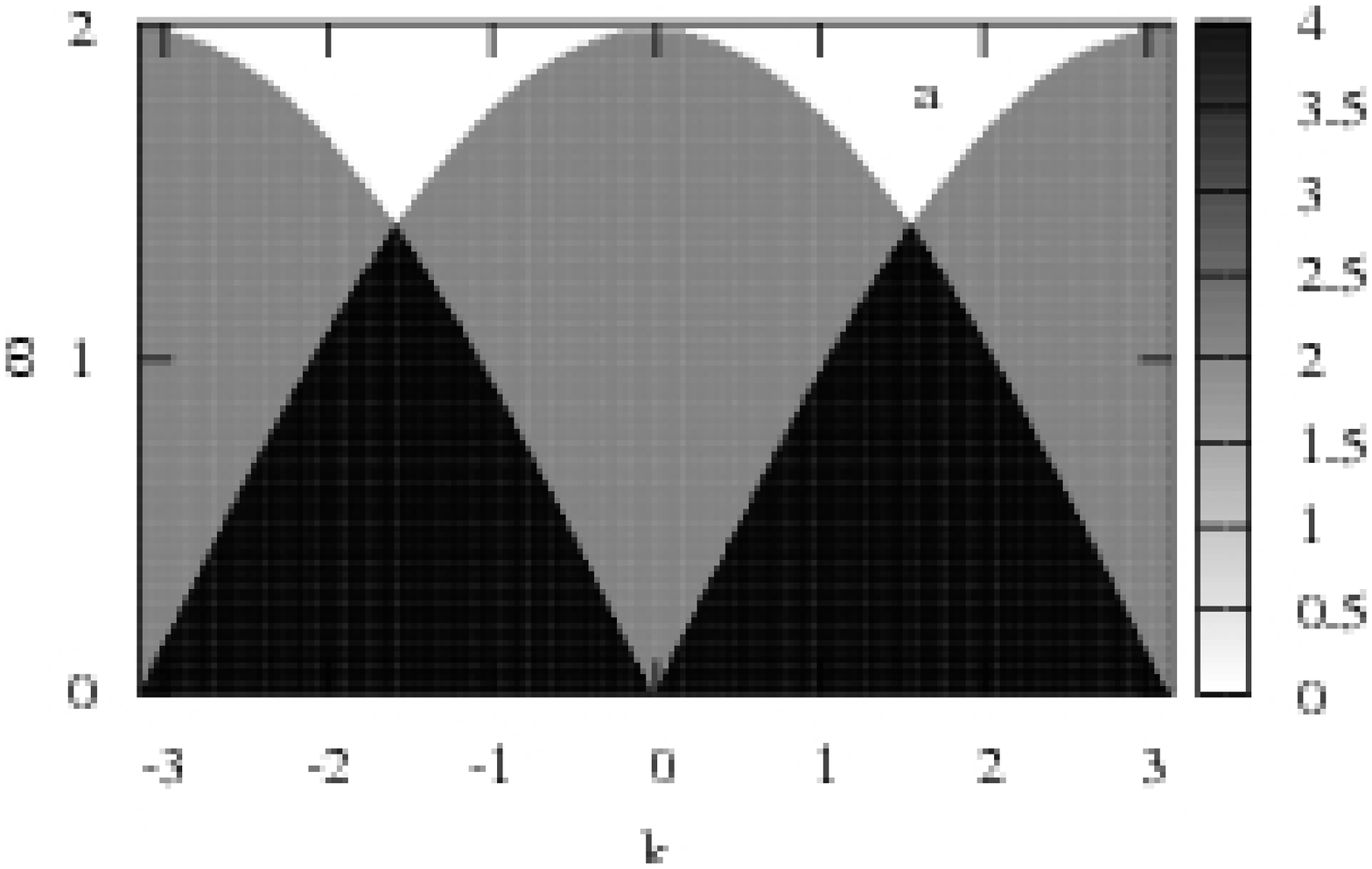,width=2.0in,angle=0}
\psfig{file=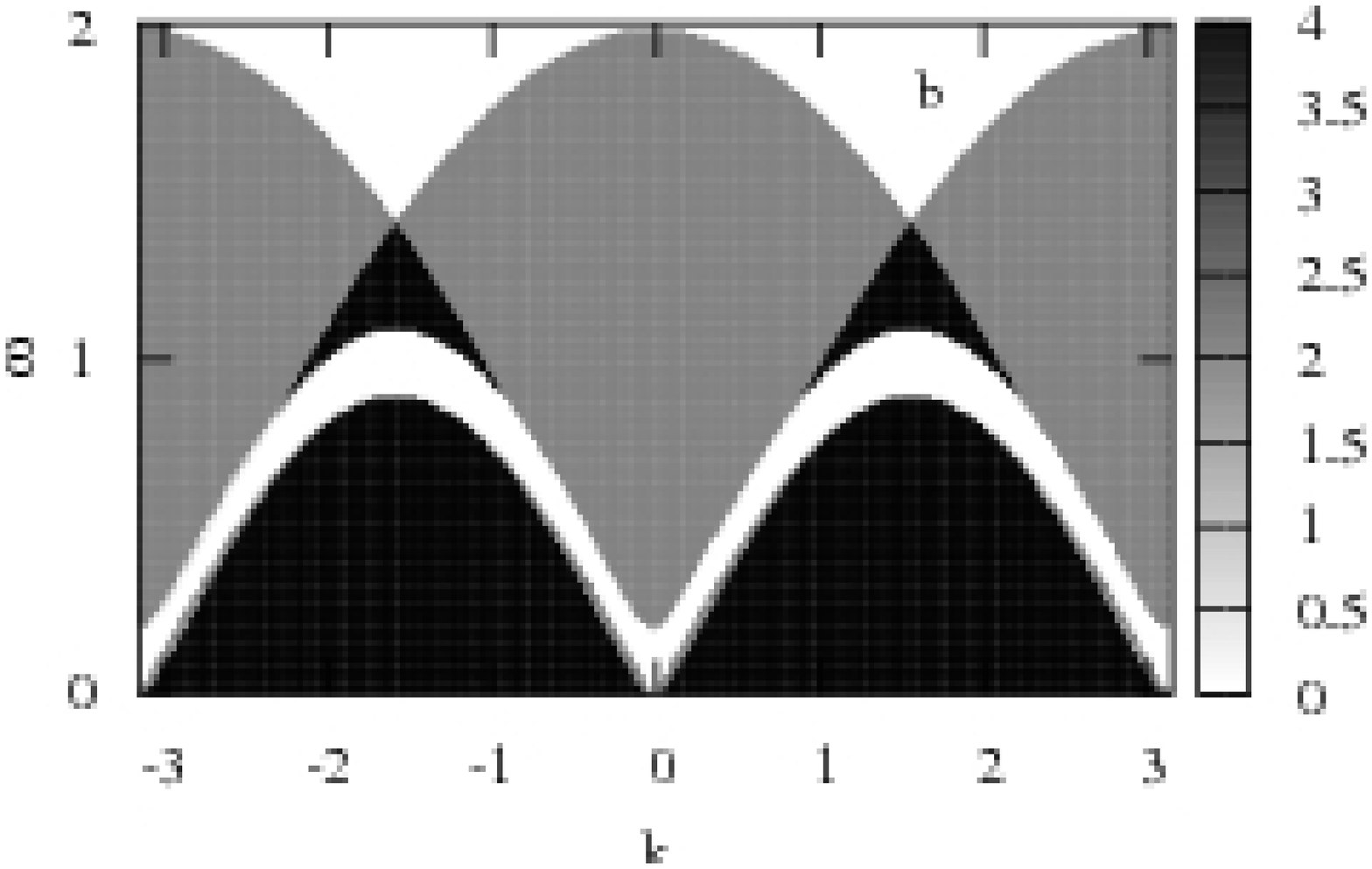,width=2.0in,angle=0}}
\vspace*{8pt}
\caption{Location of the roots of equations
$\omega=-\lambda_{k_1}+\lambda_{k_1+k}$
and
$\omega=-\lambda_{k_1}+\lambda_{k_1+k+\pi}$
($-\pi\le k_1<\pi$)
in the $k$--$\omega$ plane
for $\delta=0$ (panel a)
and $\delta=0.1$ (panel b):
light region: no roots, light gray region: two roots, dark gray region: four roots.}
\label{fig13}
\end{figure}

\begin{figure}[th]
\centerline{\psfig{file=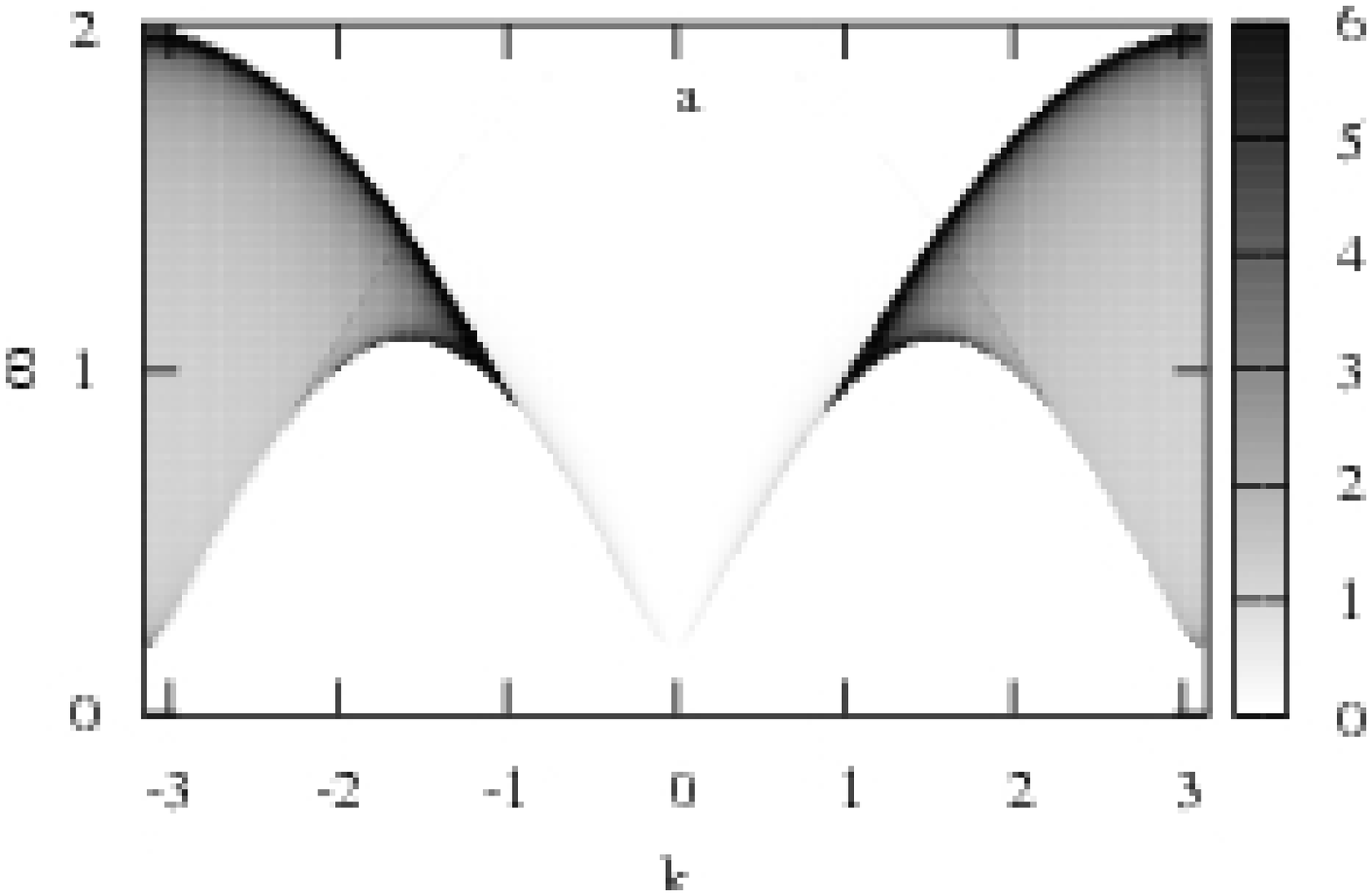,width=2.0in,angle=0}
\psfig{file=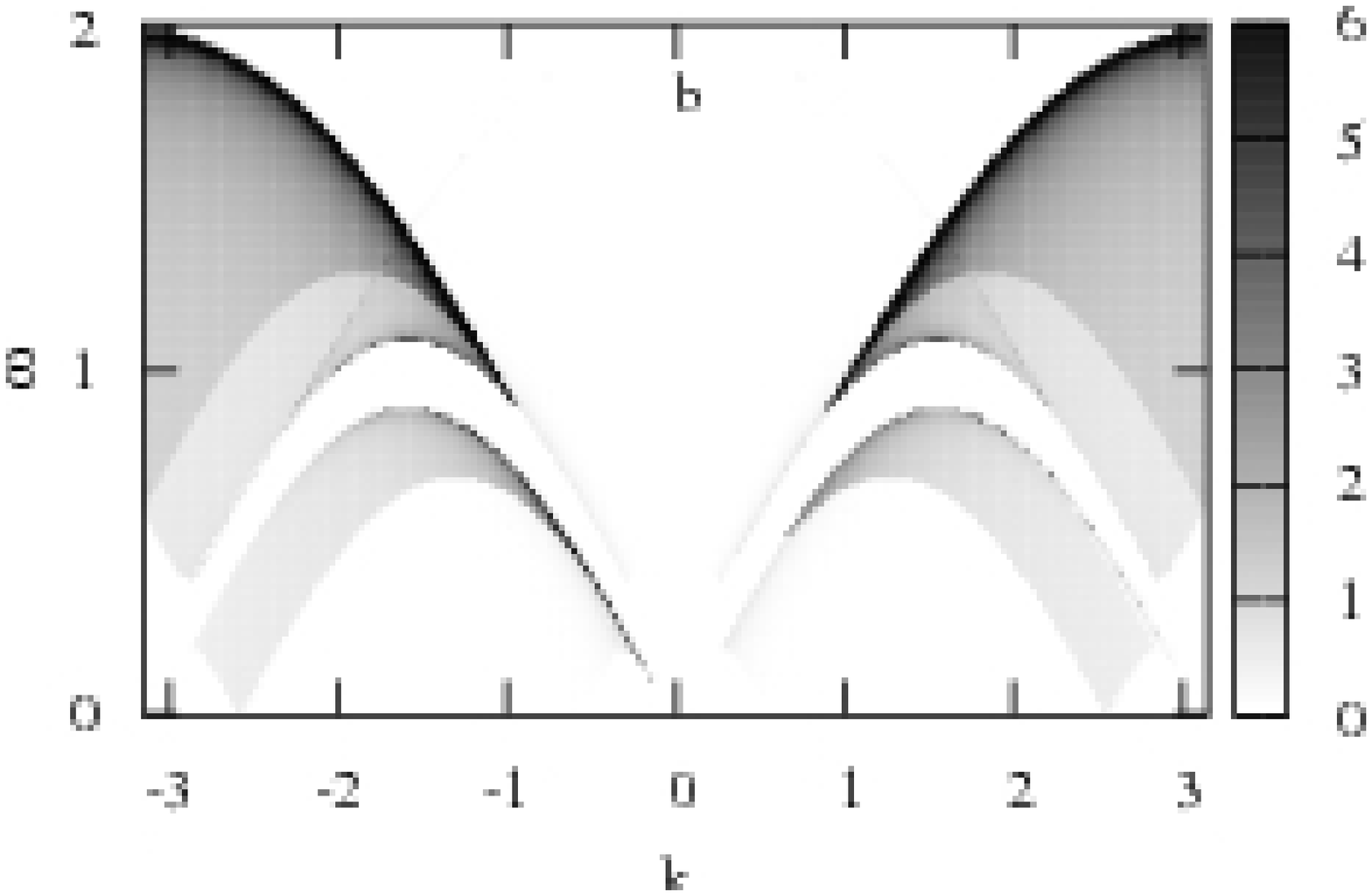,width=2.0in,angle=0}}
\centerline{\psfig{file=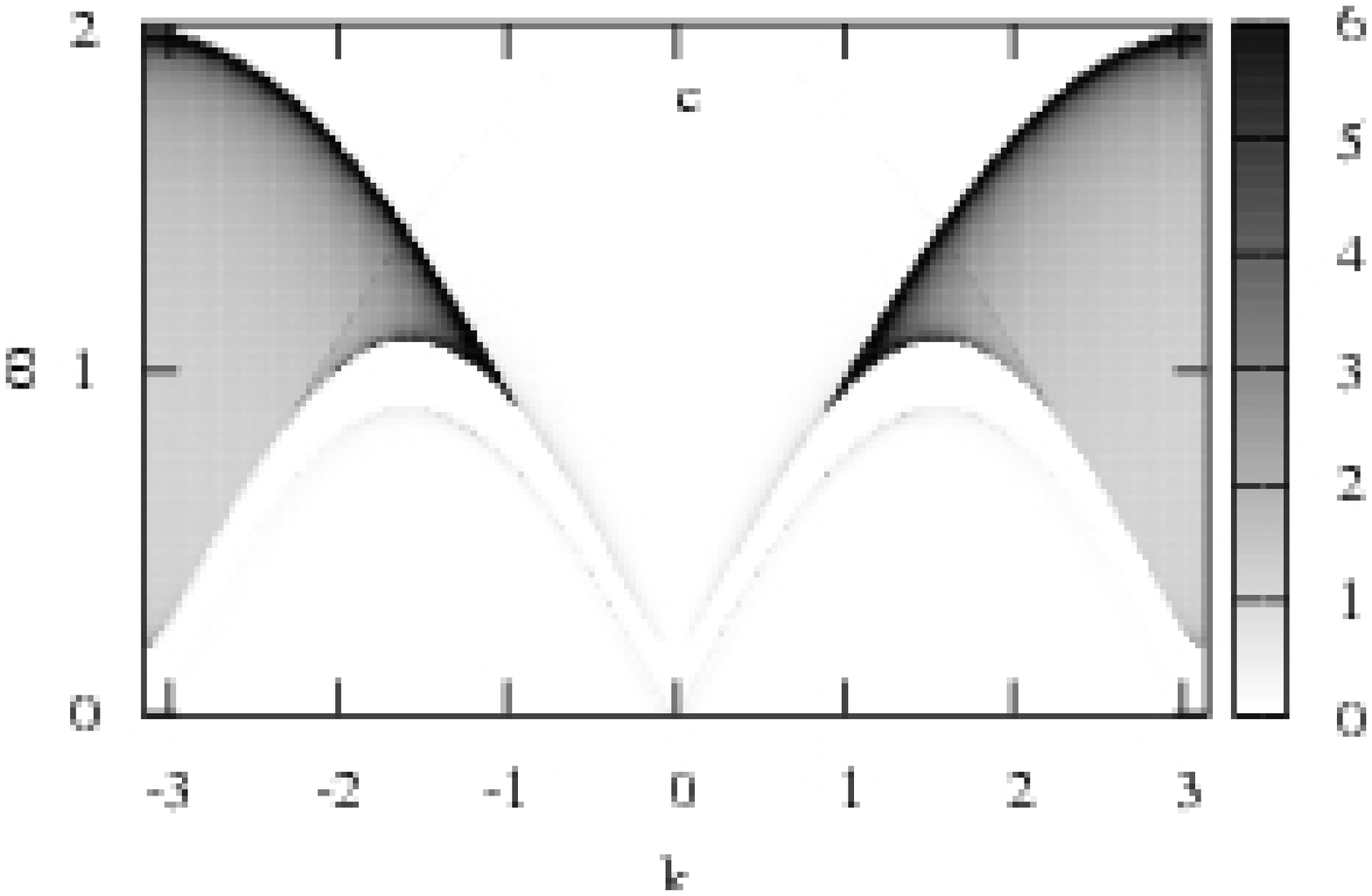,width=2.0in,angle=0}
\psfig{file=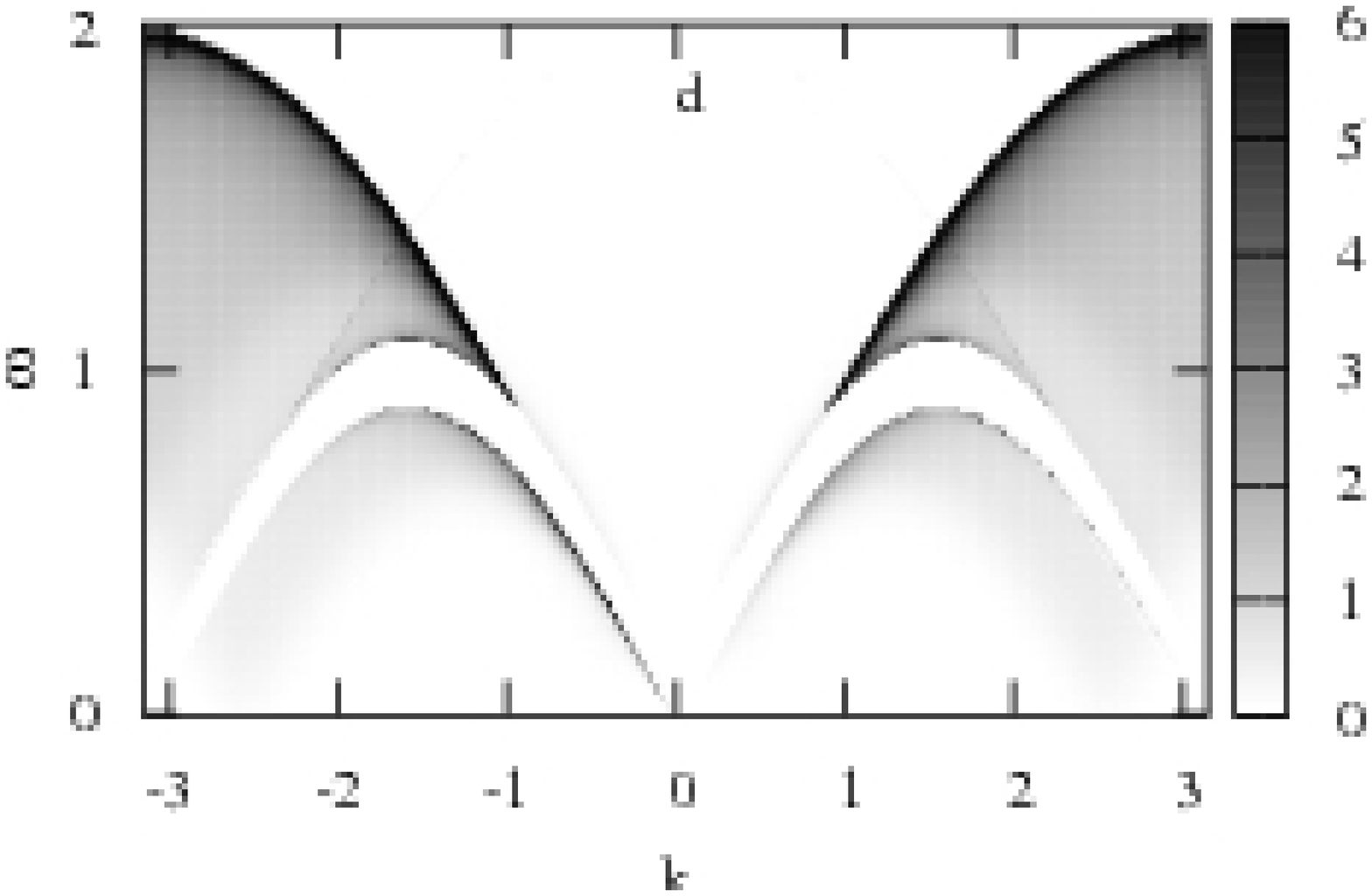,width=2.0in,angle=0}}
\centerline{\psfig{file=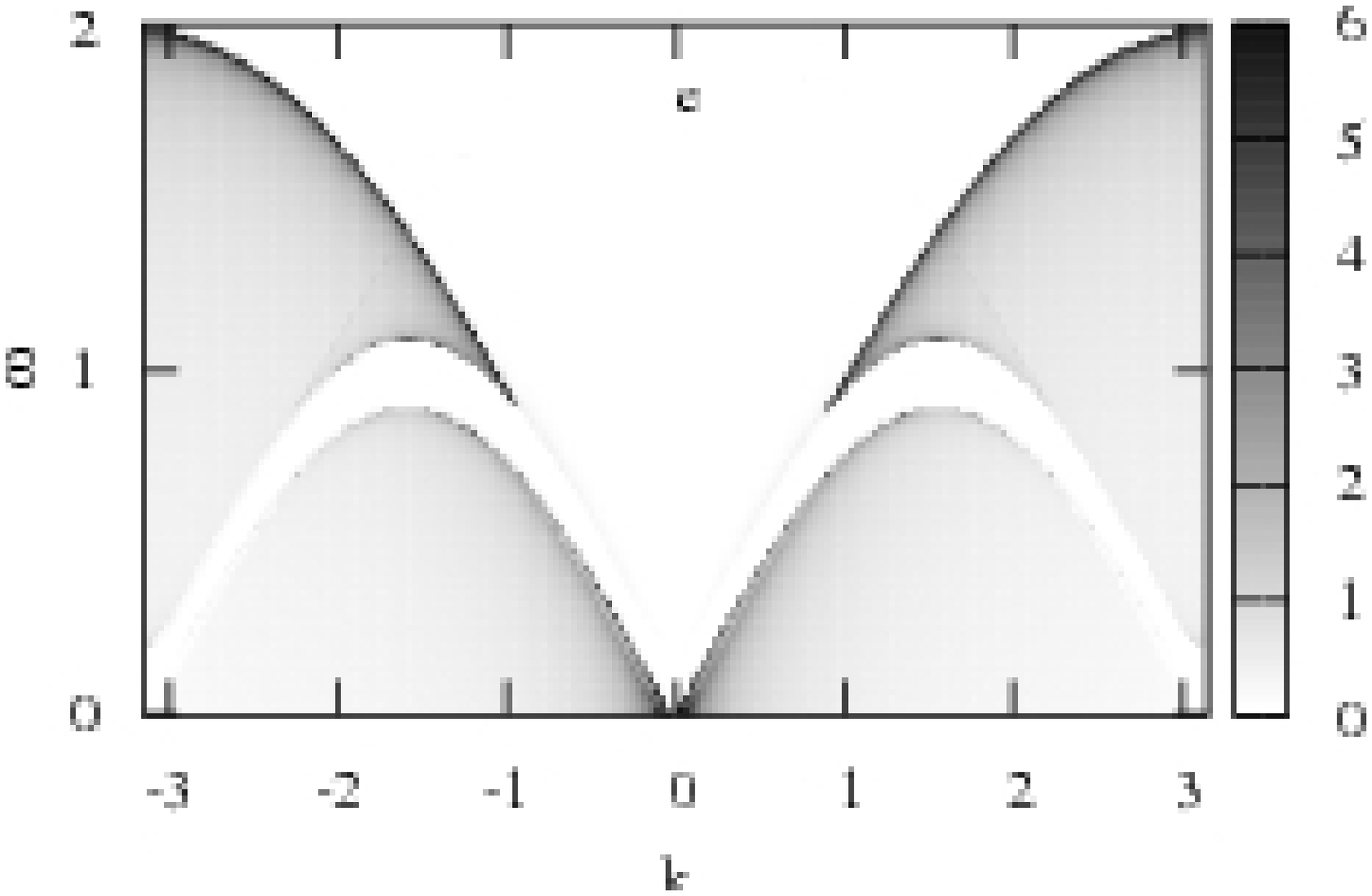,width=2.0in,angle=0}
\psfig{file=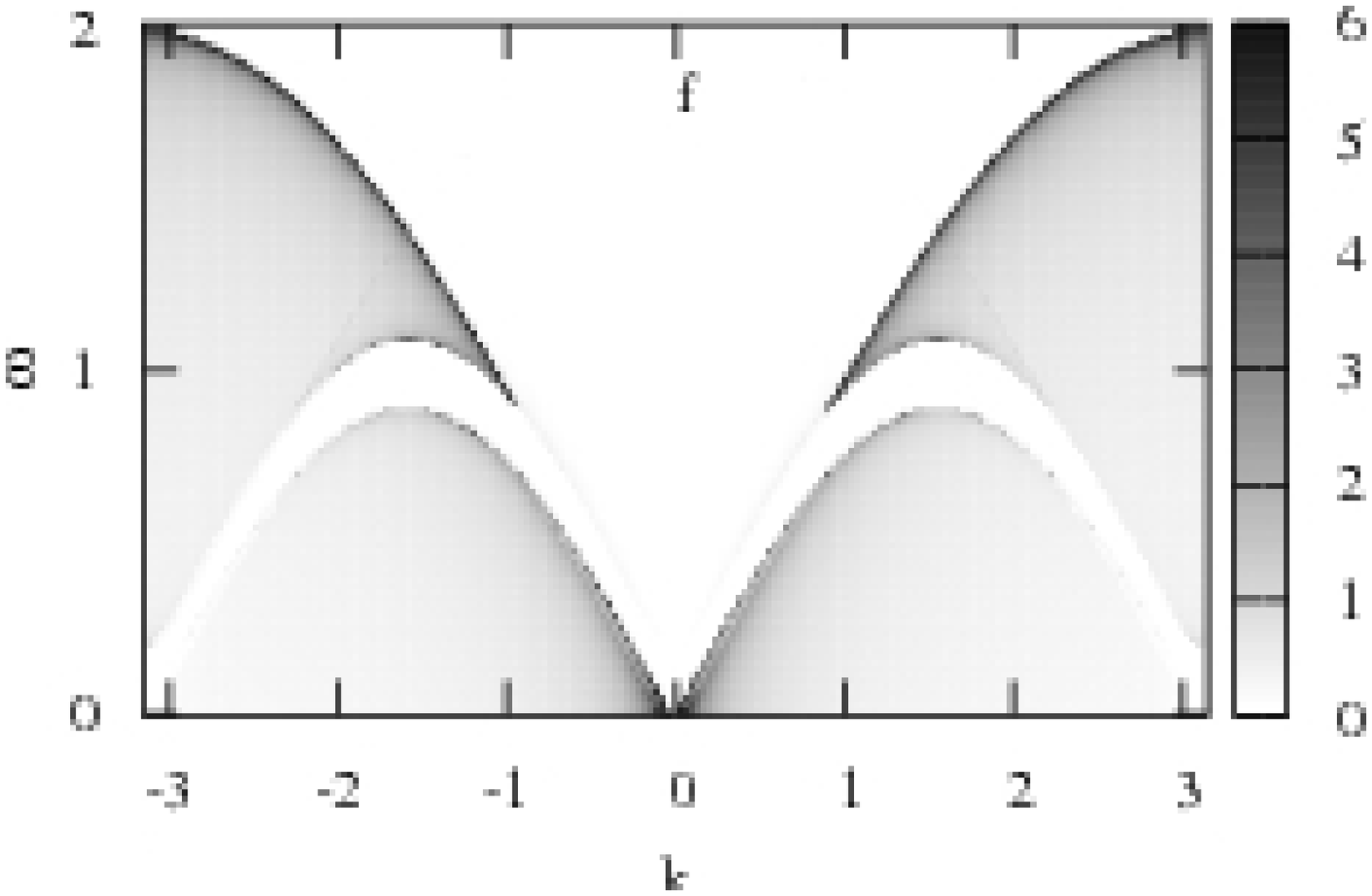,width=2.0in,angle=0}}
\vspace*{8pt}
\caption{$S_{zz}(k,\omega)$ (gray-scale plots)
for the chain (\ref{5.01}) with $J=-1$,
$\delta=0.1$ at different temperatures
$\beta=\infty$ (a, b),
$\beta=20$ (c, d),
$\beta=1$ (e, f)
for $\Omega=0$ (left panels a, c, e)
and
$\Omega=0.3$ (right panels b, d, f).
Note that the results at $\beta=1$ for $\Omega=0$ and $\Omega=0.3$
(panels e and f) are practically indistinguishable.}
\label{fig14}
\end{figure}

\begin{figure}[th]
\centerline{\psfig{file=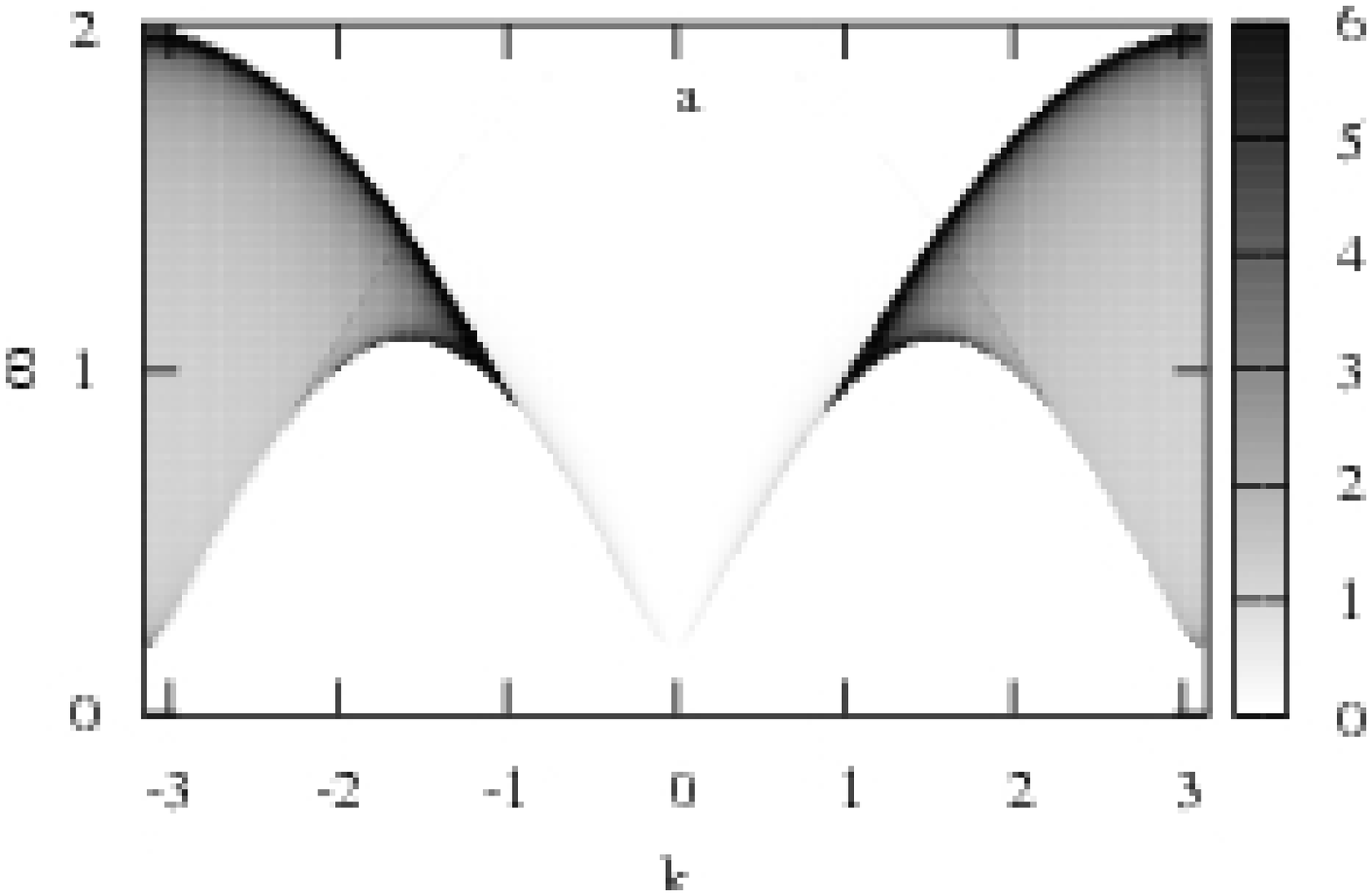,width=2.0in,angle=0}
\psfig{file=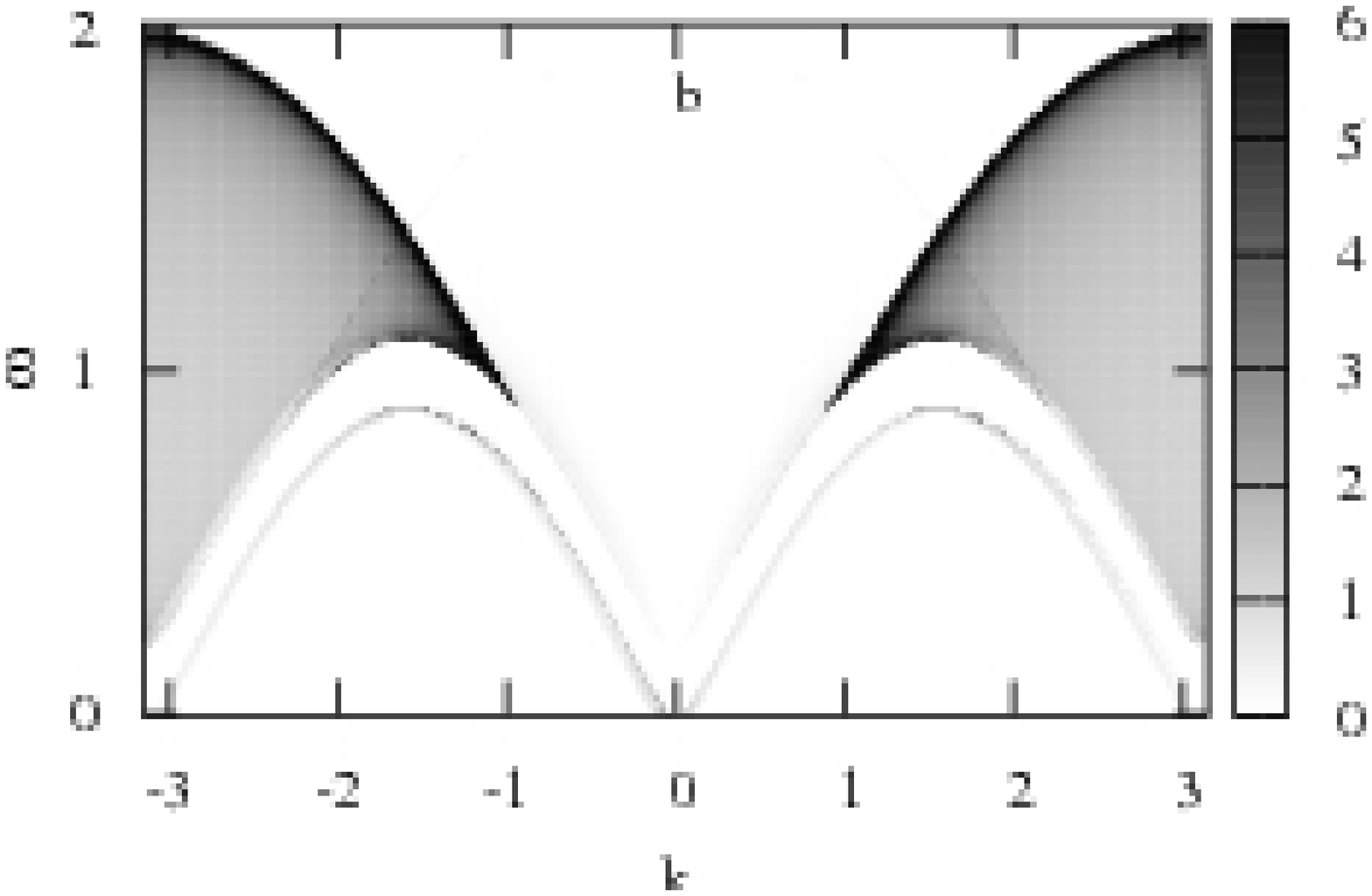,width=2.0in,angle=0}}
\centerline{\psfig{file=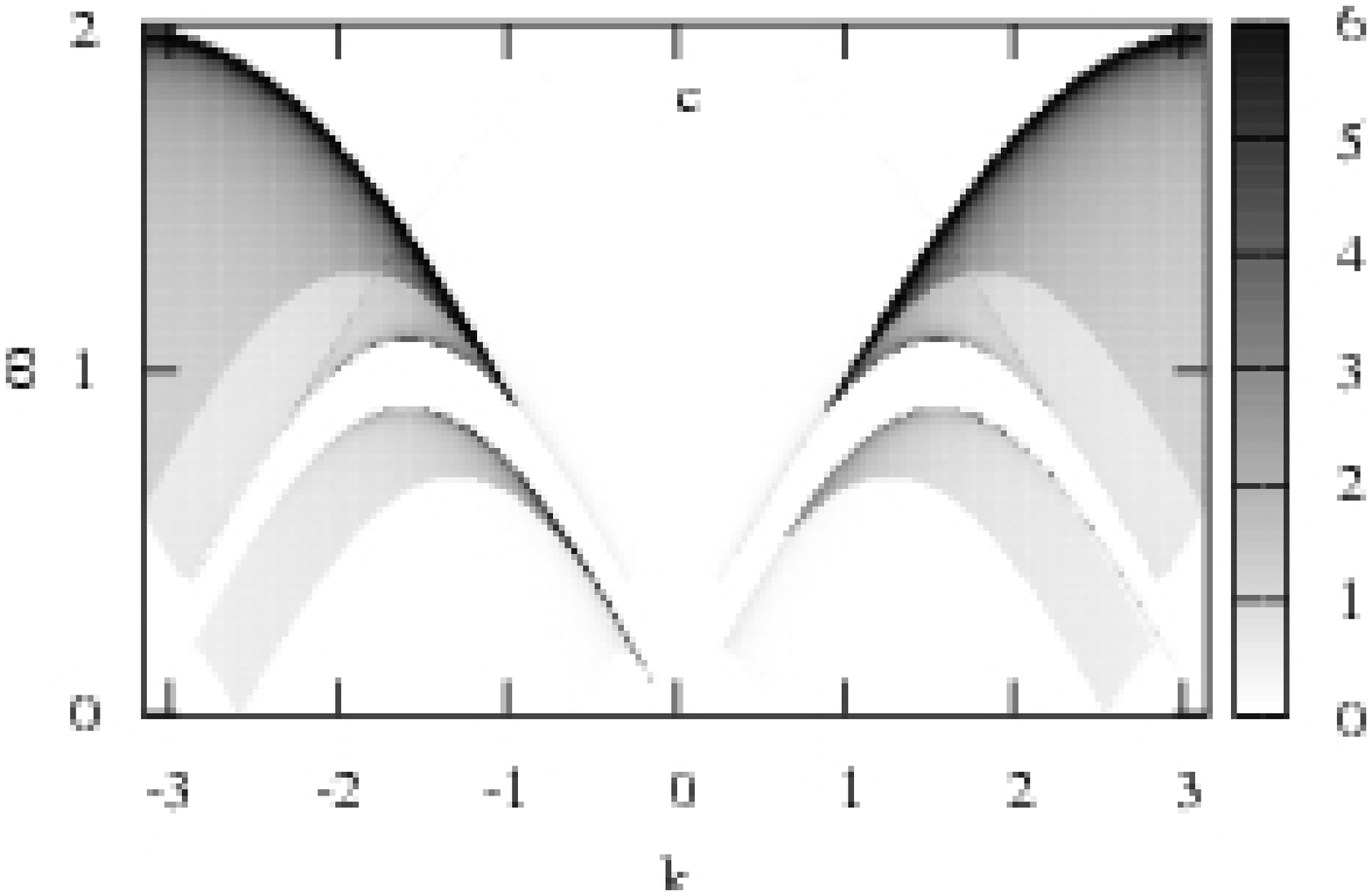,width=2.0in,angle=0}
\psfig{file=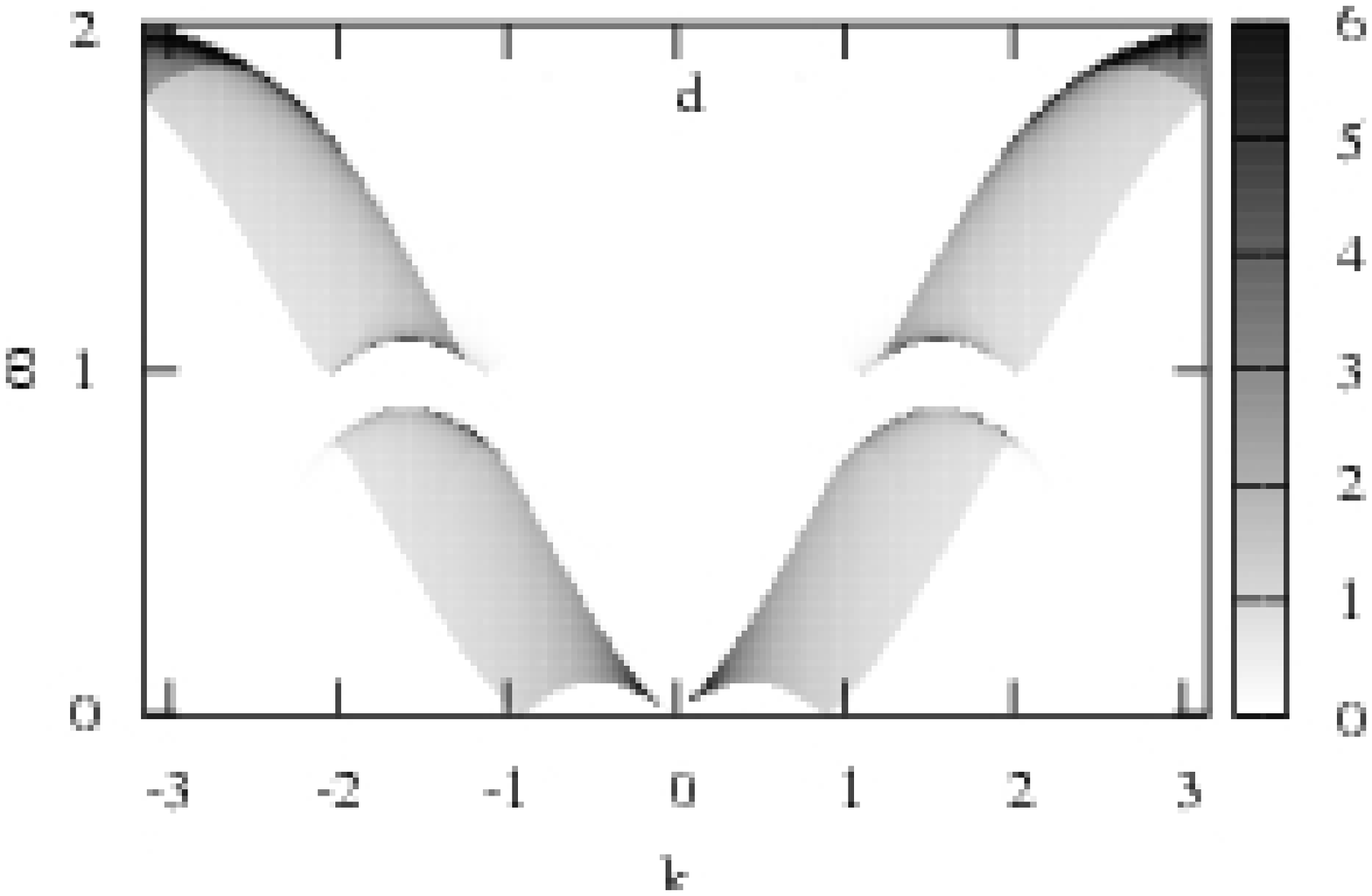,width=2.0in,angle=0}}
\vspace*{8pt}
\caption{$S_{zz}(k,\omega)$
(gray-scale plots)
for the chain (\ref{5.01}) with $J=-1$,
$\delta=0.1$ at zero temperature
$\beta=\infty$
and different values of the transverse field
$\Omega=0.1$ (a),
$\Omega=0.11$ (b),
$\Omega=0.3$ (c)
and
$\Omega=0.9$ (d).}
\label{fig15}
\end{figure}

\begin{figure}[th]
\centerline{\psfig{file=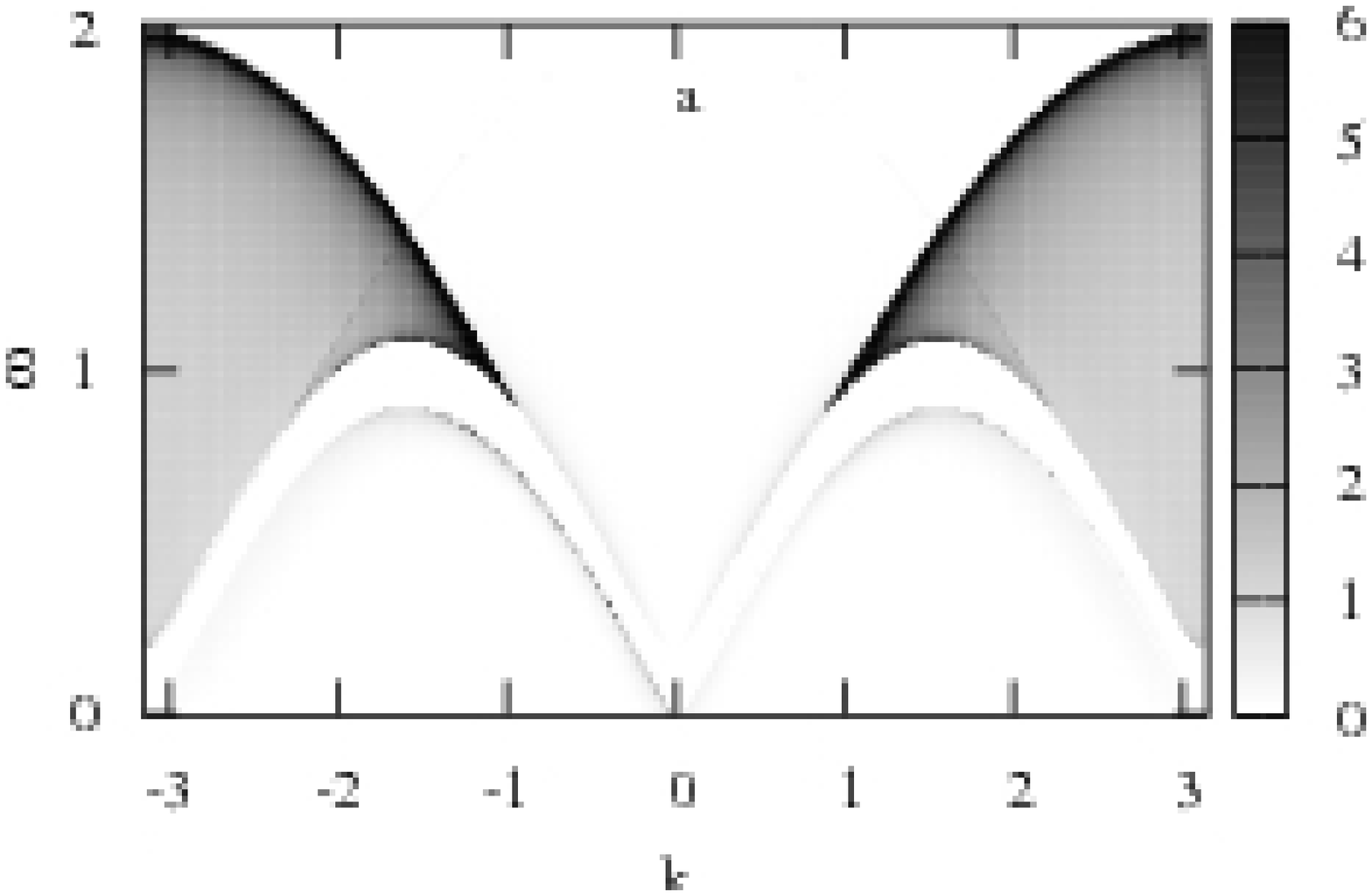,width=2.0in,angle=0}
\psfig{file=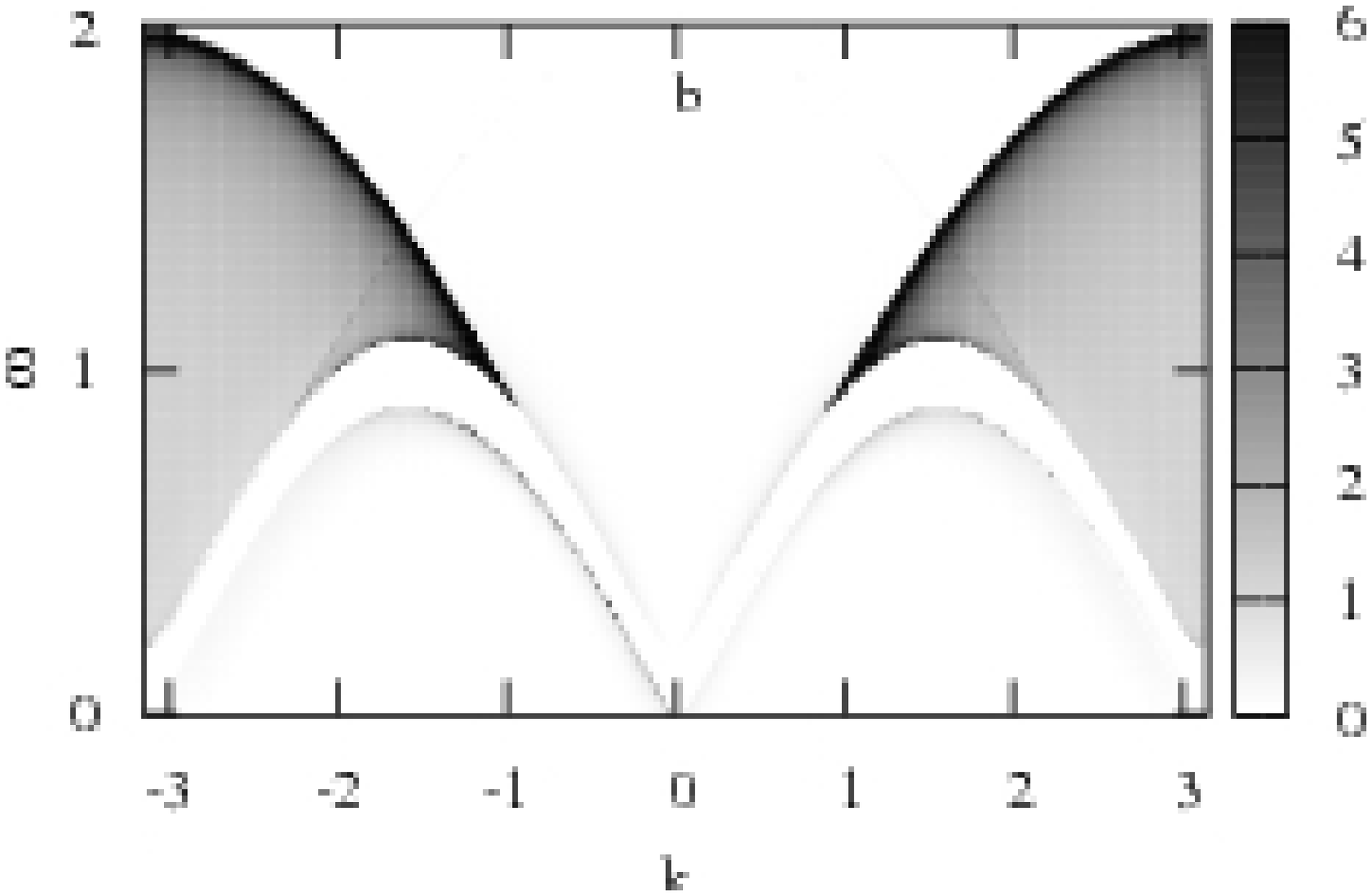,width=2.0in,angle=0}}
\centerline{\psfig{file=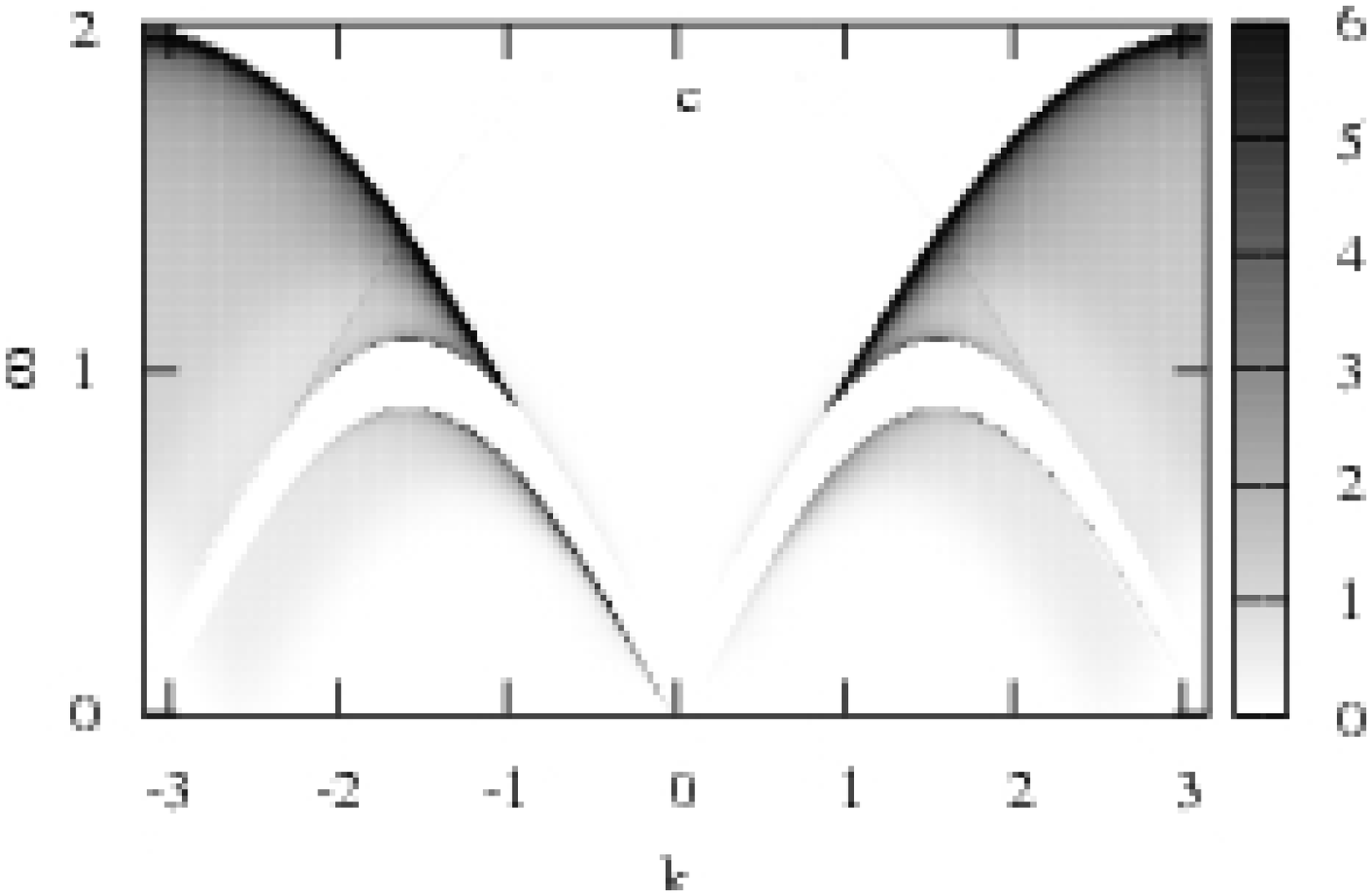,width=2.0in,angle=0}
\psfig{file=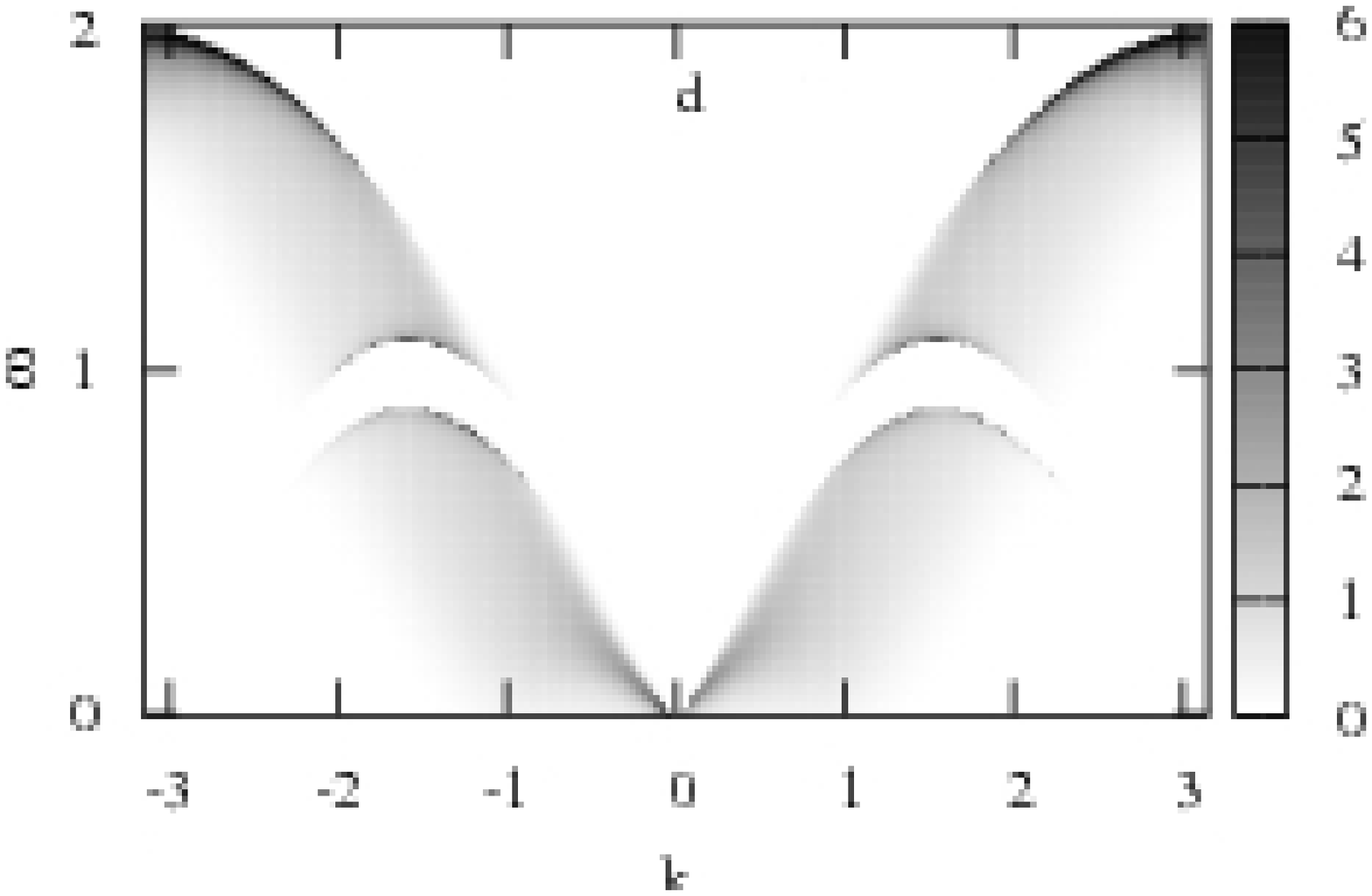,width=2.0in,angle=0}}
\vspace*{8pt}
\caption{The same as in Fig.~\ref{fig15} for $\beta=20$.}
\label{fig16}
\end{figure}

In Fig.~\ref{fig13}
we show the region of the $k$--$\omega$ plane in which the two-fermion dynamic quantity
may have nonzero values.
In Fig.~\ref{fig14}
we show the transverse dynamic structure factor at different temperatures.
Comparing Fig.~\ref{fig13}b and Fig.~\ref{fig14}
one can see the van Hove singularities and the effects of the Fermi functions
and the $C^{(1)}$- and $C^{(2)}$-functions.
In Figs.~\ref{fig15},~\ref{fig16}
we show $S_{zz}(k,\omega)$ at various values of the transverse field for two temperatures
$\beta=\infty$ and $\beta=20$.

\begin{figure}[th]
\centerline{\psfig{file=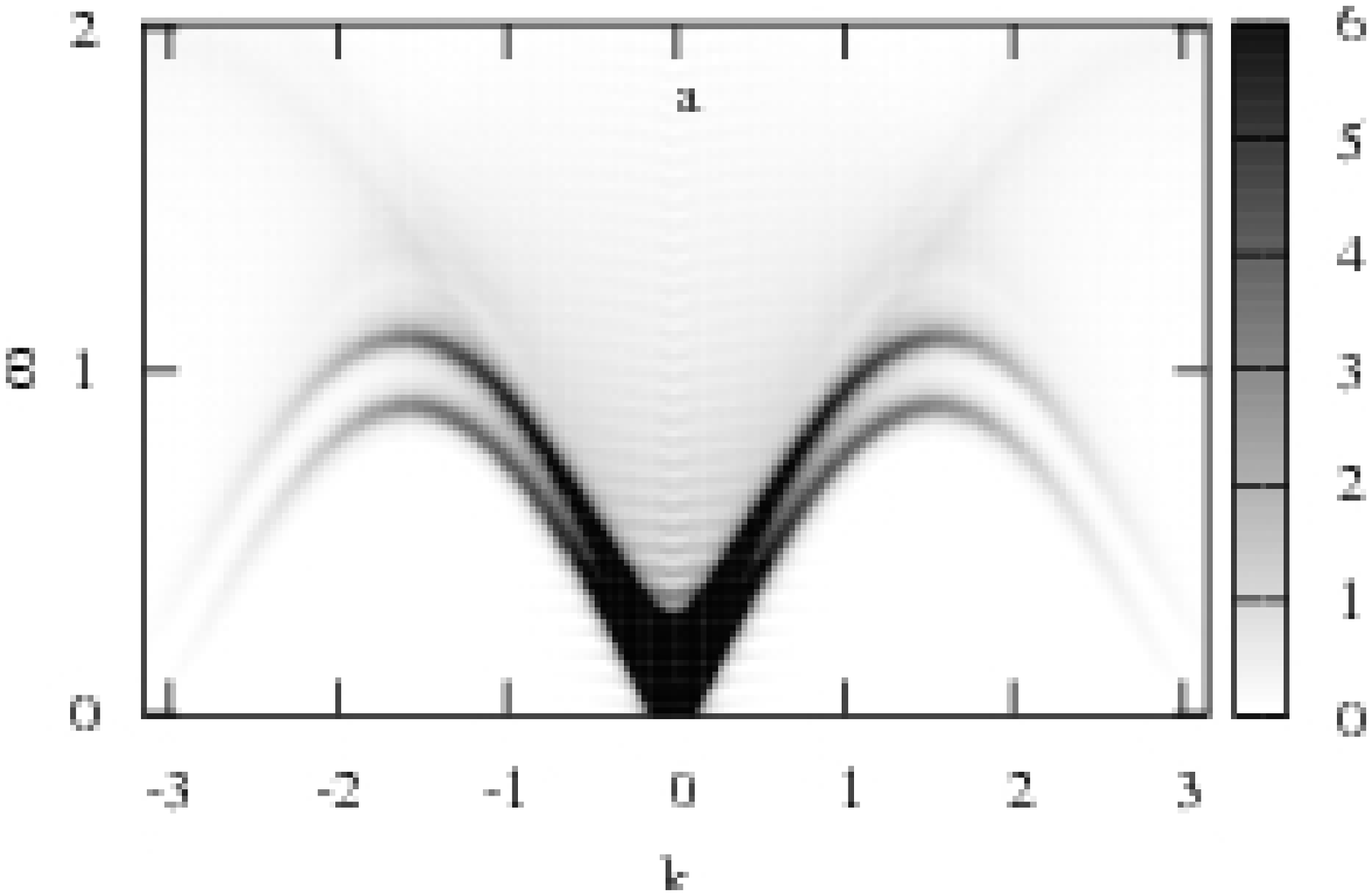,width=2.0in,angle=0}
\psfig{file=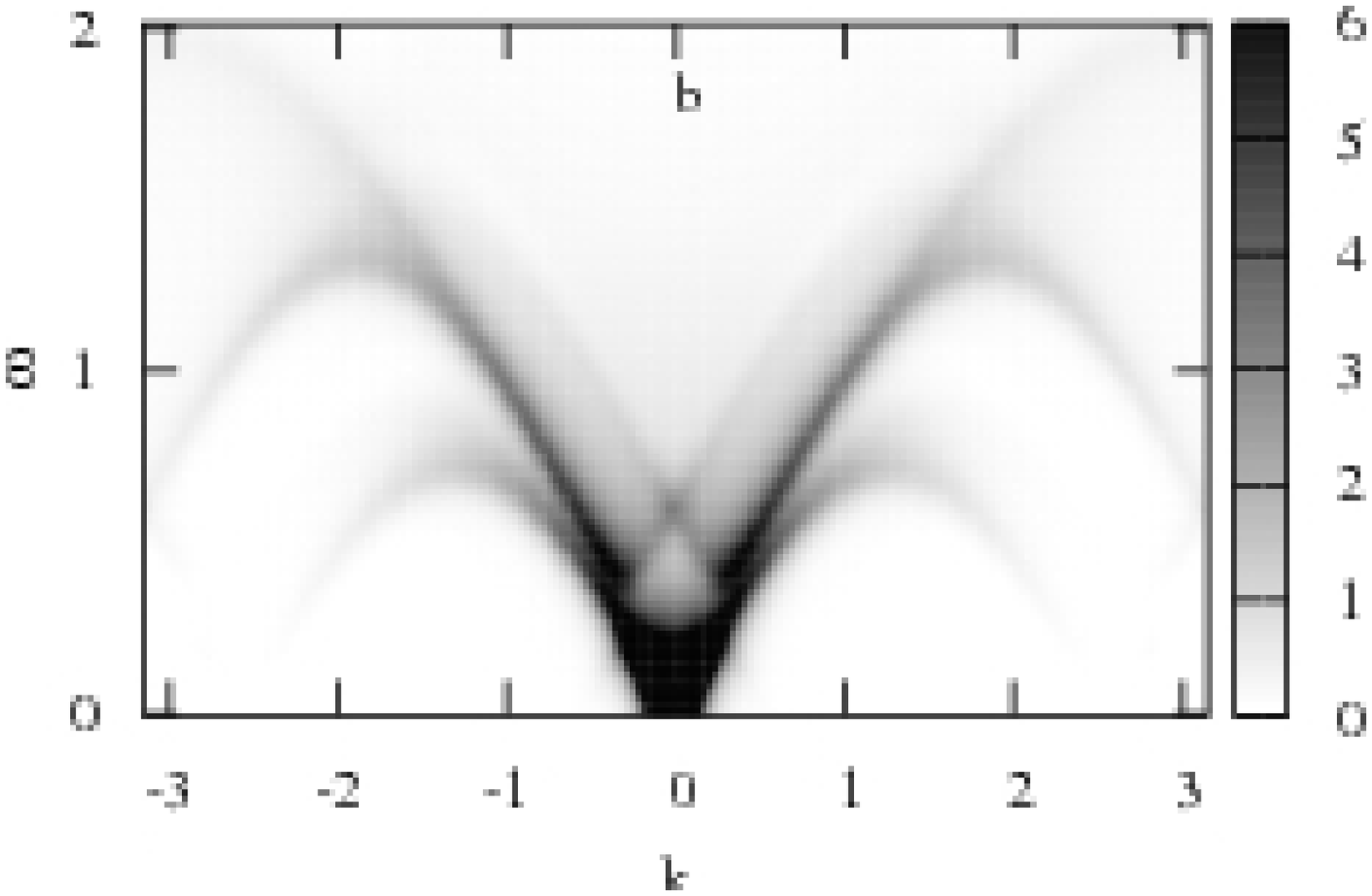,width=2.0in,angle=0}}
\centerline{\psfig{file=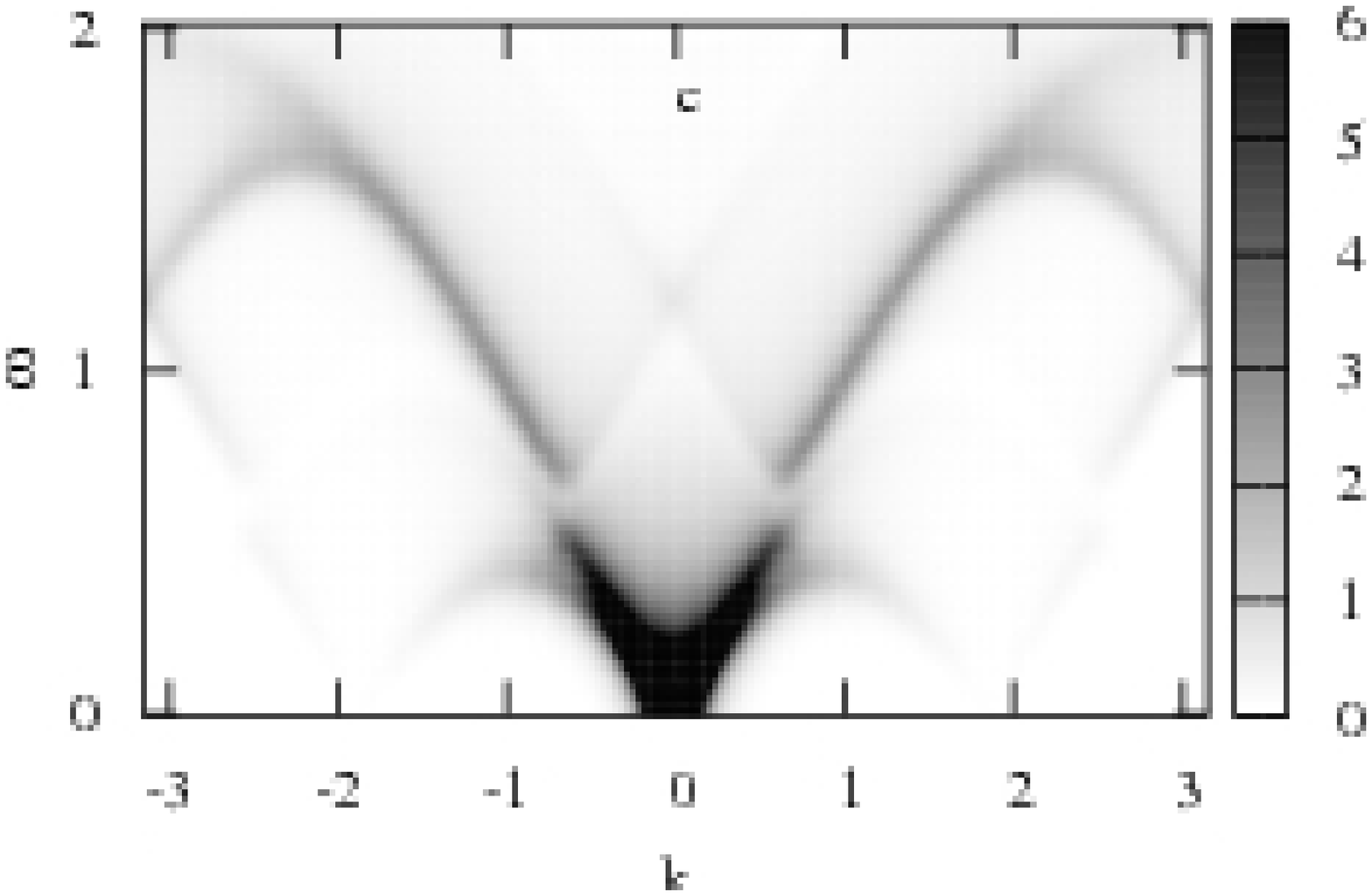,width=2.0in,angle=0}
\psfig{file=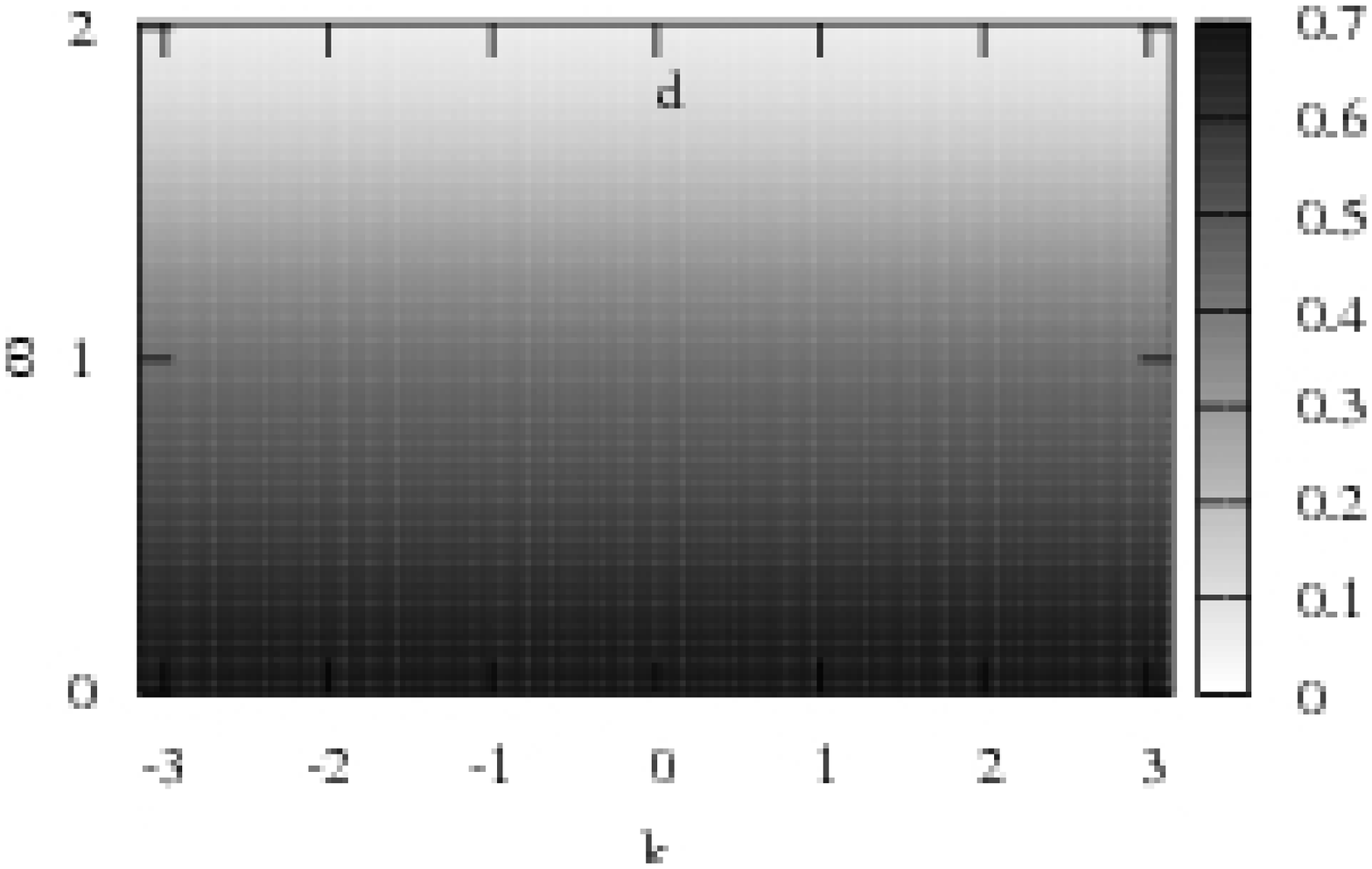,width=2.0in,angle=0}}
\vspace*{8pt}
\caption{$S_{xx}(k,\omega)$ (gray-scale plots)
for the chain (\ref{5.01}) with $J=-1$,
$\delta=0.1$ at low temperature $\beta=20$
and for $\Omega=0.1$ (a), $\Omega=0.3$ (b), $\Omega=0.6$ (c)
and in the high-temperature limit $\beta=0$ for $\Omega=0.6$ (d).}
\label{fig17}
\end{figure}

Next we pass to the $xx$ dynamic structure factor
obtained numerically.
Typically we consider chains of $N=400$ sites
assume in (\ref{1.04}) $n=41$, $l$ up to 50,
consider $t$ up to $t_c=200$
and take $\epsilon=0.001$\cite{052}.
In Fig.~\ref{fig17}
we show $S_{xx}(k,\omega)$ of the dimerized transverse $XX$ chain
at low temperatures for different values of the transverse field.

In contrast to the $zz$ dynamic structure factor which is a two-fermion dynamic quantity,
the $xx$ dynamic structure factor is a many-particle dynamic quantity
within the Jordan-Wigner fermionization approach.
Therefore,
nonzero values of $S_{xx}(k,\omega)$
far above the two-fermion excitation continua may be expected.
However, as can be seen in Fig.~\ref{fig17} the opposite is true:
At low-temperatures $S_{xx}(k,\omega)$ shows several well-defined excitation branches
which follow roughly the boundaries of the two-fermion excitation continua.
Although
we can describe the low-energy physics also using the bosonization treatment,
high-frequency features cannot be reproduced within such an approach.

Finally we note
that the dimerized $XX$ chain does not show bound-state branches;
within the fermionization approach this may be related
to the absence of interactions between fermions.
In contrast,
a particle-hole bound state can be observed
in the dimerized Heisenberg chain\cite{055}.

\section{Spin-1/2 $XY$ Chains with the Dzyaloshinskii-Moriya Interaction
         \label{secdk6}}

In this section we examine the effect of the Dzyaloshinskii-Moriya interaction
(actually, the $z$ component of the vector of the Dzyaloshinskii-Moriya interaction,
see Eq. (\ref{2.14})).
The Hamiltonian of the transverse $XX$ chain with the Dzyaloshinskii-Moriya interaction reads
\begin{eqnarray}
H=\sum_n
\left(
J\left(s_n^xs_{n+1}^x+s_n^ys_{n+1}^y\right)
+
D\left(s_n^xs_{n+1}^y-s_n^ys_{n+1}^x\right)
\right)
-h\sum_ns_n^z.
\label{6.01}
\end{eqnarray}

Interestingly,
the Dzyaloshinskii-Moriya interaction can be eliminated from the Hamiltonian (\ref{6.01})
resulting in renormalization of the isotropic $XY$ exchange interaction\cite{056}.
To see this,
consider the following spin axes rotation
\begin{eqnarray}
s_n^x\to \tilde{s}_n^x
=s_n^x\cos\phi_n +s_n^y\sin\phi_n,
\nonumber\\
s_n^y\to \tilde{s}_n^y
=-s_n^x\sin\phi_n +s_n^y\cos\phi_n,
\nonumber\\
s_n^z\to \tilde{s}_n^z
=s_n^z,
\nonumber\\
\phi_n=(n-1)\varphi,
\;\;\;
\tan\varphi=\frac{D}{J}.
\label{6.02}
\end{eqnarray}
After such a unitary transformation the Hamiltonian (\ref{6.01}) becomes
\begin{eqnarray}
H=\sum_n
\tilde{J}\left(\tilde{s}_n^x\tilde{s}_{n+1}^x+\tilde{s}_n^y\tilde{s}_{n+1}^y\right)
-h\sum_n\tilde{s}_n^z,
\nonumber\\
\tilde{J}={\rm{sgn}}(J)\sqrt{J^2+D^2}.
\label{6.03}
\end{eqnarray}

Using the unitary transformation (\ref{6.02})
we can examine the effect of the Dzyaloshinskii-Moriya interaction
using the 
dynamic quantities of the transverse $XX$ chain
without the Dzyaloshinskii-Moriya interaction discussed already in Sec.~\ref{secdk4}.
First of all we note
that the $zz$ dynamic structure factor
is given by Eq. (\ref{4.03}),
however,
with $\Lambda_k=-h+\tilde{J}\cos k$.
The formulas determining the two-fermion excitation continua boundaries
are still given by Eqs. (\ref{4.05}), (\ref{4.06}), (\ref{4.07})
but with $\tilde{J}$ instead of $J$ on the l.h.s. of these equations
and in the definition of the parameter $\alpha$.

Exploiting (\ref{6.02}) we find the following relations
between the $xx$ and $xy$ dynamic structure factors of the model (\ref{6.01})
(l.h.s. of Eq. (\ref{6.04}))
and the dynamic structure factors of the model (\ref{6.03})
(r.h.s. of Eq. (\ref{6.04}))\cite{057}
\begin{eqnarray}
S_{xx}(k,\omega)
=\frac{1}{2}
\left(
\left.S_{xx}(k-\varphi,\omega)\right\vert_{\tilde{J}}
+
\left.S_{xx}(k+\varphi,\omega)\right\vert_{\tilde{J}}
\right.
\nonumber\\
\left.
+{\rm{i}}\left(
\left.S_{xy}(k-\varphi,\omega)\right\vert_{\tilde{J}}
-
\left.S_{xy}(k+\varphi,\omega)\right\vert_{\tilde{J}}
\right)
\right),
\nonumber\\
S_{xy}(k,\omega)
=\frac{1}{2}
\left(
\left.S_{xy}(k-\varphi,\omega)\right\vert_{\tilde{J}}
+
\left.S_{xy}(k+\varphi,\omega)\right\vert_{\tilde{J}}
\right.
\nonumber\\
\left.
-{\rm{i}}\left(
\left.S_{xx}(k-\varphi,\omega)\right\vert_{\tilde{J}}
-
\left.S_{xx}(k+\varphi,\omega)\right\vert_{\tilde{J}}
\right)
\right).
\label{6.04}
\end{eqnarray}
Therefore,
using Eq. (\ref{4.30}) we obtain for the model (\ref{6.01})
\begin{eqnarray}
S_{xx}(k,\omega)
=\frac{\sqrt{\pi}}{4\tilde{J}}
\left(
\exp\left(-\frac{\left(\omega+h\right)^2}{\tilde{J}^2}\right)
+
\exp\left(-\frac{\left(\omega-h\right)^2}{\tilde{J}^2}\right)
\right),
\nonumber\\
{\rm{i}}S_{xy}(k,\omega)
=\frac{\sqrt{\pi}}{4\tilde{J}}
\left(
\exp\left(-\frac{\left(\omega+h\right)^2}{\tilde{J}^2}\right)
-
\exp\left(-\frac{\left(\omega-h\right)^2}{\tilde{J}^2}\right)
\right),
\label{6.05}
\end{eqnarray}
i.e. in the high-temperature limit
$S_{xx}(k,\omega)$ and $S_{xy}(k,\omega)$
are $k$-independent and display a single Gaussian ridge 
at $\omega=\vert h\vert$.

In  the zero-temperature and strong-field limit
($T=0$, $\vert h\vert>\sqrt{J^2+D^2}$)
according to (\ref{6.04}) and (\ref{4.32}) we find
\begin{eqnarray}
S_{xx}(k,\omega)
=-{\rm{sgn}}(h){\rm{i}}S_{xy}(k,\omega)
\nonumber\\
=\frac{\pi}{2}
\delta\left(\omega-\vert h\vert -\tilde{J}\cos\left(k+{\rm{sgn}}(h)\varphi\right)\right).
\label{6.06}
\end{eqnarray}

\begin{figure}[th]
\centerline{\psfig{file=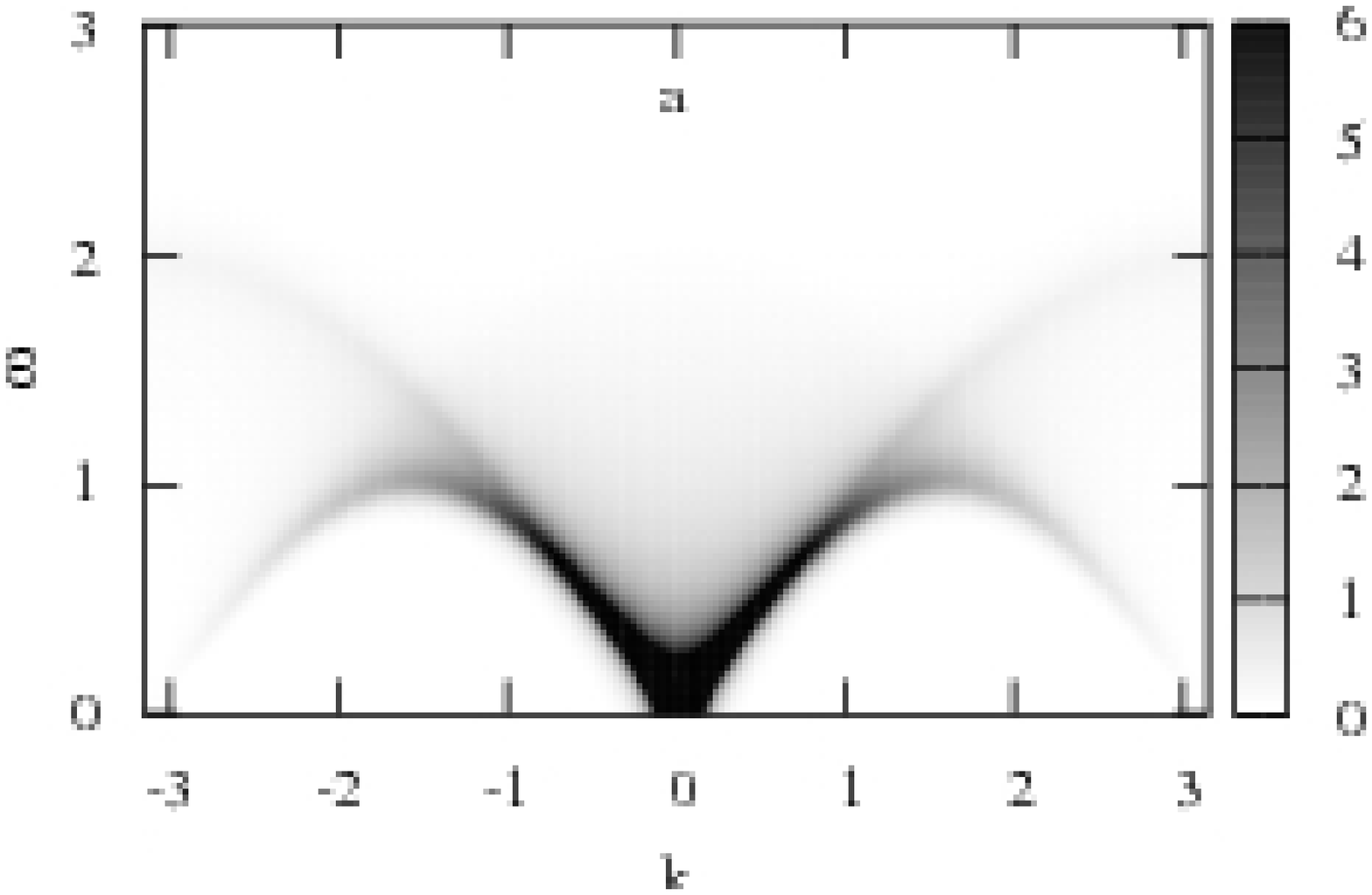,width=2.0in,angle=0}
\psfig{file=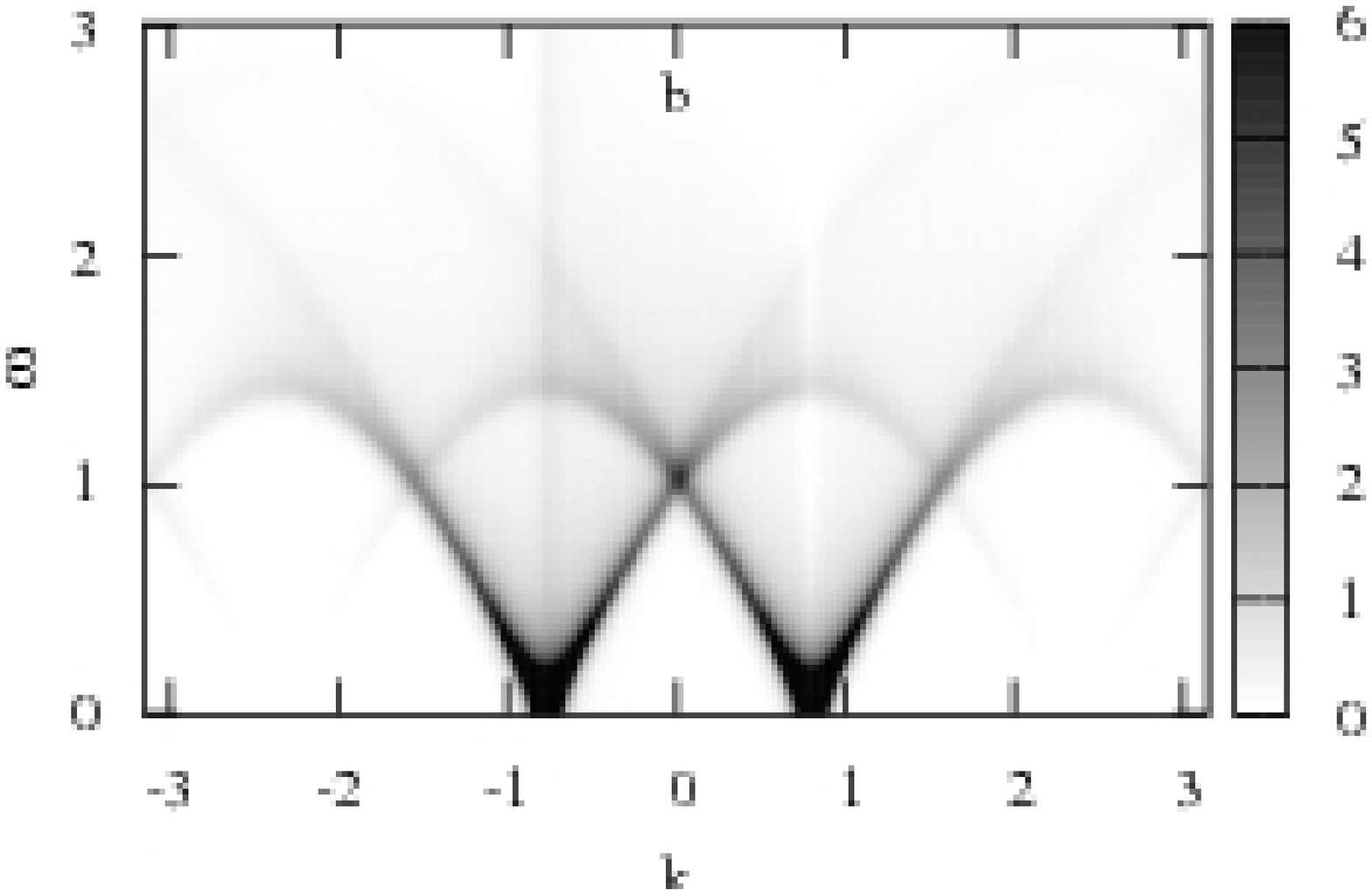,width=2.0in,angle=0}}
\centerline{\psfig{file=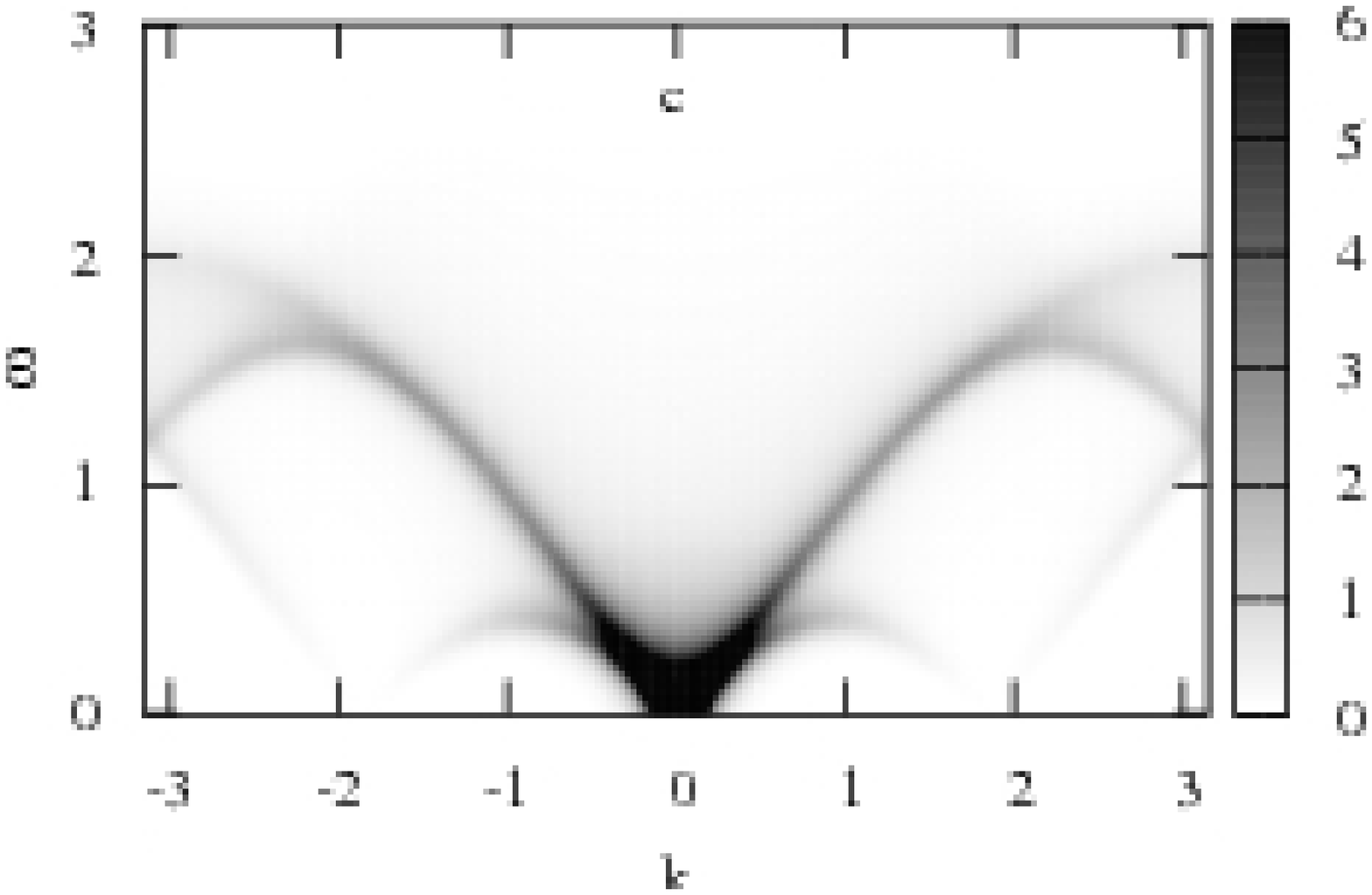,width=2.0in,angle=0}
\psfig{file=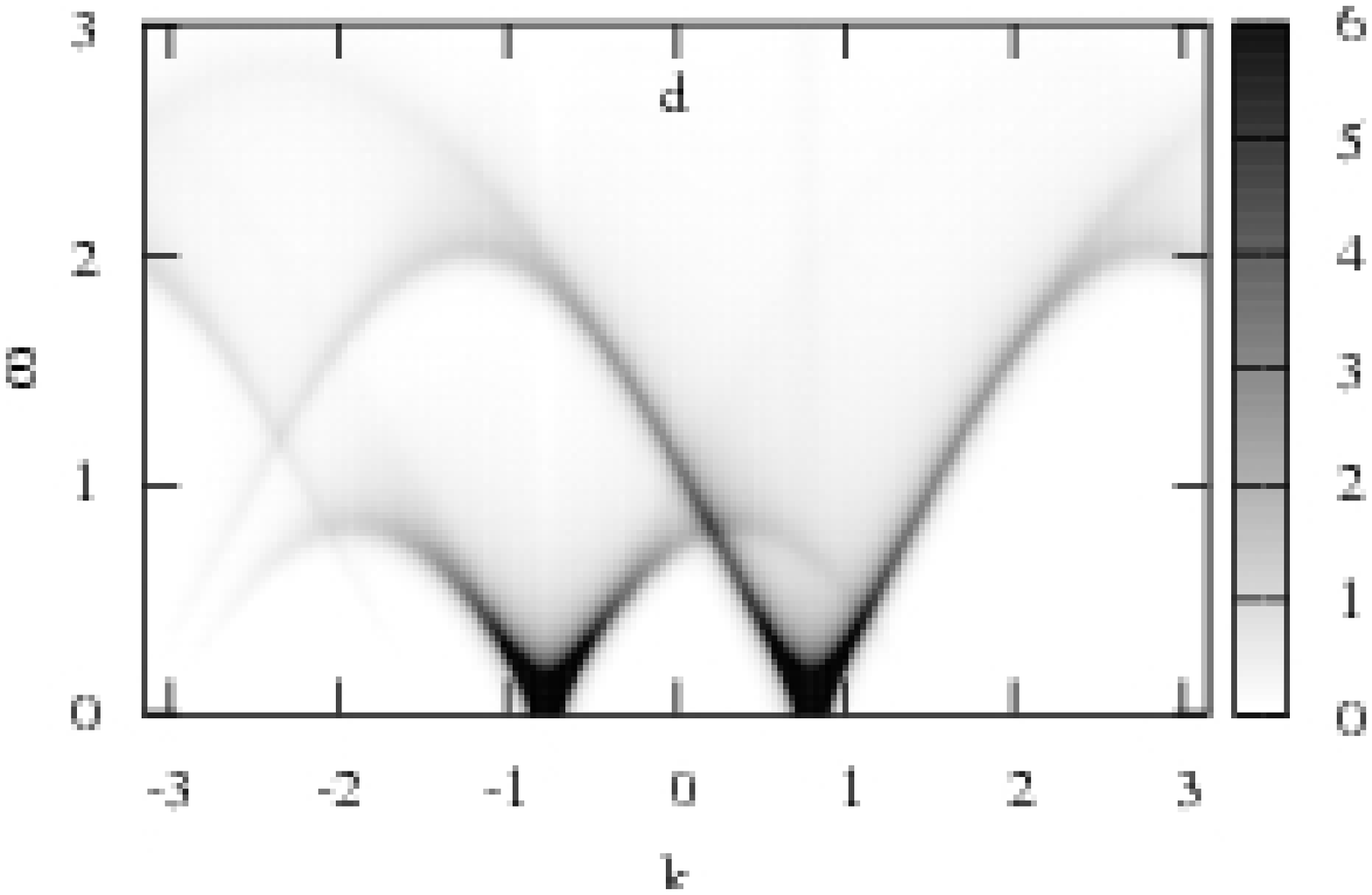,width=2.0in,angle=0}}
\vspace*{8pt}
\caption{$S_{xx}(k,\omega)$ (gray-scale plots) for the model (\ref{6.01})
with $J=-1$,
$D=0$ (left panels a, c) and $D=1$ (right panels b, d)
for $h=0$ (upper panels a, b) and $h=-0.6$ (lower panels c, d)
at low temperature $\beta=20$.}
\label{fig18}
\end{figure}

For arbitrary values of temperature and transverse field we use Eq. (\ref{6.04})
and numerical results
for the $xx$ and $xy$ dynamic structure factors
of the transverse $XX$ chain (\ref{6.03})
(see Sec.~\ref{secdk4})
to reveal the effect of the Dzyaloshinskii-Moriya interaction.
Some of our findings are plotted in Fig.~\ref{fig18}
where we show $S_{xx}(k,\omega)$ for $D=0$ (left panels) and $D\ne 0$ (right panels)
at different values of the transverse field $h$.

We recall that in the low-temperature limit
when $J<0$ and $D=0$
$S_{xx}(k,\omega)$ and $S_{xy}(k,\omega)$ are concentrated in the $k$--$\omega$ plane
along the curves (\ref{4.05}), (\ref{4.06}), (\ref{4.07})
which determine the boundaries of the two-fermion excitation continuum
$\omega_l(k)$, $\omega_m(k)$ and $\omega_u(k)$ (see Sec.~\ref{secdk4}, Figs.~\ref{fig10},~\ref{fig11}).
[For the antiferromagnetic sign of $XX$ exchange interaction $J>0$
these dynamic quantities
are concentrated along the curves
$\omega_l(k\pm\pi)$,
$\omega_m(k\pm\pi)$
and $\omega_u(k\pm\pi)$
as it follows from simple symmetry arguments.]
In the case when the Dzyaloshinskii-Moriya interaction is present,
$D\ne 0$,
the two-fermion excitation continuum splits into two continua
(see Fig.~\ref{fig18}),
the `left' one
with the boundaries
$\omega_l(k-\varphi)$, $\omega_m(k-\varphi)$ and $\omega_u(k-\varphi)$
and
the `right' one
with the boundaries
$\omega_l(k+\varphi)$, $\omega_m(k+\varphi)$ and $\omega_u(k+\varphi)$.
(The `left' and the `right' continua are connected by symmetry operation.)
The larger $D$ is the larger is the splitting controlled by $\varphi=\arctan(D/J)$.
At fixed $D\ne 0$ and $h=0$ the spectral weight is equally distributed
between the left and the right continua
(panel b in Fig.~\ref{fig18}).
While $\vert h\vert$ increases from 0 to $\sqrt{J^2+D^2}$
the spectral weight moves from one continuum to another continuum
(panel d in Fig.~\ref{fig18}).

We note in passing
that the discussed peculiarities of the $xx$ dynamic structure factor
may be used for an unambiguous determination of the Dzyaloshinskii-Moriya interaction
in chain compounds
for example, in resonance experiments\cite{058,048,057,059}.

In the case of the anisotropic $XY$ chain
the Dzyaloshinskii-Moriya interaction cannot be eliminated by the transformation (\ref{6.02}).
Now we face the Hamiltonian
\begin{eqnarray}
H=\sum_n
\left(
J^xs_n^xs_{n+1}^x+J^ys_n^ys_{n+1}^y
+
D\left(s_n^xs_{n+1}^y-s_n^ys_{n+1}^x\right)
\right)
+\Omega\sum_ns_n^z
\nonumber\\
\to
\sum_n
\left(
\frac{J+{\rm{i}}D}{2}c_n^{\dagger}c_{n+1}
-
\frac{J-{\rm{i}}D}{2}c_nc_{n+1}^{\dagger}
+\frac{\gamma}{2}\left(c_n^{\dagger}c_{n+1}^{\dagger}-c_nc_{n+1}\right)
\right.
\nonumber\\
\left.
+\Omega\left(c_n^{\dagger}c_n-\frac{1}{2}\right)
\right)
\label{6.07}
\end{eqnarray}
with $J=(J^x+J^y)/2$ and $\gamma=(J^x-J^y)/2$.
This Hamiltonian can be put into a diagonal form
by performing the Fourier transformation,
$c_k=\left(1/\sqrt{N}\right)\sum_n\exp\left({\rm{i}}kn\right) c_n$,
$c_n=\left(1/\sqrt{N}\right)\sum_k\exp\left(-{\rm{i}}kn\right) c_k$,
and the Bogolyubov transformation,
$c_k=-{\rm{i}}u_{k}\beta_k+v_k\beta_{-k}^{\dagger}$,
$\beta_k={\rm{i}}u_kc_k+v_kc_{-k}^{\dagger}$,
$u_k={\rm{sgn}}\left(\gamma\sin k\right)\left(1/\sqrt{2}\right)
\sqrt{1+\left(\Omega+J\cos k\right)/\lambda_k}$,
$v_k=\left(1/\sqrt{2}\right)\sqrt{1-\left(\Omega+J\cos k\right)/\lambda_k}$,
$\lambda_k=\sqrt{\left(\Omega+J\cos k\right)^2+\gamma^2\sin^2k}$.
The final result reads
$H=\sum_k\Lambda_k\left(\beta_k^{\dagger}\beta_k-1/2\right)$,
$\Lambda_k=D\sin k+\lambda_k$.
We notice that the elementary excitation energy spectrum is gapless
when $\Omega^2\le J^2+D^2-\gamma^2$ and $\gamma^2\le D^2$
or
when $\Omega^2=J^2$ and $\gamma^2>D^2$.

\begin{figure}[th]
\centerline{\psfig{file=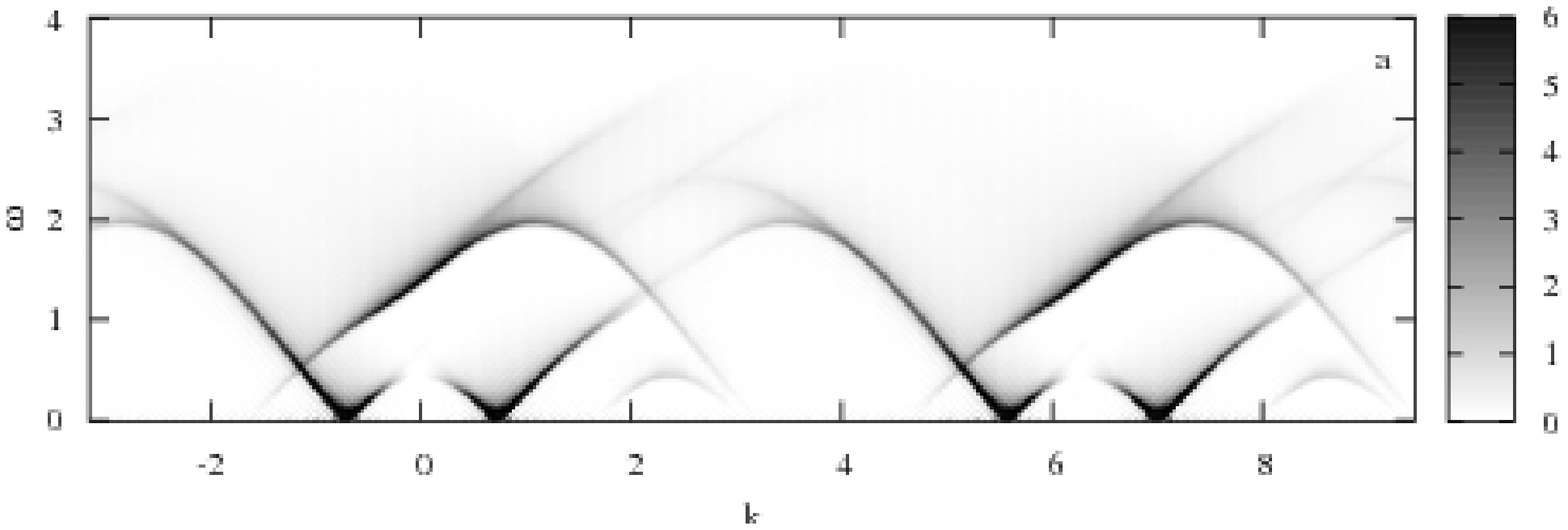,width=4.0in,angle=0}}
\centerline{\psfig{file=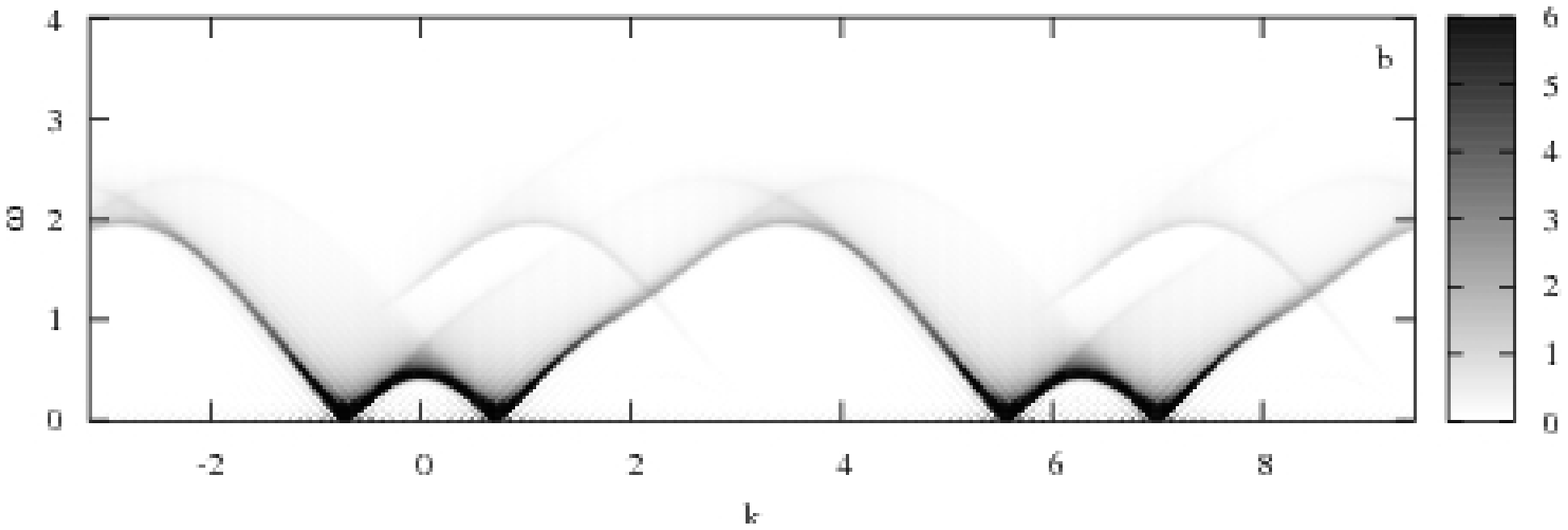,width=4.0in,angle=0}}
\centerline{\psfig{file=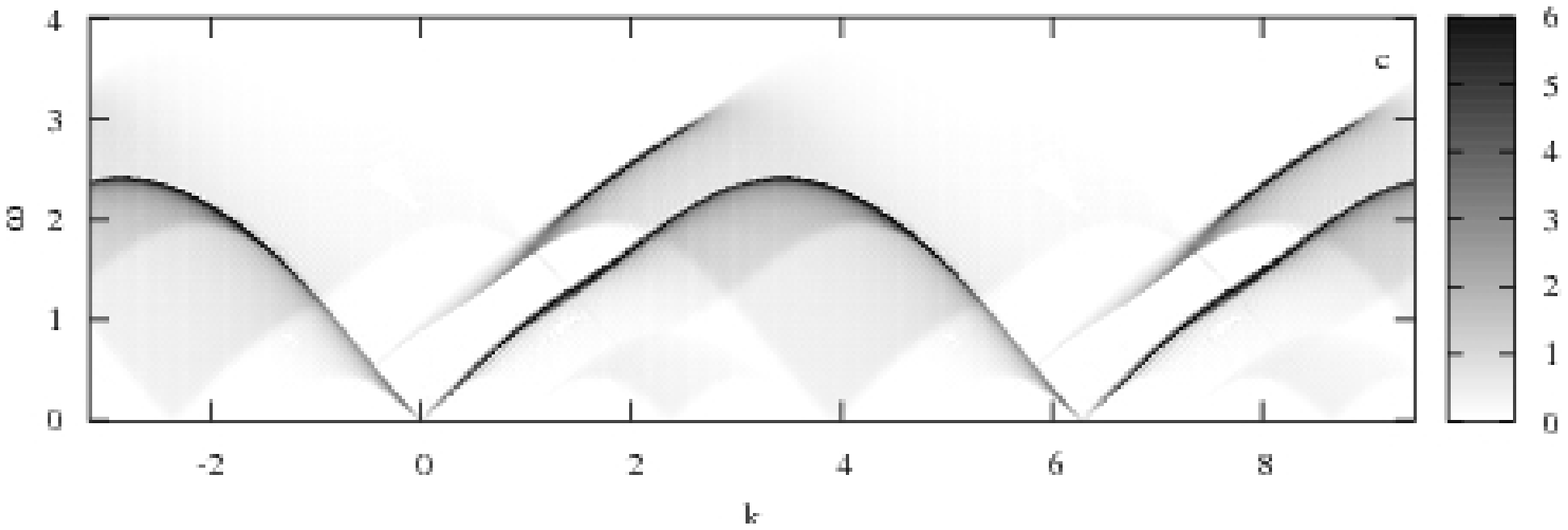,width=4.0in,angle=0}}
\vspace*{8pt}
\caption{$S_{xx}(k,\omega)$ (a),
$S_{yy}(k,\omega)$ (b),
$S_{zz}(k,\omega)$ (c)
for the spin chain (\ref{6.07})
with $J=-1$, $\gamma=0.5$, $D=1$, $\Omega=0.5$
at low temperature $\beta=50$.
Note that these quantities are shown for $k$ that varies from $-\pi$ to $3\pi$.}
\label{fig19}
\end{figure}

The calculation of the transverse dynamic structure factor
follows the lines explained in some detail in Sec.~\ref{secdk4}
and ends up with\cite{060}
\begin{eqnarray}
S_{zz}(k,\omega)=\sum_{j=1}^3S_{zz}^{(j)}(k,\omega),
\nonumber\\
S_{zz}^{(j)}(k,\omega)
=
\int_{-\pi}^{\pi}{\rm{d}}k_1B^{(j)}(k_1,k)C^{(j)}(k_1,k)
\delta\left(\omega-E^{(j)}(k_1,k)\right),
\nonumber\\
B^{(1)}(k_1,k)=B^{(3)}(k_1,k)=\frac{1-f(k_1,k)}{4},
\;\;\;
B^{(2)}(k_1,k)=\frac{1+f(k_1,k)}{2},
\nonumber\\
f(k_1,k)
=
\nonumber\\
\frac{\left(\Omega+J\cos\left(k_1-\frac{k}{2}\right)\right)
\left(\Omega+J\cos\left(k_1+\frac{k}{2}\right)\right)
-\gamma^2\sin\left(k_1-\frac{k}{2}\right)\sin\left(k_1+\frac{k}{2}\right)}
{\lambda_{k_1-\frac{k}{2}}\lambda_{k_1+\frac{k}{2}}},
\nonumber\\
C^{(1)}(k_1,k)=\left(1-n_{k_1+\frac{k}{2}}\right)\left(1-n_{-k_1+\frac{k}{2}}\right),
\nonumber\\
C^{(2)}(k_1,k)=\left(1-n_{k_1+\frac{k}{2}}\right)n_{k_1-\frac{k}{2}},
\nonumber\\
C^{(3)}(k_1,k)=n_{k_1-\frac{k}{2}}n_{-k_1-\frac{k}{2}},
\nonumber\\
E^{(1)}(k_1,k)=\Lambda_{k_1+\frac{k}{2}}+\Lambda_{-k_1+\frac{k}{2}},
\nonumber\\
E^{(2)}(k_1,k)=\Lambda_{k_1+\frac{k}{2}}-\Lambda_{k_1-\frac{k}{2}},
\nonumber\\
E^{(3)}(k_1,k)=-\Lambda_{k_1-\frac{k}{2}}-\Lambda_{-k_1-\frac{k}{2}}.
\label{6.08}
\end{eqnarray}
The transverse dynamic factor,
as it follows from Eq. (\ref{6.08}),
is shown in panel c in Fig.~\ref{fig19}
for a typical set of parameters.

\begin{figure}[th]
\centerline{\psfig{file=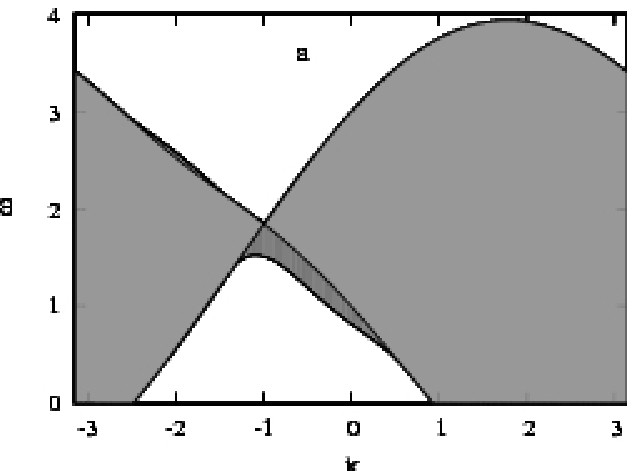,width=2.0in,angle=0}
\psfig{file=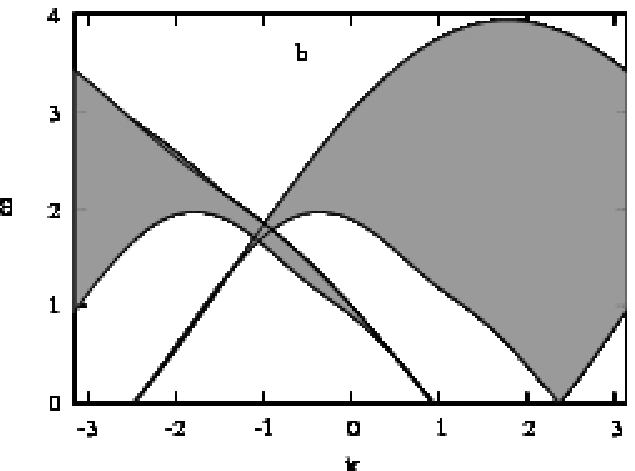,width=2.0in,angle=0}}
\centerline{\psfig{file=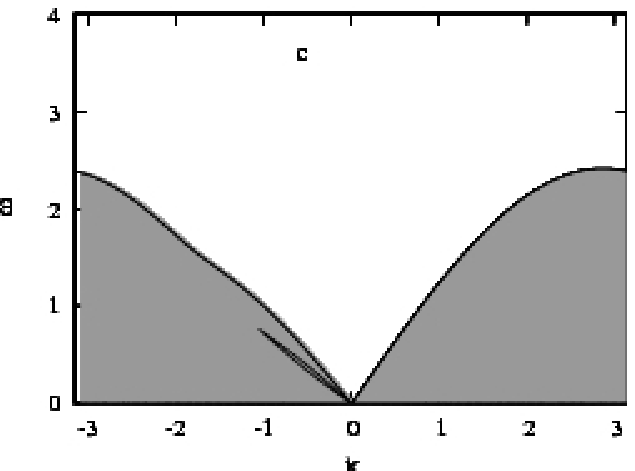,width=2.0in,angle=0}
\psfig{file=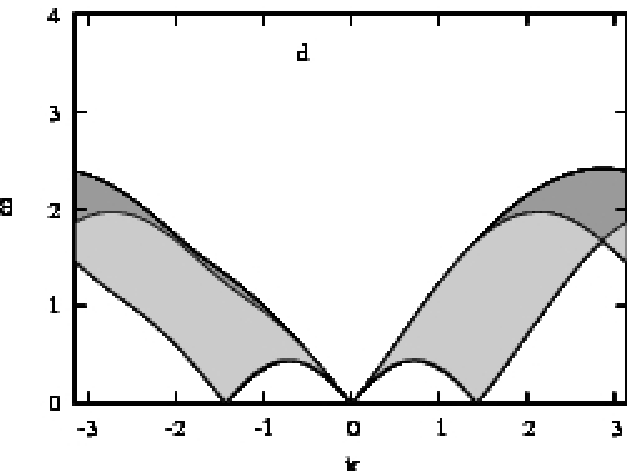,width=2.0in,angle=0}}
\centerline{\psfig{file=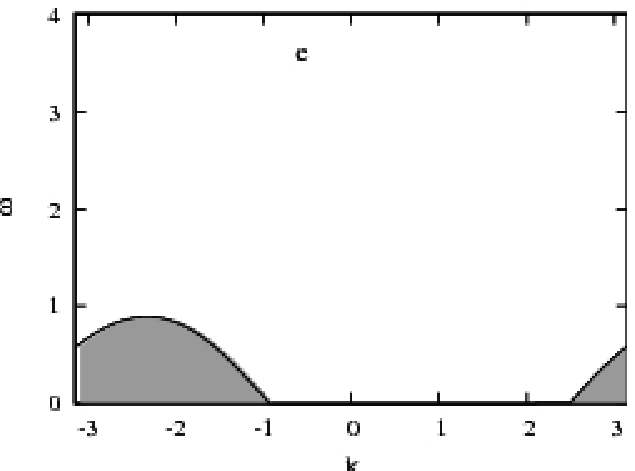,width=2.0in,angle=0}
\psfig{file=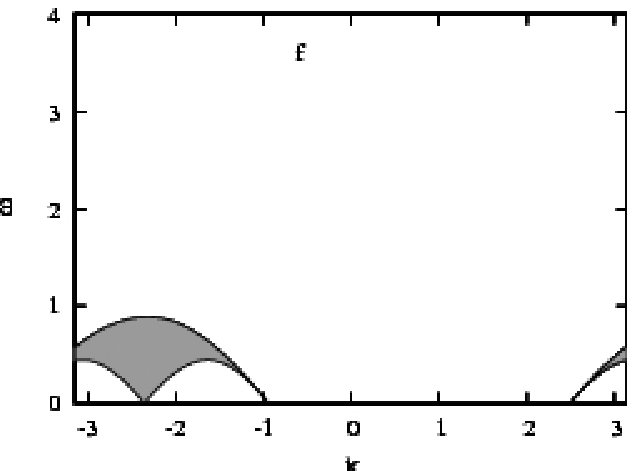,width=2.0in,angle=0}}
\vspace*{8pt}
\caption{Two-fermion excitation continua
($j=1$ (a, b), $j=2$ (c, d), $j=3$ (e, f))
for $J=-1$, $\gamma=0.5$, $D=1$, $\Omega=0.5$.
Left panels: $T\to\infty$;
right panels: $T=0$.}
\label{fig20}
\end{figure}

From Eq. (\ref{6.08}) we see that the transverse dynamic structure factor
is governed by three two-fermion excitation continua.
Let us discuss some properties of these continua
(see Fig.~\ref{fig20}
where we show two-fermion excitation continua for a specific set of parameters
$J=-1$, $\gamma=0.5$, $D=1$, $\Omega=0.5$).
We begin with the high-temperature limit
when the Fermi factors are not essential
(left panels in Fig.~\ref{fig20}).
The two-fermion dynamic structure factor may have nonzero values in the $k$--$\omega$ plane
if the equation $\omega-E^{(j)}(k_1,k)=0$ has at least one solution
$k_1^{\star}$, $-\pi \le k_1^{\star}<\pi$.
Next,
the lower and the upper boundaries are given by
\begin{eqnarray}
\omega_{l}^{(j)}(k)
=\min_{-\pi\le k_1<\pi}
\left\{0,E^{(j)}(k_1,k)\right\},
\nonumber\\
\omega_{u}^{(j)}(k)
=\max_{-\pi\le k_1<\pi}
\left\{E^{(j)}(k_1,k)\right\}.
\label{6.09}
\end{eqnarray}
The two-fermion  dynamic quantities may exhibit van Hove singularities
along the line $\omega^{(j)}_s(k)=E^{(j)}(k_1,k)$
where $k_1$ satisfies the equation
\begin{eqnarray}
\frac{\partial}{\partial k_1}E^{(j)}(k_1,k)
=0.
\label{6.10}
\end{eqnarray}
If for the solution of Eq. (\ref{6.10})
we also have
$\partial^2 E^{(j)}(k_1,k)/\partial k_1^2\ne 0$
the two-fermion dynamic structure factor
shows the well-known square-root singularity.
However, it may happen that
$\partial^2 E^{(j)}(k_1,k)/\partial k_1^2= 0$
but
$\partial^3 E^{(j)}(k_1,k)/\partial k_1^3\ne 0$.
Then a van Hove singularity is characterized by the exponent 2/3.
That is really the case,
for example,
for $J=1$, $\gamma=0.5$,  $D=1$, $\Omega=0.5$
for $j=2$ at $k=1.0784\ldots$.
We have
$\partial E^{(2)}(k_1,k)/\partial k_1
=
\partial^2 E^{(2)}(k_1,k)/\partial k_1^2= 0$,
$\partial^3 E^{(2)}(k_1,k)/\partial k_1^3\ne 0$
at $k_1=k_1^{\star}=2.1648\ldots$.
Therefore in the $\epsilon$-vicinity of $\omega=0.7859\ldots$
the two-fermion dynamic structure factor
should be proportional to $\vert\epsilon\vert^{-2/3}$.
We mention that the singularity with this exponent
is also present for $D=0$\cite{060}.

In the zero-temperature case
the effect of the Fermi functions
involved in the $C^{(j)}$-functions
(see Eq. (\ref{6.08}))
becomes important
(right panels in Fig.~\ref{fig20}).
As a result the region of possible values of $k_1$ is contracted.
Further details can be found in Ref.~\cite{060}.

Finally, we mention the role of the $B^{(j)}$-functions 
(see Eq. (\ref{6.08}))
for the two-fermion dynamic structure factors
which are responsible for some specific features of the transverse dynamic structure factor
(compare panels b, d, f in Fig.~\ref{fig20}
and panel c in Fig.~\ref{fig19}).

We pass to the $xx$ and $yy$ dynamic structure factors
(see panels a and b in Fig.~\ref{fig19}).
These dynamic structure factors are many-fermion dynamic quantities
and although they are not restricted to some region in the $k$--$\omega$ plane, 
they are rather small outside the two-fermion excitation continua
(compare panels a and b with panel c in Fig.~\ref{fig19}).
In the low-temperature limit the $xx$ and $yy$ dynamic structure factors
show several washed-out excitation branches
which are in correspondence with characteristic lines
of the two-fermion excitation continua.
In the high-temperature limit $S_{xx}(k,\omega)$ and $S_{yy}(k,\omega)$
become $k$-independent.

It should be stressed
that the constant frequency or constant wave-vector scans of the dynamic structure factors
clearly manifest the presence of the Dzyaloshinskii-Moriya interaction
and some easily recognized features of these quantities
can be used for determining the Dzyaloshinskii-Moriya interaction.

\section{Square-Lattice Spin-1/2 Isotropic $XY$ Model
         \label{secdk7}}

Let us discuss what kind of results for spin models can be obtained in two dimensions
after applying the Jordan-Wigner transformation (Sec.~\ref{secdk3}).
We consider the spin-1/2 isotropic $XY$ model on a spatially anisotropic square lattice
with the Hamiltonian
\begin{eqnarray}
H=\sum_{i=0}^{\infty}\sum_{j=0}^{\infty}
\left(
\frac{J}{2}\left(s_{i,j}^+s_{i+1,j}^- + s_{i,j}^-s_{i+1,j}^+\right)
\right.
\nonumber\\
\left.
+
\frac{J_{\perp}}{2}\left(s_{i,j}^+s_{i,j+1}^- + s_{i,j}^-s_{i,j+1}^+\right)
\right),
\label{7.01}
\end{eqnarray}
where $J$ and $J_{\perp}$ are the $XX$ exchange interactions
in the horizontal and vertical directions.
Our aim is to calculate the transverse dynamic structure factor
\begin{eqnarray}
S_{zz}({\bf{k}},\omega)
=\sum_{p=0}^{\infty}\sum_{q=0}^{\infty}
\exp\left({\rm{i}}\left(k_xp+k_yq\right)\right)
\int_{-\infty}^{\infty}
{\rm{d}}t\exp\left({\rm{i}}\omega t\right)
\nonumber\\
\cdot
\left(
\langle s_{n,m}^z(t)s_{n+p,m+q}^z\rangle
-
\langle s_{n,m}^z\rangle \langle s_{n+p,m+q}^z\rangle
\right).
\label{7.02}
\end{eqnarray}
We notice that the transverse dynamic structure factor $S_{zz}({\bf{k}},\omega)$ (\ref{7.02})
for the spin model (\ref{7.01})
is related to the density-density dynamic structure factor of hard-core bosons on a square-lattice
\cite{061}.

We apply the two-dimensional Jordan-Wigner transformation (\ref{3.01}), (\ref{3.03})
to the spin Hamiltonian (\ref{7.01}).
Moreover, we adopt the mean-field approach for the phase factors
and change the gauge leaving the flux per elementary plaquette $\Phi_0$
to be equal to $\pi$.
As a result we arrive at the Hamiltonian like (\ref{3.07}),
i.e.
\begin{eqnarray}
H=\sum_{i=0}^{\infty}\sum_{j=0}^{\infty}
\left(
\frac{J}{2}\left(-1\right)^{i+j}
\left(d_{i,j}^{\dagger}d_{i+1,j} - d_{i,j}d_{i+1,j}^{\dagger}\right)
\right.
\nonumber\\
\left.
+
\frac{J_{\perp}}{2}\left(d_{i,j}^{\dagger}d_{i,j+1} - d_{i,j}d_{i,j+1}^{\dagger}\right)
\right).
\label{7.03}
\end{eqnarray}
The Hamiltonian (\ref{7.03}) contains the correct results in the one-dimensional limit
when either $J_{\perp}=0$ or $J=0$
(in the former case to recover the one-dimensional Hamiltonian (\ref{2.12}) (with $\Omega=0$)
one has to perform in addition a gauge transformation
$d_{i,j}^{\dagger}=\exp\left({\rm{i}}\pi\psi_i\right)f_{i,j}^{\dagger}$,
$\psi_0=0$,
$\psi_{i+1}=\psi_{i}+i$).
The Hamiltonian (\ref{7.03}) can be diagonalized by performing
1) the Fourier transformation,
$d_{i,j}=\left(1/\sqrt{N_xN_y}\right)
\sum_{k_x,k_y}\exp\left({\rm{i}}\left(k_xi+k_yj\right)\right)d_{k_x,k_y}$,
$k_\alpha=2\pi n_\alpha/N_\alpha$,
$n_\alpha=-N_\alpha/2,-N_\alpha/2+1,\ldots,N_\alpha/2-1$,
$\alpha=x,y$,
$N_x=N_y=\sqrt{N}\to\infty$ is even,
which yields
\begin{eqnarray}
H=\frac{1}{2}
\sum_{{\bf{k}}}
\vert E_{\bf{k}}\vert
\left(
\cos\gamma_{\bf{k}}
\left(b_{\bf{k}}^\dagger b_{\bf{k}}-a_{\bf{k}}^\dagger a_{\bf{k}}\right)
+
{\rm{i}}\sin\gamma_{\bf{k}}
\left(b_{\bf{k}}^\dagger a_{\bf{k}}-a_{\bf{k}}^\dagger b_{\bf{k}}\right)
\right),
\nonumber\\
\vert E_{\bf{k}}\vert=\sqrt{J_{\perp}^2\cos^2k_y+J^2\sin^2k_x},
\nonumber\\
\cos\gamma_{\bf{k}}=\frac{J_{\perp}\cos k_y}{\vert E_{\bf{k}}\vert},
\;\;\;
\sin\gamma_{\bf{k}}=\frac{J\sin k_x}{\vert E_{\bf{k}}\vert}
\label{7.04}
\end{eqnarray}
with
$b_{\bf{k}}=d_{k_x,k_y}$
and
$a_{\bf{k}}=d_{k_x\pm \pi,k_y\pm \pi}$
and
2) the Bogolyubov transformation,
$\alpha_{\bf{k}}=\cos\left(\gamma_{\bf{k}}/2\right)b_{\bf{k}}
+{\rm{i}}\sin\left(\gamma_{\bf{k}}/2\right)a_{\bf{k}}$,
$\beta_{\bf{k}}=\sin\left(\gamma_{\bf{k}}/2\right)b_{\bf{k}}
-{\rm{i}}\cos\left(\gamma_{\bf{k}}/2\right)a_{\bf{k}}$,
which yields
\begin{eqnarray}
H={\sum_{{\bf{k}}}}^{\prime}
\Lambda_{\bf{k}}
\left(
\alpha_{\bf{k}}^{\dagger}\alpha_{\bf{k}}
-
\beta_{\bf{k}}^{\dagger}\beta_{\bf{k}}
\right),
\nonumber\\
\Lambda_{\bf{k}}=\vert E_{\bf{k}}\vert=\sqrt{J_{\perp}^2\cos^2k_y+J^2\sin^2k_x}\ge 0
\label{7.05}
\end{eqnarray}
(the prime near the sum in Eq. (\ref{7.05}) means
that ${\bf {k}}$ varies in the thermodynamic limit in the region
$-\pi\le k_y\le \pi$,
$-\pi+\vert k_y\vert\le k_x\le \pi-\vert k_y\vert$).

The calculation of the transverse dynamic structure factor repeats the steps
elaborated in some detail for the one-dimensional case.
First, we use the Wick-Bloch-de Dominicis theorem
to obtain the $zz$ time-dependent spin correlation function
\begin{eqnarray}
\langle s_{n,m}^z(t)s_{n+p,m+q}^z\rangle-
\langle s_{n,m}^z\rangle\langle s_{n+p,m+q}^z\rangle
\nonumber\\
=
\langle d^{\dagger}_{n,m}(t)d_{n+p,m+q}\rangle
\langle d_{n,m}(t)d_{n+p,m+q}^{\dagger}\rangle,
\nonumber\\
\langle d^{\dagger}_{n,m}(t)d_{n+p,m+q}\rangle
=
\frac{1}{2N}\sum_{\bf{k}}\exp\left({\rm{i}}\left(k_xp+k_yq\right)\right)
\nonumber\\
\cdot
\left(
\left(\cos^2\frac{\gamma_{\bf{k}}}{2}-{\rm{i}}\left(-1\right)^{n+m}
\left((-1)^{p+q}-1\right)\cos\frac{\gamma_{\bf{k}}}{2}\sin\frac{\gamma_{\bf{k}}}{2}
+(-1)^{p+q}\sin^2\frac{\gamma_{\bf{k}}}{2}\right)
\right.
\nonumber\\
\left.
\cdot
n_{\bf{k}}\exp\left({\rm{i}}\Lambda_{\bf{k}}t\right)
\right.
\nonumber\\
\left.
+
\left(\sin^2\frac{\gamma_{\bf{k}}}{2}+{\rm{i}}\left(-1\right)^{n+m}
\left((-1)^{p+q}-1\right)\cos\frac{\gamma_{\bf{k}}}{2}\sin\frac{\gamma_{\bf{k}}}{2}
+(-1)^{p+q}\cos^2\frac{\gamma_{\bf{k}}}{2}\right)
\right.
\nonumber\\
\left.
\cdot
\left(1-n_{\bf{k}}\right)\exp\left({-\rm{i}}\Lambda_{\bf{k}}t\right)
\right),
\nonumber\\
\langle d_{n,m}(t)d_{n+p,m+q}^{\dagger}\rangle
=\frac{1}{2N}\sum_{\bf{k}}\exp\left(-{\rm{i}}\left(k_xp+k_yq\right)\right)
\nonumber\\
\cdot
\left(
\left(\cos^2\frac{\gamma_{\bf{k}}}{2}+{\rm{i}}\left(-1\right)^{n+m}
\left((-1)^{p+q}-1\right)\cos\frac{\gamma_{\bf{k}}}{2}\sin\frac{\gamma_{\bf{k}}}{2}
+(-1)^{p+q}\sin^2\frac{\gamma_{\bf{k}}}{2}\right)
\right.
\nonumber\\
\left.
\cdot
\left(1-n_{\bf{k}}\right)\exp\left(-{\rm{i}}\Lambda_{\bf{k}}t\right)
\right.
\nonumber\\
\left.
+
\left(\sin^2\frac{\gamma_{\bf{k}}}{2}-{\rm{i}}\left(-1\right)^{n+m}
\left((-1)^{p+q}-1\right)\cos\frac{\gamma_{\bf{k}}}{2}\sin\frac{\gamma_{\bf{k}}}{2}
+(-1)^{p+q}\cos^2\frac{\gamma_{\bf{k}}}{2}\right)
\right.
\nonumber\\
\left.
\cdot
n_{\bf{k}}\exp\left({\rm{i}}\Lambda_{\bf{k}}t\right)
\right).
\label{7.06}
\end{eqnarray}
Then we plug Eq. (\ref{7.06}) into Eq. (\ref{7.02})
to obtain the following expression for the $zz$ dynamic structure factor
\begin{eqnarray}
S_{zz}({\bf{k}},\omega)
=
\pi\int_{-\pi}^{\pi}\frac{{\rm{d}}k_{1y}}{2\pi}
\int_{-\pi}^{\pi}\frac{{\rm{d}}k_{1x}}{2\pi}
\nonumber\\
\cdot
\left(
\cos^2\frac{\gamma_{{\bf{k}}_1+{\bf{k}}}-\gamma_{{\bf{k}}_1}}{2}
\left(n_{{\bf{k}}_1}\left(1-n_{{\bf{k}}_1+{\bf{k}}}\right)
\delta\left(\omega+\Lambda_{{\bf{k}}_1}-\Lambda_{{\bf{k}}_1+{\bf{k}}}\right)
\right.
\right.
\nonumber\\
\left.
\left.
+\left(1-n_{{\bf{k}}_1}\right)n_{{\bf{k}}_1+{\bf{k}}}
\delta\left(\omega-\Lambda_{{\bf{k}}_1}+\Lambda_{{\bf{k}}_1+{\bf{k}}}\right)
\right)
\right.
\nonumber\\
\left.
+\sin^2\frac{\gamma_{{\bf{k}}_1+{\bf{k}}}-\gamma_{{\bf{k}}_1}}{2}
\left(n_{{\bf{k}}_1}n_{{\bf{k}}_1+{\bf{k}}}
\delta\left(\omega+\Lambda_{{\bf{k}}_1}+\Lambda_{{\bf{k}}_1+{\bf{k}}}\right)
\right.
\right.
\nonumber\\
\left.
\left.
+\left(1-n_{{\bf{k}}_1}\right)\left(1-n_{{\bf{k}}_1+{\bf{k}}}\right)
\delta\left(\omega-\Lambda_{{\bf{k}}_1}-\Lambda_{{\bf{k}}_1+{\bf{k}}}\right)
\right)
\right).
\label{7.07}
\end{eqnarray}
One can easily convince oneself that Eq. (\ref{7.07}) contains the correct result
in the one-dimensional limit (\ref{4.03}) (with $\Omega=0$)
when either $J_\perp=0$ or $J=0$.
In the two-dimensional case Eq. (\ref{7.07}) is an approximate formula
for the transverse dynamic structure factor of the spin-1/2 isotropic $XY$ model
on a spatially anisotropic square lattice.

\begin{figure}[th]
\centerline{\psfig{file=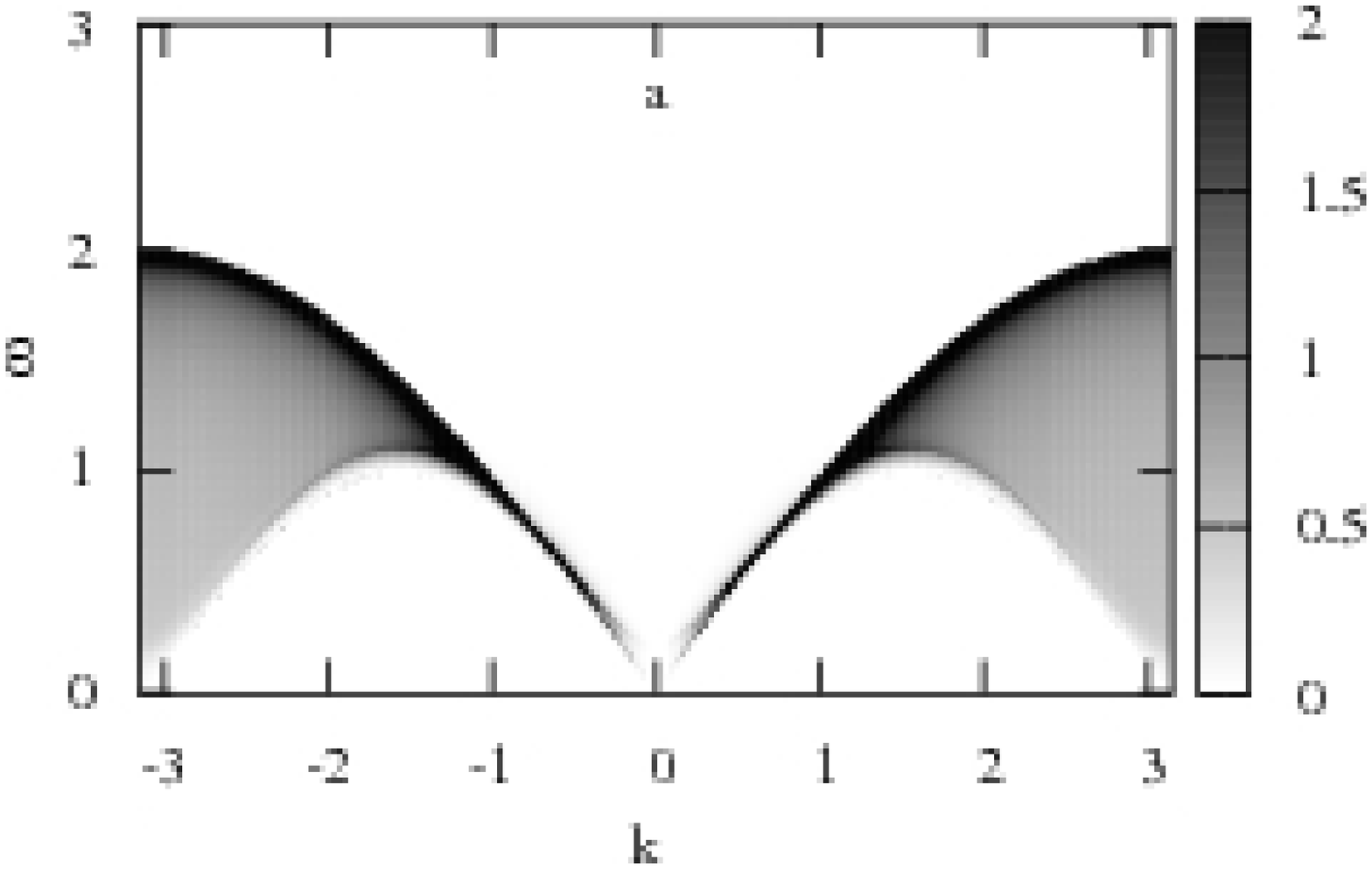,width=2.0in,angle=0}
\psfig{file=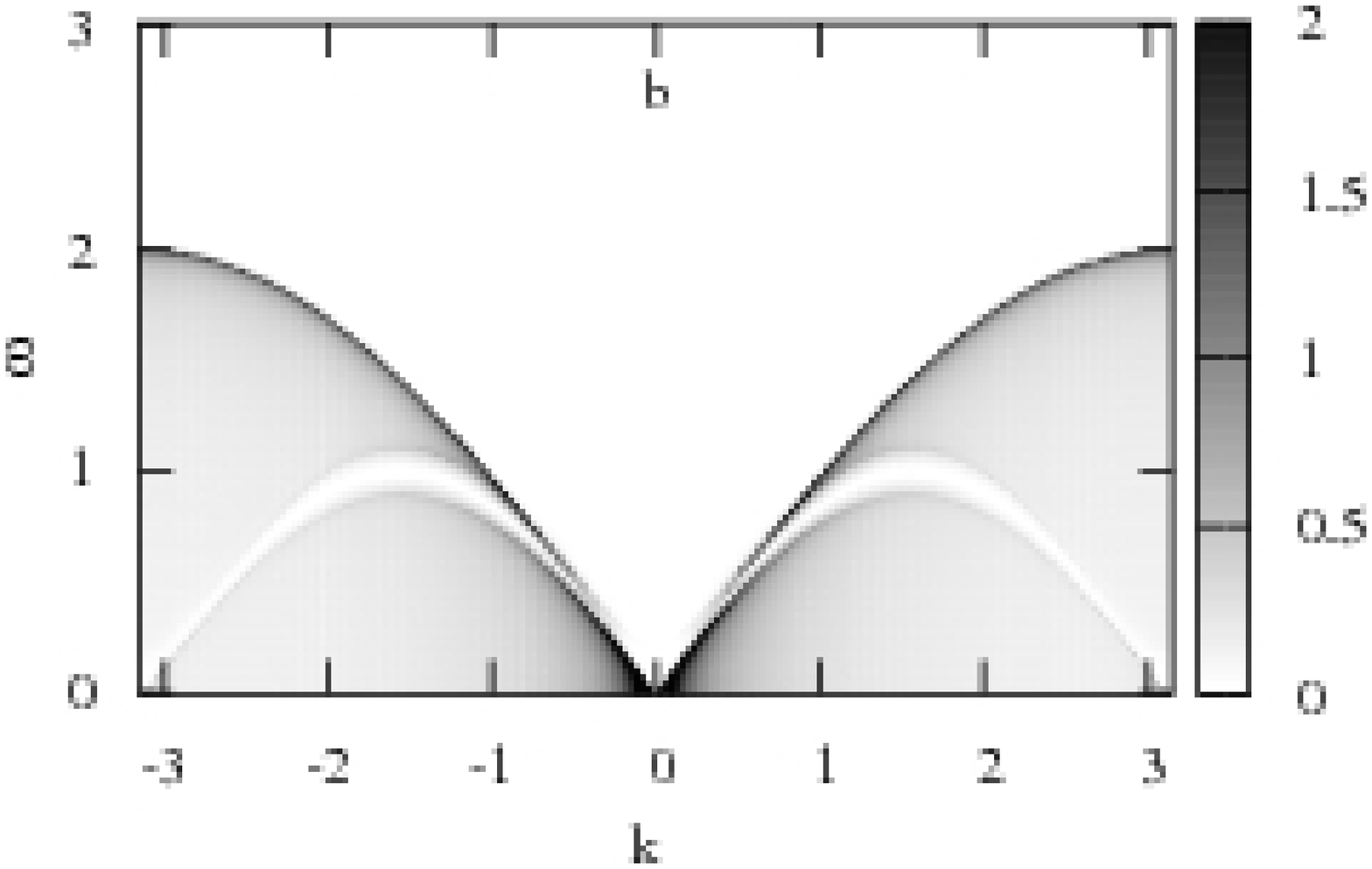,width=2.0in,angle=0}}
\centerline{\psfig{file=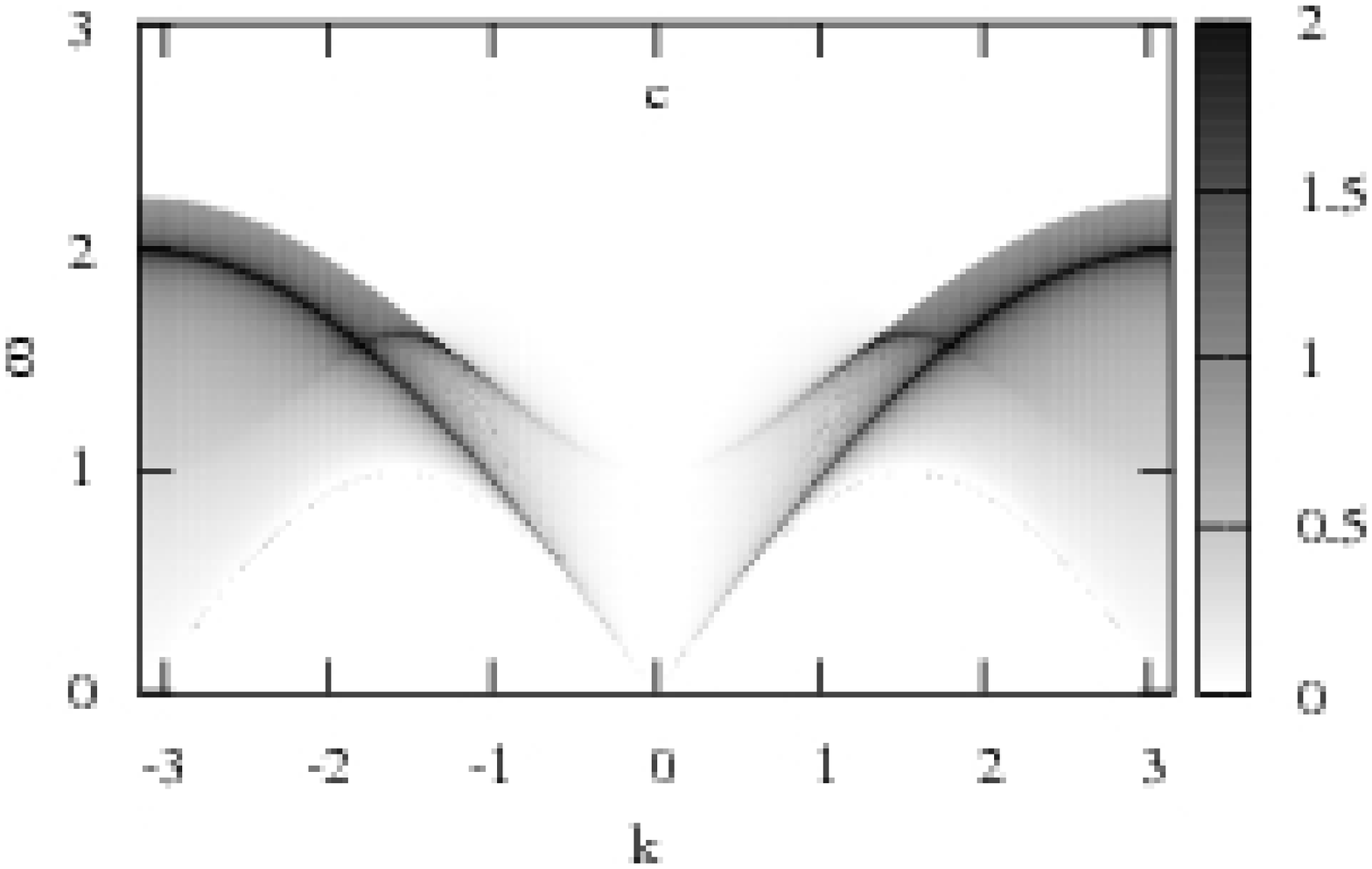,width=2.0in,angle=0}
\psfig{file=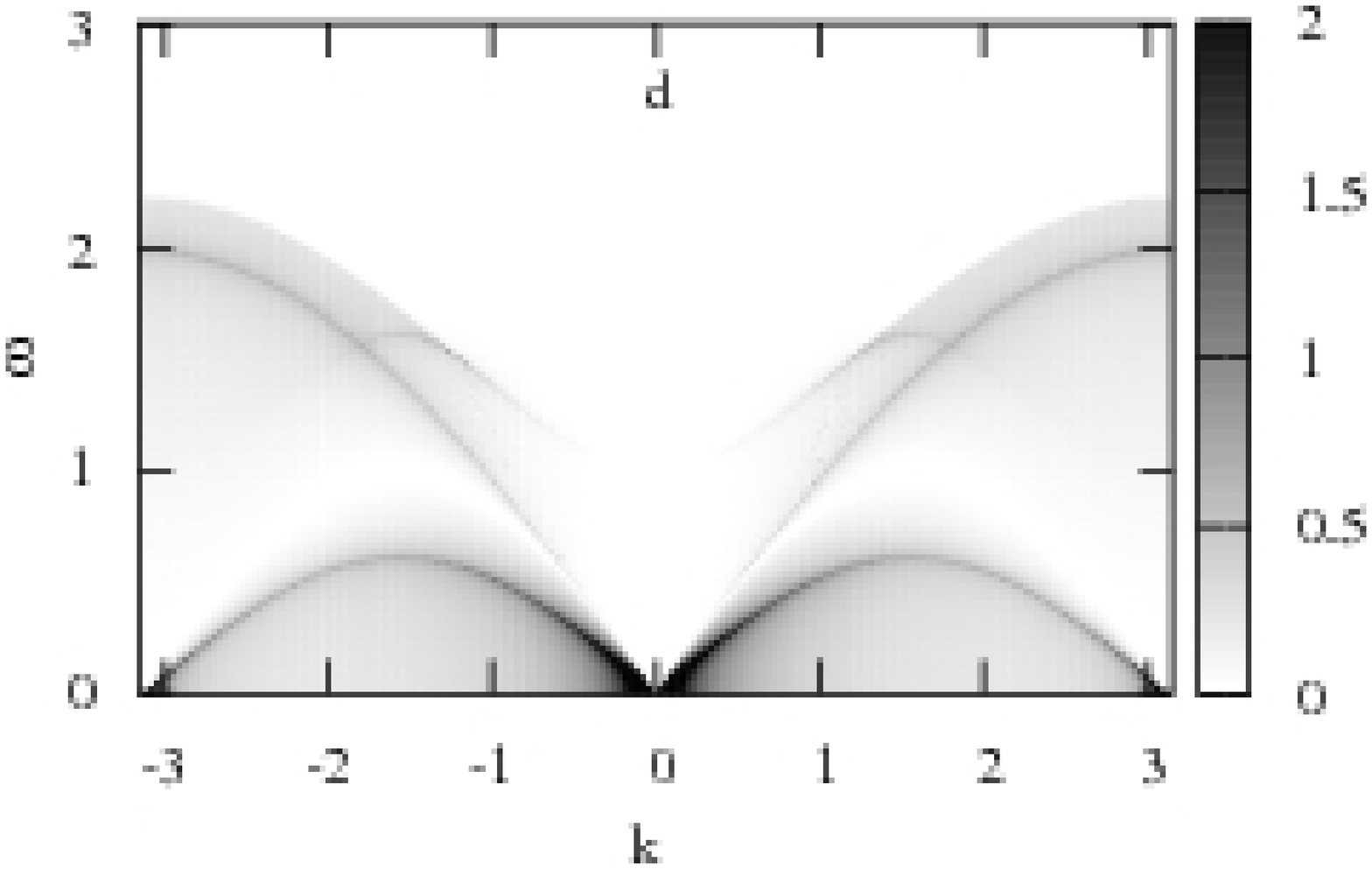,width=2.0in,angle=0}} \vspace*{8pt}
\caption{The $zz$ dynamic structure factor $S_{zz}(k_x,0,\omega)$
(gray-scale plots)
for the square-lattice $s=1/2$ $XX$ model (\ref{7.01})
as it follows from Eq. (\ref{7.07}) at $T=0$ (left column) and at $T=10$ (right column).
$J=-1$, $J_{\perp}=-0.1$ (a, b), $J_{\perp}=-0.5$ (c, d).}
\label{fig21}
\end{figure}

\begin{figure}[th]
\centerline{\psfig{file=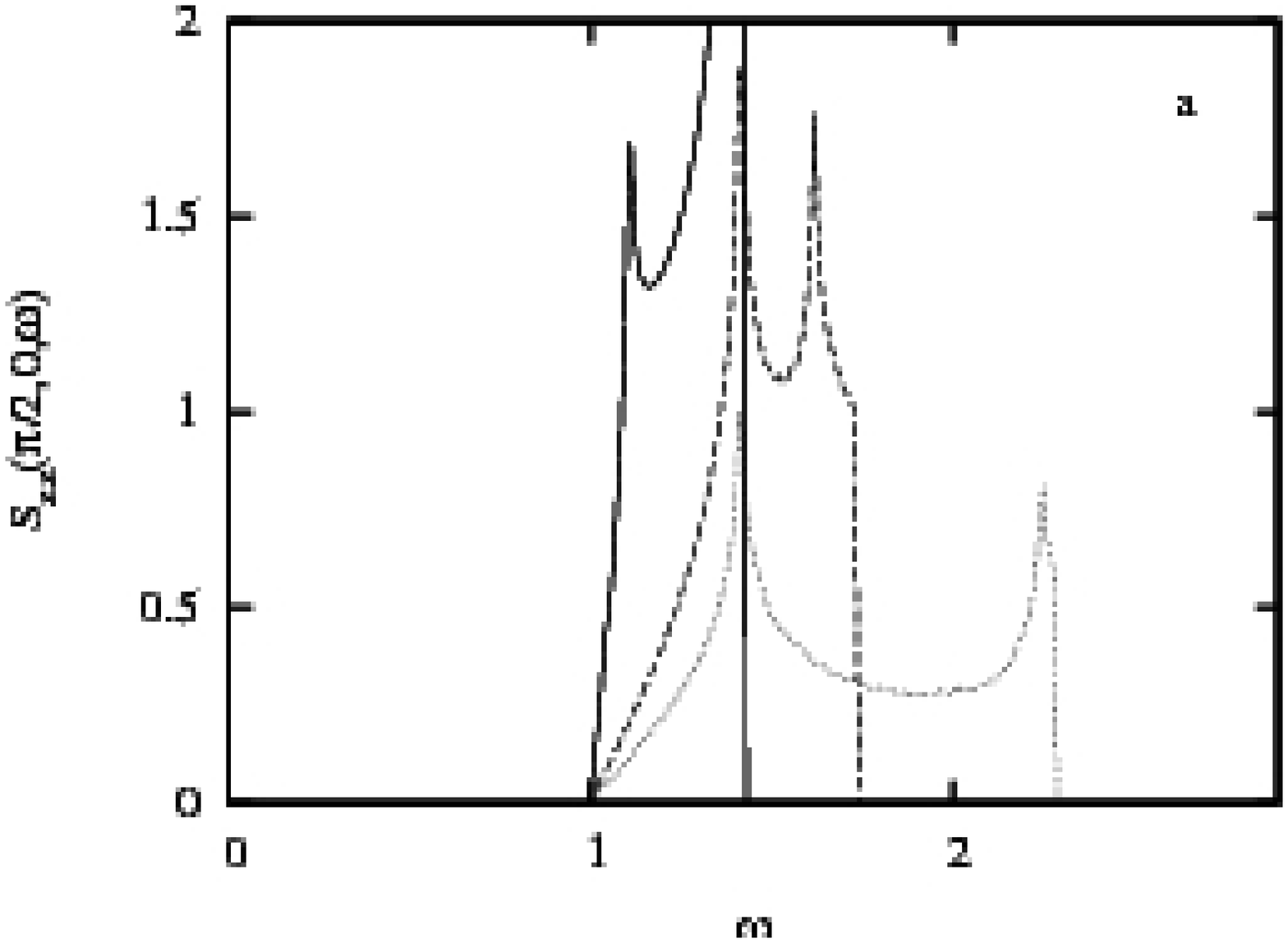,width=2.0in,angle=0}
\psfig{file=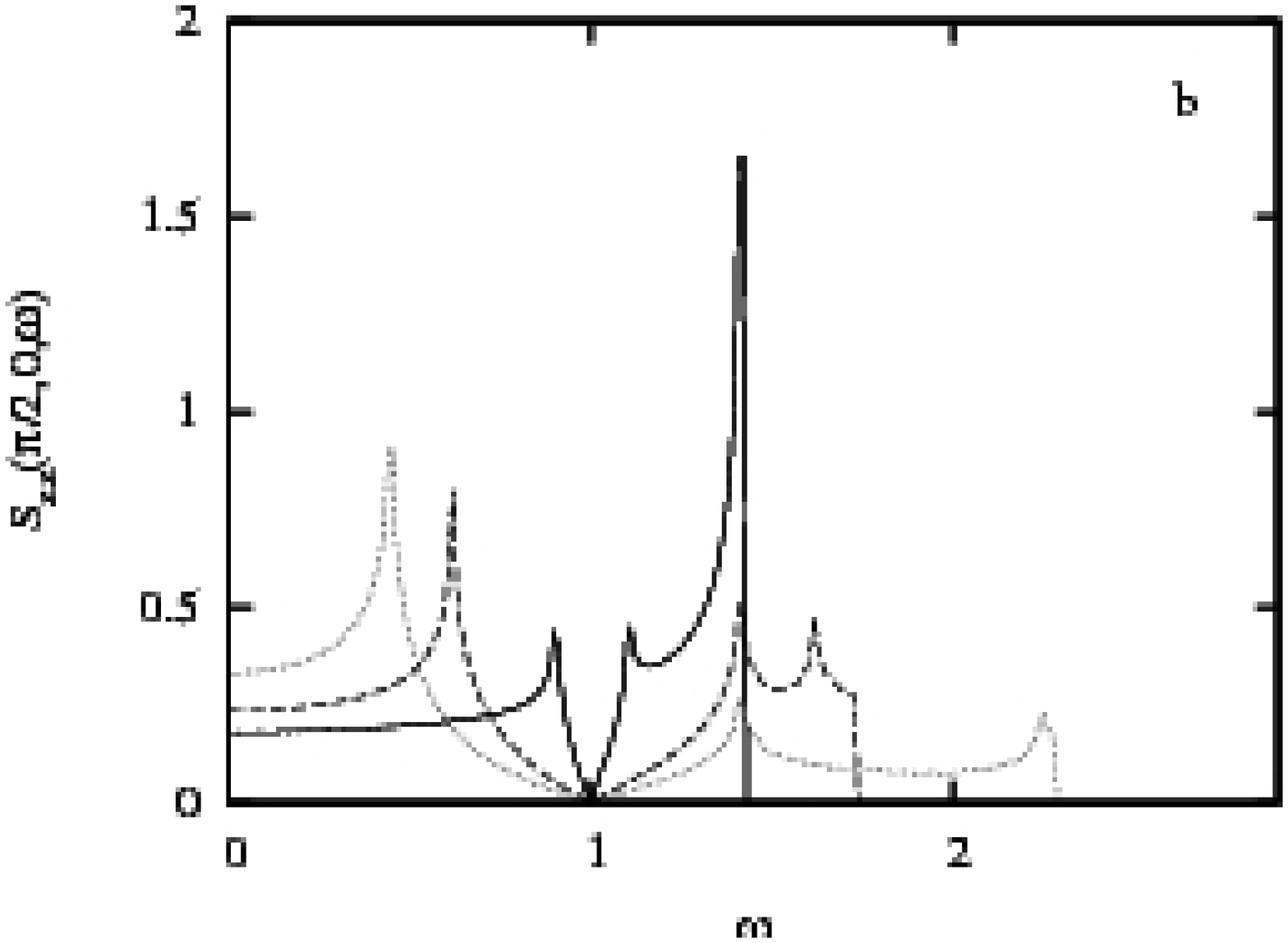,width=2.0in,angle=0}}
\vspace*{8pt}
\caption{Frequency dependence of the $zz$ dynamic structure factor (\ref{7.07})
for momentum transfer along the chain $k_x=\pi/2$
as the interchain interaction changes
($J=-1$, $J_{\perp}=-0.1$ (solid curves),
$J=-1$, $J_{\perp}=-0.5$ (dashed curves),
$J=-1$, $J_{\perp}=-0.9$ (dotted curves))
at zero temperature $T=0$ (a) and high temperature $T=10$ (b).}
\label{fig22}
\end{figure}

Let us discuss the dynamic quantity obtained in some detail\cite{062}.
In Fig.~\ref{fig21}
we show gray-scale plots for $S_{zz}({\bf{k}},\omega)$
and in Fig.~\ref{fig22}
we show frequency profiles of $S_{zz}({\bf{k}},\omega)$
for a representative set of parameters.
Formula (\ref{7.07}) implies the interpretation of $S_{zz}({\bf{k}},\omega)$
as a two-fermion excitation quantity.
As can be seen from Figs.~\ref{fig21},~\ref{fig22}
$S_{zz}({\bf{k}},\omega)$ exhibits several washed-out excitation branches
which can be generated by two spinless fermions in accordance with (\ref{7.07}).
We begin with the low-temperature limit
when only the fourth term in Eq. (\ref{7.07})
(which contains
$\left(1-n_{{\bf{k}}_1}\right)\left(1-n_{{\bf{k}}_1+{\bf{k}}}\right)$)
survives.
Consider, for example,
two fermions with
${\bf{k}}_1=(0,\pi/2)-{\bf{k}}/2$
and
${\bf{k}}_1+{\bf{k}}=(0,\pi/2)+{\bf{k}}/2$
with the energy of the pair
\begin{eqnarray}
\omega_{{\bf{k}}}
=2\sqrt{J^2\sin^2\frac{k_x}{2}+J_{\perp}^2\sin^2\frac{k_y}{2}}
\label{7.08}
\end{eqnarray}
or two fermions with
${\bf{k}}_1=(\pi/2,0)-{\bf{k}}/2$
and
${\bf{k}}_1+{\bf{k}}=(\pi/2,0)+{\bf{k}}/2$
with the energy of the pair
\begin{eqnarray}
\omega_{{\bf{k}}}
=2\sqrt{J^2\cos^2\frac{k_x}{2}+J_{\perp}^2\cos^2\frac{k_y}{2}}.
\label{7.09}
\end{eqnarray}
These modes are the well-known spin waves\cite{063}
clearly present at low temperatures
(panels a and c in Fig.~\ref{fig21}).
Further,
one can recognize the high frequency modes
[${\bf{k}}_1=-{\bf{k}}/2$,
${\bf{k}}_1=(\pi/2,\pi/2)-{\bf{k}}/2$]
with the dispersion relations
\begin{eqnarray}
\omega_{{\bf{k}}}
=2\sqrt{J^2\sin^2\frac{k_x}{2}+J_{\perp}^2\cos^2\frac{k_y}{2}},
\label{7.10}
\\
\omega_{{\bf{k}}}
=2\sqrt{J^2\cos^2\frac{k_x}{2}+J_{\perp}^2\sin^2\frac{k_y}{2}}.
\label{7.11}
\end{eqnarray}
Another set of high-frequency modes
[${\bf{k}}_1=0$,
${\bf{k}}_1=(\pi/2,\pi/2)$]
have the dispersion relations
\begin{eqnarray}
\omega_{{\bf{k}}}
=J_{\perp}
+\sqrt{J^2\sin^2 k_x+J_{\perp}^2\cos^2 k_y},
\label{7.12}
\\
\omega_{{\bf{k}}}
=J
+\sqrt{J^2\cos^2 k_x+J_{\perp}^2\sin^2\ k_y}.
\label{7.13}
\end{eqnarray}
The low-frequency mode
[${\bf{k}}_1=(0,\pi/2)$]
with the dispersion relation
\begin{eqnarray}
\omega_{{\bf{k}}}
=\sqrt{J^2\sin^2k_x+J_{\perp}^2\sin^2k_y}
\label{7.14}
\end{eqnarray}
(it is composed of two fermions, the energy of one of which equals zero)
forms the low-frequency cutoff at zero temperature.
Comparing left and right panels in Figs.~\ref{fig21} and \ref{fig22}
one can also see the modes
which become visible only as temperature increases
(at zero temperature they are forbidden because of the Fermi factors in Eq. (\ref{7.07})).
Putting $k_{1x}=-k_x$, $k_{1y}=\pi/2-k_x$ for $k_y=0$
and $k_{1x}=k_y$, $k_{1y}=\pi/2-k_y$ for $k_x=0$ we get
\begin{eqnarray}
\omega_{k_x,0}
=\left(\sqrt{J^2+J_{\perp}^2}-J_{\perp}\right)\vert\sin k_x\vert,
\nonumber\\
\omega_{0,k_y}
=\left(\sqrt{J^2+J_{\perp}^2}-J\right)\vert\sin k_y\vert.
\label{7.15}
\end{eqnarray}
This excitation branch contains most of the spectral weight at high temperatures
(see panel d in Fig.~\ref{fig21}
and panel b in Fig.~\ref{fig22}).

The established modes
(\ref{7.08}) -- (\ref{7.15})
manifest themselves as peaks, cusps or cutoffs in the frequency or wave-vector profiles
of $S_{zz}({\bf{k}},\omega)$.
The frequency profiles depicted in Fig.~\ref{fig22} 
may be almost symmetric or asymmetric, they may resemble $\delta$-peaks
or result from two coalesced peaks,
they may gradually disappear or may be abruptly cut off.

It is worthwhile to mention here
some experimental studies on dynamic properties of two-dimensional quantum spin models,
in particular, the neutron scattering experiments on Cs$_2$CuCl$_4$\cite{064}
(for a theory of dynamic correlations in the spin-liquid phase in Cs$_2$CuCl$_4$
see Ref.~\cite{065}).
Cs$_2$CuCl$_4$ is a two-dimensional low-exchange quantum magnet.
It has a layered crystal structure;
in each layer the exchange paths form a triangle lattice
with nonequivalent interactions along chains $J=0.374(5)$ meV
and along zig-zag bonds $J^{\prime}=0.34(3)J$.
The interlayer coupling is small $J^{\prime\prime}=0.045(5)J$
and it stabilizes the long-range magnetic order below $T_{\rm{N}}=0.62(1)$ K.
The neutron scattering measurements in the spin-liquid phase
(i.e. above $T_{\rm{N}}$ but below $J$, $J^\prime$ when the two-dimensional magnetic layers are decoupled)
clearly indicate that the dynamic correlations are dominated
by highly dispersive excitation continua
which is a characteristic signature of fractionalization of spin-1 spin waves
into pairs of deconfined spin-1/2 spinons.
Linear spin-wave theory including one- and two-magnon processes cannot describe the continuum scattering.
The proposed theories\cite{065} are based
either on a quasi-one-dimensional approach
(that immediately introduces spinon language)
or on the explicitly two-dimensional resonating-valence-bond picture.

As a final remark we recall that Eq. (\ref{7.07})
contains the exact result (\ref{4.03}) in the one-dimensional limit.
On the other hand, Eq. (\ref{7.07}) gives an approximate result in the two-dimensional case
because of the mean-field description of the phase factors
which arise after fermionization.
The adopted mean-field treatment neglects a complicated interaction between spinless fermions.
In the case of the $XXZ$ Heisenberg model
the interaction between fermions is present even within the adopted mean-field procedure
due to the interaction between $z$ spin components.
The quartic terms in the fermionic Hamiltonian may be treated after making further approximation
(see references cited in Ref.~\cite{027} and also Ref.~\cite{066}).

\section{Conclusions
         \label{secdk8}}

The Jordan-Wigner transformation which realizes a spin-to-fermion mapping
was suggested as a rigorous framework for the description of spin-1/2 $XY$ chains 
in the early 1960s.
In general,
the Jordan-Wigner fermionization permits to map a system of interacting spins $s=1/2$
onto a system of spinless fermions.
It may happen that the spinless fermions are noninteracting.
In this case this approach reveals an exactly solvable spin model.
However, even for exactly solvable spin models
not all `simple' quantities of interest in spin language
remain simple in fermionic language.
For example,
the $z$ spin component attached to the site $j$, $s^z_j$,
becomes the product of two Fermi operators attached to this site, $c_j^{\dagger}c_j-1/2$.
In contrast,
the local spin operators
$s^x_j$, $s_j^y$, $s_j^{\pm}$
become nonlocal objects in fermionic description involving a string of sites $1,2, \ldots, j$
(see Eqs. (\ref{2.08}), (\ref{2.09})).
This leads to some complications in studying the dynamics of fluctuations of these operators:
the dynamics of fluctuations of operators which seem to be rather simple in spin language
may be governed by many-particle correlations in fermionic language.
As we have discussed in sections \ref{secdk4}, \ref{secdk5}, \ref{secdk6},
the Jordan-Wigner fermionization approach permits to establish
a number of rigorous results for the dynamics of spin-1/2 $XY$ chains.
Especially easy are the cases of two- and four-fermion dynamic quantities
which are amenable mostly to analytical calculations.
The case of many-fermion dynamic quantities is more complicated,
however, these quantities can be examined numerically at very high precision.

For more realistic spin-1/2 $XXZ$ Heisenberg chains
the Jordan-Wigner fermionization approach leads to a system of interacting spinless fermions.
The simplest way to proceed in this case
is to apply Hartree-Fock-like approximations\cite{015}.
If we are interested in low-energy physics only
it might be helpful to apply the bosonization approach\cite{016,017,018}.

The results for one-dimensional quantum spin systems obtained using the Jordan-Wigner fermionization
can be compared with the outcomes of alternative approaches:
field-theoretic bosonization techniques\cite{016,017,018} valid in the low-energy limit
(see Fig.~\ref{fig12}),
Bethe ansatz calculations
(for calculation of dynamic structure factors of spin-1/2 $XXZ$ chains
see Refs.~\cite{067,068})
or exact diagonalization computations which, however, are restricted to small finite systems.

For two-dimensional quantum spin  models
achievements are rather modest.
In this case
the Jordan-Wigner fermionization approach can provide an approximate theory;
the simplest one treats in the mean-field spirit the phase factors which arise after fermionization.

We end up with a brief comment
about the experimental relevance of some of the dynamic quantities calculated.
They may be used for interpretation of
the energy absorption in electron spin resonance (ESR) experiments\cite{058}.
Consider an ESR experiment
in which the static magnetic field is directed along the $z$ axis
and the electromagnetic wave with the polarization in $\alpha\perp z$ direction
(say $\alpha=x$)
are applied to a magnetic system which is described as a spin-1/2 $XY$ chain
(ESR experiment in the standard Faraday configuration).
In such an ESR  experiment
one measures the intensity of the radiation absorption $I(\omega)$
as a function of frequency $\omega>0$ of the electromagnetic wave.
Within the linear response theory the absorption intensity is written as
\begin{eqnarray}
I(\omega)\propto \omega\Im \chi_{\alpha\alpha}(0,\omega),
\label{8.01}
\end{eqnarray}
where $\Im\chi_{\alpha\alpha}(0,\omega)$
is the imaginary part of the $\alpha\alpha$ component of the dynamic susceptibility
$\chi_{\alpha\alpha}(k,\omega)$ at zero wave-vector $k=0$.
We notice that
\begin{eqnarray}
\Im \chi_{\alpha\alpha}(0,\omega)
=\frac{1-\exp\left(-\beta\omega\right)}{2}
S_{\alpha\alpha}(0,\omega),
\label{8.02}
\end{eqnarray}
where the dynamic structure factor is defined by Eq. (\ref{1.04}).
Thus,
the peculiarities of the dynamic structure factor $S_{\alpha\alpha}(0,\omega)$
caused,
e.g.,
by the $XY$ exchange interaction anisotropy, Dzyaloshinskii-Moriya interaction or dimerization
should manifest themselves in ESR experiments.
The time-dependent spin correlation functions taken at the same site or at the neighboring sites
manifest themselves in the spin-lattice relaxation rate $1/T_1$
measured by nuclear magnetic resonance (NMR)\cite{059}.

The activity in the field of the Jordan-Wigner fermionization approach
has expanded much over the last few decades.
Despite some limitations,
the Jordan-Wigner fermionization approach has a wide range of applicability.
Particularly attractive
is that it allows one to handle complicated problems of low-dimensional quantum spin systems
armed with relatively simple tools.
It thus seems quite likely that it will continue to be used successfully in the coming years.

\section*{Acknowledgments}
\addcontentsline{toc}{section}{Acknowledgments}

The author would like to thank T.~Krokhmalskii, 
T.~Verkholyak, J.~Stolze, G.~M\"{u}ller and H.~B\"{u}ttner
in collaboration with whom
the study of the dynamics
was performed.
He is grateful to T.~Krokhmalskii for preparing all figures for the paper, 
many interesting conversations and helpful comments and suggestions.
He thanks J.~Stolze and T.~Verkholyak for a critical reading of the manuscript.
NATO support is acknowledged
(the grant reference number CBP.NUKR.CLG 982540,
project ``Dynamic Probes of Low-Dimensional Quantum Magnets'').
The author acknowledges kind hospitality of the Organizers
of the 43rd Karpacz Winter School of Theoretical Physics
``Condensed Matter Physics in the Prime of XXI Century:
Phenomena, Materials, Ideas, Methods''
in L\c{a}dek Zdr\'{o}j in February 2007.

\end{document}